\documentclass[a4paper,11pt]{article}
\pdfoutput=1 
\usepackage{jcappub} 
\usepackage[T1]{fontenc} 
\usepackage{amsmath}
\usepackage{physics}
\usepackage{tikz}
\usetikzlibrary{positioning}
\usepackage{amssymb}
 \usepackage{booktabs}
\usepackage{mathtools}
\usepackage{yhmath}
\usepackage{graphicx}
\usepackage{subcaption}
\usepackage{aas_macros}

\title{\boldmath Parity conservation of the isotropic galaxy 4PCF in redshift space}

\author[a,b,c]{C. Kongprachaya}
\author[a,d,e]{, N. Bartolo}
\author[a,d,e]{, D. Bertacca}
\author[a,d,e,f]{,\\S. Matarrese}

\affiliation[a]{Dipartimento di Fisica e Astronomia “Galileo Galilei”, Università degli Studi di Padova, Via Marzolo 8, I-35131, Padova, Italy}
\affiliation[b]{Dipartimento di Fisica e Astronomia “Augusto Righi”, Università di Bologna, via Gobetti 93/2, I-40129 Bologna, Italy}
\affiliation[c]{INAF -- Istituto di Radioastronomia, Via P.~Gobetti 101, 40129 Bologna, Italy}
\affiliation[d]{INAF, Osservatorio Astronomico di Padova, Vicolo dell’Osservatorio 5, I-35122 Padova, Italy}
\affiliation[e]{INFN, Sezione di Padova, Via Marzolo 8, I-35131 Padova, Italy}
\affiliation[f]{Gran Sasso Science Institute, Viale F. Crispi 7, I-67100 L’Aquila, Italy}

\emailAdd{chanasorn.kongprachaya@inaf.it}
\emailAdd{nicola.bartolo@pd.infn.it}
\emailAdd{daniele.bertacca@unipd.it}
\emailAdd{sabino.matarrese@pd.infn.it}

\abstract{
Recent observational investigation of cosmological Parity Violation (PV) through the galaxy Four-point Correlation Function (4PCF) has motivated a careful theoretical examination of potential systematic contributions. In this work, we demonstrate that parity-odd contributions to the isotropic 4PCF estimator induced by Redshift Space Distortion (RSD), up to the leading order $\mathcal{H}/k$ of general relativistic corrections, vanish upon averaging over all the Line-of-Sight (LOS). We analyse the perturbative relativistic galaxy number density, combining the biased tracer expansion and the dark matter gravitational dynamics within the framework of Standard Eulerian Perturbation Theory (EPT) to third order in the density perturbation. Building on prior studies of the parity-odd mode of the trispectrum (the Fourier-space counterpart of the 4PCF), we extend existing results in three directions: incorporating the wide-angle effect by adopting a local LOS definition for each galaxy in the 4PCF, accounting for tidal bias in the galaxy population, and generalising the assessment to an arbitrary cosmological background. Expanding the 4PCF in spherical harmonics, we show that angular momentum selection rules, encoded in the Wigner 3-j symbols, enforce that only parity-even contributions survive the integration over all LOS orientations, both in the plane-parallel approximation and in the full wide-angle case. This result is crucial for large-scale statistical analyses using isotropic estimators. It supports interpreting possible PV detections as a signal not due to a systematic RSD artefact. We conclude with a discussion of the caveats of our results and identify potential sources of PV.
}

\begin{document}
\maketitle
\flushbottom
\section{Introduction}\label{sec:intro}

The research on Parity Violation (PV) in the galaxy distribution, as a Large-Scale Structure (LSS) probe, was proposed by \cite{Cahn2023PhRvL.130t1002C}, who introduced the Four-point Galaxy Correlation Function (4PCF) as a statistical tool to test parity symmetry on cosmological scales. Subsequently, the same group reported up to $7\sigma$-detection of a non-zero parity-odd mode in the observed 4PCF of the Baryon Oscillation Spectroscopic Survey (BOSS) galaxy catalogue of the Sloan Digital Sky Survey (SDSS) \cite{Hou:2022wfj}. In the same year, \cite{Philcox:2022hkh} also performed an independent analysis on the same dataset and reported a $2.9\sigma$-detection. Such intriguing claims spurred further studies, and the subsequent analyses did not support evidence for parity violation in LSS  \cite{Philcox:2024mmz,Krolewski:2024paz,Adari:2024vkf}. On the other hand, using a sample of the Luminous Red Galaxies (LRGs) from the Dark Energy Spectroscopic Instrument (DESI), the authors of \cite{Slepian:2025kbb} recently found a potential $4{-}10\sigma$ detection. Taken together, these results constitute a promising step for LSS analyses of possible parity-violating signatures on cosmological scales, and they motivate theoretical investigations of inflationary mechanisms that can generate such signals. Since GR conserves parity, any such mechanism must involve physics beyond GR. The simplest and most studied example is Chern-Simons gravity \cite{Lue:1998mq,Jackiw:2003pm} (see \cite{Salvarese} for the first attempt), where a parity-violating term coupled to a scalar field is added to the GR Lagrangian (see also \cite{Bartolo:2017szm_PVGWs} and references therein). In a broader landscape, \cite{Bartolo:2020gsh,Moretti:2024fzb} (see also references therein) study chirality in more general scalar-tensor models, which include parity-violating operators with higher derivatives of the inflaton field. Beyond modified gravity theories, parity-odd correlation functions can also be generated within the Effective Field Theory (EFT) of inflation, where the no-go theorems and yes-go examples are derived in \cite{Cabass:2022rhr}. In addition, \cite{Maresuke2016PhRvD..94h3503S,Liu:2019fag,Niu:2022fki,Fujita2024JCAP...05..127F,Reinhard:2024evr} compute the parity-odd trispectrum arising from the matter sector of Axion inflation (see also, e.g., \cite{Coulton:2023oug} and \cite{Cabass:2022rhr}, for the discussion of other possibilities).

However, before attributing these detections to a fundamental symmetry breaking, it is essential to assess whether observational or systematic effects could mimic a PV signal. For instance, \cite{Krolewski:2024paz} explored the possible bias in the parity-conserving mock data set used in \cite{Hou:2022wfj}. A particularly relevant observational effect in this context is Redshift Space Distortion (RSD), first introduced by \cite{Kaiser:1987qv_RSD}, where the peculiar velocity of the galaxy contributes to the redshift factor via the relativistic Doppler effect. This perturbation on top of the cosmological redshift distorts the position of the galaxy in redshift space, where the magnitude of distortion depends on the projection of its velocity on the observer's Line-of-Sight (LOS). Since the cosine factor from the projection changes sign under parity transformation of the peculiar velocity, this dependence may lead to PV. \cite{Leonard:2024xvy} develops a model for the 4PCF with RSD effects in the plane-parallel limit, using perturbative solutions for dark matter in the Einstein-de Sitter (EdS) universe and a power-law biased tracer expansion. They find that the classical RSD cannot source the parity-odd components of the 4PCF isotropic estimator. This is because the integration over all possible LOS, a crucial step for the isotropic estimator, averages out the parity-odd terms from the RSD effect. This analysis was then extended by considering RSD in the context of GR. \cite{Paul:2024uim, Paul:2024xff, Paul2025arXiv250803249P} computed the RSD galaxy trispectrum (Fourier counterpart of the 4PCF) with the leading-order relativistic correction and concluded that those corrections could induce PV in the observed 4PCF.

In this paper, we examine whether parity-odd signals from the RSD effect with the leading-order GR correction survive the averaging over all LOS or not. In conclusion, we demonstrate that those parity-odd contributions, as computed in \cite{Paul:2024uim, Paul:2024xff, Paul2025arXiv250803249P}, are suppressed upon integration over all LOS orientations, in direct analogy with the classical result of \cite{Leonard:2024xvy}. We further extend this assessment beyond the plane-parallel approximation. Instead of assuming a common LOS for all four points in the 4PCF, we assign each point a {\it local} LOS. Through this definition, LOS depend on the separation between each point in the 4PCF, thereby capturing wide-angle effects. In addition, we adopt a more general bias expansion that includes tidal field contributions at second and third order \cite{McDonald2009JCAP...08..020M}. Throughout, we compute the result on an arbitrary cosmological background.

The paper is organised as follows. In Section~\ref{sec: Synopsis: the Observational Implication and its Validity}, we summarise the key results, their implications for galaxy surveys, and the justification of our analysis. In Section~\ref{sec: The Relativistic Galaxy Number Density: Redshift Space Distortion}, we review the study of the relativistic galaxy number density \cite{DiDio2019JCAP...04..050D,DiDio2020JCAP...09..058D} (see also \cite{Yoo2009PhRvD..80h3514Y,Yoo2010PhRvD..82h3508Y,Bouvin2011PhRvD..84f3505B,Jeong2012PhRvD..85b3504J,Bertacca:2012tp_GRcorrect,Bertacca2014JCAP...09..037B,Gervani:2025wmh}), focusing on the RSD effect. In Section~\ref{sec: Kernel Solutions of the Observed Galaxy Number Density}, we incorporate the Eulerian Perturbation Theory (EPT) solution for the dark matter on an arbitrary cosmological background \cite{Bernardeau2002PhR...367....1B, Takahashi2008PThPh.120..549T, Fasiello2016PhRvD..94f3516F} and the bias expansion for galaxies \cite{McDonald2009JCAP...08..020M} (see Appendices~\ref{sec: Perturbative Solution of the Large-scale Structure} and \ref{sec: The Bias Expansion}) into the relativistic galaxy number density up to third order and analyse the resulting kernel solutions. In Section~\ref{sec: Probing Parity Violation With 4PCF}, we compute the 4PCF isotropic estimator and assess its parity properties both in the plane-parallel limit and in the wide-angle case.
\section{Executive summary and observational implications}\label{sec: Synopsis: the Observational Implication and its Validity}

This section provides a self-contained summary of the main results of the paper, intended for readers who wish to grasp the key findings without following all the technical details. We present the essential steps leading to our main result and discuss its practical implications for real observations. We write the observed galaxy number density contrast in the perturbative regime as a function of redshift ($z$) and the unit vector along the LOS ($\vb*{\hat{n}}$),
\begin{equation}
    \Delta_g(\vb*{\hat{n}},z)=\sum^\infty_{n=1}\Delta_g^{(n)}(\vb*{\hat{n}},z),
\end{equation}
where we show that, up to third order of density perturbation $\delta$, the $n$th-order solution can be written as a linear combination of the kernel integrals,
\begin{equation}
    \Delta_g^{(n)}(\vb*{\hat{n}},z)=\sum^{R_n}_{r=1}\mathcal{Z}_{n,r}(z)\int_{\vb*{k}}e^{-i\vb*{k}\cdot\vb*{\vb*{\hat{n}}}r}\int_{\vb*{k_1}} \dots \int_{\vb*{k_n}}\delta^{(3)}_D\left(\vb*{k}-\sum^n_{\sigma=1}\vb*{k_\sigma}\right)\mathcal{S}_{n,r}(\vb*{k_1},\dots ,\vb*{k_n};\vb*{\hat{n}})\prod^n_{\sigma=1}\delta^{(1)}_{\vb*{k_\sigma}},
\end{equation}
where $\mathcal{S}_{n,r}$ is the $r$th kernel of the $n$th-order solution, while $\mathcal{Z}_{n,r}(z)$ encodes the redshift evolution associated with the kernel $\mathcal{S}_{n,r}$. The kernel $\mathcal{S}_{n,r}$ encodes the mode-mixing and the LOS dependence introduced by the dynamics of dark matter fluid, galaxy bias, and the RSD effect, while $\mathcal{Z}_{n,r}(z)$ collects the comoving distance/redshift-dependent parts of RSD effect and the time-dependent perturbation growth rate of the LSS dynamics.\footnote{Readers can find derivation and examples of $\mathcal{Z}_{n,r}(z)$ and $\mathcal{S}_{n,r}$ in Section~\ref{sec: Kernel Solutions of the Observed Galaxy Number Density}.} $R_n$ is the number of kernels at the $n$th order, and $\delta^{(1)}_{\vb*{k_\sigma}}$ is the initial amplitude of the linear solution of the dark matter density contrast in Fourier space, i.e., the $\vb*{k}$-dependent part of equation~\eqref{eqn: linear ansatz of dm}. In the plane wave factor, $r=r(z)$ is the comoving distance from the observer to the galaxy. We use the shorthand notation $\int_{\vb*{k_\sigma}}\equiv\int\frac{d^3k_\sigma}{(2\pi)^3}$ for the three-dimensional Fourier integral. This expression generalises the relativistic galaxy number density contrast in the weak-field approximation presented in \cite{DiDio2019JCAP...04..050D, DiDio2020JCAP...09..058D} (see also \cite{Yoo2009PhRvD..80h3514Y,Yoo2010PhRvD..82h3508Y,Bouvin2011PhRvD..84f3505B,Jeong2012PhRvD..85b3504J,Bertacca:2012tp_GRcorrect,Bertacca2014JCAP...09..037B, Gervani:2025wmh})\footnote{\cite{DiDio2019JCAP...04..050D, DiDio2020JCAP...09..058D} (and this work) expressed the density perturbation in Poisson gauge, while the other references wrote it in synchronous comoving gauge. Though an accurate bias expansion is in the synchronous comoving gauge \cite{Jeong2012PhRvD..85b3504J}, we can use the expression in the Poisson gauge since the transformation of the density between the two gauges is at the order $(\mathcal{H}/k)^2$ and neglected in this work, as mentioned in \cite{DiDio2019JCAP...04..050D, DiDio2020JCAP...09..058D}.}, where the general relativistic correction is included up to order $\mathcal{H}/k$, neglecting the lensing effect (see \cite{Hui2007PhRvD..76j3502H} and references therein) and the peculiar velocity of the observer. This form is obtained by combining the perturbative solution of the dark matter density contrast $\delta$ and velocity divergence $\theta$ from the Standard Eulerian Perturbation Theory (EPT) \cite{Bernardeau2002PhR...367....1B, Takahashi2008PThPh.120..549T, Fasiello2016PhRvD..94f3516F} with the bias expansion from \cite{McDonald2009JCAP...08..020M}. Throughout this paper, we work with an arbitrary cosmological background where the growth rate $f$ is left as a free function (see Appendices~\ref{sec: Perturbative Solution of the Large-scale Structure} and \ref{sec: The Bias Expansion}). As a result, the $n$th-order observed galaxy number density contrast is a Fourier transform of $\delta^{(1)}_{\vb*{k_\sigma}}$ convolved with the corresponding kernel $\mathcal{S}_{n,r}$, which can be expressed as a polynomial in the following geometrical factors,
\begin{equation}
    \mu_{\vb*{k_\sigma}}\equiv\vb*{\hat{k}_\sigma}\cdot\vb*{\hat{n}},\quad k_\sigma\equiv\norm{\vb*{k_\sigma}}, \quad \kappa_{\rho\sigma}\equiv\vb*{\hat{k}_\rho}\cdot\vb*{\hat{k}_\sigma}=\frac{\vb*{k_\rho}\cdot\vb*{k_\sigma}}{\norm{\vb*{k_\rho}}\norm{\vb*{k_\sigma}}},
\end{equation}
where $\mu_{\vb*{k_\sigma}}$ is the cosine of the angle between the mode vector $\vb*{k_\sigma}$ and LOS, $k_\sigma$ is the magnitude of $\vb*{k_\sigma}$, and $\kappa_{\rho\sigma}$ is the cosine of the angle between the two mode vectors $\vb*{k_\rho}$ and $\vb*{k_\sigma}$. We therefore write a general form of the kernel as
\begin{equation}
    S_{n,r}(\vb*{k_1},\dots,\vb*{k_n};\vb*{\hat{n}})=\sum_{\{\alpha_n\},\{\beta_n\},\{\gamma_{n\neq n}\}} C_{\{\alpha_n\},\{\beta_n\},\{\gamma_{n\neq n}\}} \prod_{\sigma=1}^n(i\mu_{\vb*{k_\sigma}})^{\alpha_\sigma} k_\sigma^{\beta_\sigma}\prod_{\rho\neq\sigma}^n\kappa_{\rho\sigma}^{\gamma_{\rho\sigma}},
\end{equation}
where $\{\alpha_n\},\{\beta_n\},\{\gamma_{n\neq n}\}$ are sets of integers and $C_{\{\alpha_n\},\{\beta_n\},\{\gamma_{n\neq n}\}}$ is a real coefficient assigned to each monomial. The exponents $\alpha_\sigma$ and $\gamma_{\rho\sigma}$ are zero or positive integers, while $\beta_\sigma$ can be negative because some term in the kernel $S_{n,r}$ contains the inverse norm of mode vectors. The indices $\rho$ and $\sigma$ cannot be equal since $\kappa_{\rho\rho}=\vb*{\hat{k}_\rho}\cdot\vb*{\hat{k}_\rho}=1$. Under a parity transformation, the direction of the mode vector flips, $\vb*{k_\sigma}\rightarrow-\vb*{k_\sigma}$. However, we fix the LOS direction, which is defined by the observer, to be the same, $\vb*{\hat{n}}\rightarrow\vb*{\hat{n}}$. Even though $\vb*{\hat{n}}\rightarrow-\vb*{\hat{n}}$ in a general parity transformation, as a vector in configuration space, the practical observation \cite{Hou:2022wfj, Philcox:2022hkh, Slepian:2025kbb} probes and compares the 4PCF and its parity counterpart in the same angular patch on the sky, i.e., the same LOS. Thus, we keep $\vb*{\hat{n}}$ unchanged to demonstrate the PV from the RSD effect in the case where the statistics are probed from one direction ($\vb*{\hat{n}}$) only and not the opposite LOS ($-\vb*{\hat{n}}$).\footnote{See \cite{Bouvin2014PhRvD..89h3535B} as an example of the setup where $\vb*{\hat{n}}$ is fixed. However, this choice of LOS parity property does not affect the result of this paper, where the isotropic 4PCF estimator averages over all LOS.} Therefore, the parity transform of the factor $\mu_{\vb*{k_\sigma}}\equiv\vb*{\hat{k}_\sigma}\cdot\vb*{\hat{n}}$ is $\mu_{\vb*{k_\sigma}}\rightarrow-\mu_{\vb*{k_\sigma}}$. Thus, the factor $\mu_{\vb*{k_\sigma}}$ is parity-odd. Meanwhile, $k_\sigma$ and $\kappa_{\rho\sigma}$ are parity-even. Consequently, any monomial in $S_{n,r}$ with $\sum^n_{\sigma=1}\alpha_\sigma=\text{odd}$ is parity-odd, and parity-even otherwise. For example, the term proportional to $\mu_{\vb*{k_1}}\mu_{\vb*{k_2}}^2$ ($\sum^2_{\sigma=1}\alpha_\sigma=1+2=3$) is parity odd, while the term proportional to $\mu_{\vb*{k_1}}\mu_{\vb*{k_2}}^3$ ($\sum^2_{\sigma=1}\alpha_\sigma=1+3=4$) is parity even. Therefore, the kernel $S_{n,r}$ is written as a sum of the parity-odd and -even contributions. A similar 4PCF kernel definition appears in \cite{Leonard:2024xvy, Paul:2024uim, Paul:2024xff, Paul2025arXiv250803249P}.

With this kernel solution, we can compute a prediction for the observational 4PCF isotropic estimator, $\hat{\zeta}$, through \cite{Philcox2022MNRAS.509.2457P}
\begin{equation}\label{eqn: observational estimator of 4PCF}
    \hat{\zeta}(\vb*{r_1},\vb*{r_2},\vb*{r_3}) = \int \frac{d^3s}{V_s}  \Delta_g(\vb*{s})\Delta_g(\vb*{s}+\vb*{r_1})\Delta_g(\vb*{s}+\vb*{r_2})\Delta_g(\vb*{s}+\vb*{r_3}),
\end{equation}
which estimates the statistics with the spatial average. $V_s$ is the survey volume, $\vb*{s}$ is the comoving position of the reference galaxy, $\vb*{r_1}$, $\vb*{r_2}$, and $\vb*{r_3}$ are the relative positions of three other points in 4PCF with respect to the reference galaxy. This is an observational estimate of the 4PCF used in \cite{Hou:2022wfj}, which relies on the cosmological principle: statistical \emph{homogeneity} and \emph{isotropy}. The homogeneity assumption implies that the measured $\hat{\zeta}$ is independent of the spatial position from which we build the 4PCF configuration, i.e., $\vb*{s}$. The estimated statistics only depend on the relative positions between galaxies, $\vb*{r_1}$, $\vb*{r_2}$, and $\vb*{r_3}$, which are the internal geometry of the 4PCF. The isotropy assumption implies that the measured $\hat{\zeta}$ depends only on the magnitudes $r_1=\norm{\vb*{r_1}}$, $r_2=\norm{\vb*{r_2}}$, $r_3=\norm{\vb*{r_3}}$, and the angles between them; the measured statistic does not depend on the absolute rotation of the 4PCF configuration.

Throughout this paper, we assume that $V_s$ is the total survey volume, which is a spherical shell around the observer with uniform thickness (redshift range) in every line of sight. This is not the case in real galaxy surveys, as the samples are drawn from a masked sky. In particular, the observable sky is restricted by the field of view of the telescope. Some patches are contaminated by, for example, bright foregrounds and fiber collisions, which are removed by a veto mask. Therefore, the survey footprint does not necessarily cover the whole angular space of the sky in practice. Moreover, the radial selection function inherent in the survey also distorts the observed sample distribution. These masks introduce survey edges and regions where the galaxies are under-counted. This effect can be accounted for by using the edge correction matrix \cite{Slepian2015MNRAS.454.4142S,Philcox2022MNRAS.509.2457P}. The correction and potential systematics induced by the survey geometry are assessed in \cite{Hou:2022wfj}. Therefore, we ignore these effects in this work. In addition, we discuss the impact of spherical survey sky incompleteness in Section~\ref{sec: Discussion and Potential Sources of Violation}.

On the other hand, the theoretical prediction fundamentally comes in the form of an ensemble average,
\begin{equation}
    \zeta(\vb*{s};\vb*{r_1},\vb*{r_2},\vb*{r_3}) = \expval{\Delta_g(\vb*{s})\Delta_g(\vb*{s}+\vb*{r_1})\Delta_g(\vb*{s}+\vb*{r_2})\Delta_g(\vb*{s}+\vb*{r_3})},
\end{equation}
where $\expval{.}$ is an average over all statistical realizations of the field, $\delta$. We emphasise that $\Delta_g$ is not a statistically isotropic and homogeneous random field, even though the underlying real-space matter distribution $\delta$ is assumed to be so via the cosmological principle. $\Delta_g$ represents the galaxy distribution as mapped into redshift space, and this mapping is corrupted by the RSD effect. As the contamination depends on the LOS direction, the translational invariance of $\zeta$ is broken. Therefore, the configuration position $\vb*{s}$ is retained in the argument of $\zeta$. Crucially, however, RSD is an observer-relative anisotropy, not a violation of the underlying cosmological principle: it preserves the rotational symmetry of the correlation functions about the observer \cite{Hamilton:1997zq_wideangle}, so its effect on any statistic is fully determined by the angle between $\vb*{r_1}$, $\vb*{r_2}$, $\vb*{r_3}$ and the LOS, rather than by absolute direction in space.

The isotropic estimator $\hat{\zeta}$ is constructed under this idealisation: the survey-volume average over reference points $\vb*{s}$ effectively also averages over the many distinct LOS orientations subtended by different regions of the survey, so that the explicit LOS dependence introduced by RSD is folded into the average rather than appearing as a residual free argument. The estimator is thus only isotropic in the sense that it depends purely on the internal configuration ($\vb*{r_1}$, $\vb*{r_2}$, $\vb*{r_3}$, and the angle between them) after this orientation averaging has been carried out; it does not imply that $\Delta_g$ itself is a statistically isotropic field.

The relation between the isotropic estimator and the ensemble-averaged 4PCF is obtained by taking the ensemble average of equation~\eqref{eqn: observational estimator of 4PCF},
\begin{equation}
    \begin{split}
        \expval{\hat{\zeta}(\vb*{r_1},\vb*{r_2},\vb*{r_3})} &= \int \frac{d^3s}{V_s}  \expval{\Delta_g(\vb*{s})\Delta_g(\vb*{s}+\vb*{r_1})\Delta_g(\vb*{s}+\vb*{r_2})\Delta_g(\vb*{s}+\vb*{r_3})} \\
        &=\int \frac{d^3s}{V_s} \zeta(\vb*{s};\vb*{r_1},\vb*{r_2},\vb*{r_3}).
    \end{split}
\end{equation}
The ensemble average $\expval{.}$ does not depend on the position $\vb*{s}$, so it can pass through the integral $\int d^3s$. Lastly, \emph{ergodicity} is assumed, under which the isotropic estimator converges to the true ensemble average as $V_s$ increases,
\begin{equation}
    \hat{\zeta}(\vb*{r_1},\vb*{r_2},\vb*{r_3}) \xrightarrow{V_s\rightarrow\,\infty}  \expval{\hat{\zeta}(\vb*{r_1},\vb*{r_2},\vb*{r_3})}
\end{equation}
This is true under the \emph{fair sample hypothesis}, which states that samples of galaxies drawn from regions of the universe separated by a sufficiently large distance are accurate realizations of the same physical process. Equivalently, the spatially averaged quantity converges to the ensemble average as the survey volume increases. \cite{Sarkar2009MNRAS.399L.128S, Scrimgeour2012MNRAS.425..116S} show that this hypothesis is satisfied when the separation is over $70$ and $80$~$h^{-1}$Mpc for SDSS DR6 and WiggleZ Dark Energy Survey, respectively. Therefore, for large volume surveys, we can write
\begin{equation}
    \hat{\zeta}(\vb*{r_1},\vb*{r_2},\vb*{r_3})=\int \frac{d^3s}{V_s} \zeta(\vb*{s};\vb*{r_1},\vb*{r_2},\vb*{r_3})=\frac{1}{V_s}\int ds\,s^2\int d\Omega_s \, \zeta(\vb*{s};\vb*{r_1},\vb*{r_2},\vb*{r_3}),
\end{equation}
where the second equality came from projecting the infinitesimal volume element on the spherical coordinates, $d^3s=s^2dsd\Omega_s$. Given that the linear matter density contrast, $\delta^{(1)}$, is a Gaussian random field, the leading contributions of the connected 4PCF are
\begin{equation}
    \begin{split}
        \zeta(\vb*{s};\vb*{r_1},\vb*{r_2},\vb*{r_3}) & = \expval{\Delta_g(\vb*{s})\Delta_g(\vb*{s}+\vb*{r_1})\Delta_g(\vb*{s}+\vb*{r_2})\Delta_g(\vb*{s}+\vb*{r_3})} \\
        & =\expval{\Delta_g^{(1)}(\vb*{s})\Delta_g^{(1)}(\vb*{s}+\vb*{r_1})\Delta_g^{(1)}(\vb*{s}+\vb*{r_2})\Delta_g^{(3)}(\vb*{s}+\vb*{r_3})}+3\;\text{perm.} \\
        &+\expval{\Delta_g^{(1)}(\vb*{s})\Delta_g^{(1)}(\vb*{s}+\vb*{r_1})\Delta_g^{(2)}(\vb*{s}+\vb*{r_2})\Delta_g^{(2)}(\vb*{s}+\vb*{r_3})}+5\;\text{perm.} \\
        &+\mathcal{O}((\delta^{(1)})^8),
    \end{split}
\end{equation}
where $\expval{\;}$ is the ensemble average, and 3 and 5 perm. refer to terms obtained by distinct permutations (see Appendix~\ref{apx: Leading Contribution of the Connected 4PCF}).

From this point, we assume an equal redshift $\bar{z}$ for all $\Delta_g$. For the analysis in \cite{Hou:2022wfj, Philcox:2022hkh,Slepian:2025kbb}, the separation between galaxies in the 4PCF configuration is restricted to $20\,h^{-1}<r<164\,h^{-1}$ Mpc, which can introduce a redshift difference of up to $\Delta z\lesssim 0.05$. This corresponds to $\lesssim 5\%$ error in the final prediction of 4PCF. We discuss the effect of unequal redshift on PV in Section~\ref{sec: Discussion and Potential Sources of Violation}. Under the equal redshift approximation, these leading contributions are proportional to the Fourier transform of the product of kernels and the initial linear matter power spectra $P_m(k)$,
\begin{equation}\label{eqn: summary 1}
    \begin{split}
        &\expval{\Delta_g^{(1)}(\vb*{s})\Delta_g^{(1)}(\vb*{s}+\vb*{r_1})\Delta_g^{(1)}(\vb*{s}+\vb*{r_2})\Delta_g^{(3)}(\vb*{s}+\vb*{r_3})}=\sum_{r_1=1}^{R_1}\sum_{r_2=1}^{R_1}\sum_{r_3=1}^{R_1}\sum_{r_4=1}^{R_3} \\
        &\mathcal{Z}_{1,r_1}(\bar{z})\mathcal{Z}_{1,r_2}(\bar{z})\mathcal{Z}_{1,r_3}(\bar{z})\mathcal{Z}_{3,r_4}(\bar{z})\int_{\vb*{k_1}}\int_{\vb*{k_2}}\int_{\vb*{k_3}}\int_{\vb*{k_4}} \exp\left(i\sum_{\sigma=1}^3\vb*{k_\sigma}\cdot\vb*{r_\sigma}\right)\delta_D^{(3)}\left(\vb*{k_{1234}}\right) \\
        &\underbrace{\mathcal{S}_{1,r_1}(\vb*{k_1};\vb*{\hat{n}_1})\mathcal{S}_{1,r_2}(\vb*{k_2};\vb*{\hat{n}_2})\mathcal{S}_{1,r_3}(\vb*{k_3};\vb*{\hat{n}_3})\mathcal{S}_{3,r_4}(\vb*{k_1},\vb*{k_2},\vb*{k_3};\vb*{\hat{n}_4})P_m(k_1)P_m(k_2)P_m(k_3)}_{\mathcal{T}_{1113}(\vb*{k_1},\vb*{k_2},\vb*{k_3};\vb*{k_4};\vb*{\hat{n}_1},\vb*{\hat{n}_2},\vb*{\hat{n}_3};\vb*{\hat{n}_4})},
    \end{split}
\end{equation}
and
\begin{equation}\label{eqn: summary 2}
    \begin{split}
         &\expval{\Delta_g^{(1)}(\vb*{s})\Delta_g^{(1)}(\vb*{s}+\vb*{r_1})\Delta_g^{(2)}(\vb*{s}+\vb*{r_2})\Delta_g^{(2)}(\vb*{s}+\vb*{r_3})}=\sum_{r_1=1}^{R_1}\sum_{r_2=1}^{R_1}\sum_{r_3=1}^{R_2}\sum_{r_4=1}^{R_2} \\
        &\mathcal{Z}_{1,r_1}(\bar{z})\mathcal{Z}_{1,r_2}(\bar{z})\mathcal{Z}_{2,r_3}(\bar{z})\mathcal{Z}_{2,r_4}(\bar{z})\int_{\vb*{k_1}}\int_{\vb*{k_2}}\int_{\vb*{k_3}}\int_{\vb*{k_4}} \exp\left(i\sum_{\sigma=1}^3\vb*{k_\sigma}\cdot\vb*{r_\sigma}\right)\delta_D^{(3)}\left(\vb*{k_{1234}}\right) \\
        &\mathcal{S}_{1,r_1}(\vb*{k_1};\vb*{\hat{n}_1})\mathcal{S}_{1,r_2}(\vb*{k_2};\vb*{\hat{n}_2})\Big[\mathcal{S}_{2,r_3}(\vb*{k_1},\vb*{k_{24}};\vb*{\hat{n}_3})\mathcal{S}_{2,r_4}(\vb*{k_2},\vb*{k_{13}};\vb*{\hat{n}_4})P_m(k_{13}) \\
        &\underbrace{+\mathcal{S}_{2,r_3}(\vb*{k_1},\vb*{k_{23}};\vb*{\hat{n}_3})\mathcal{S}_{2,r_4}(\vb*{k_2},\vb*{k_{14}};\vb*{\hat{n}_4})P_m(k_{14})\Big]P_m(k_1)P_m(k_2)\quad \quad \quad \quad}_{\mathcal{T}_{1122}(\vb*{k_1},\vb*{k_2};\vb*{k_3},\vb*{k_4};\vb*{\hat{n}_1},\vb*{\hat{n}_2};\vb*{\hat{n}_3},\vb*{\hat{n}_4})},
    \end{split}
\end{equation}
where $\vb*{\hat{n}_1}$, $\vb*{\hat{n}_2}$, $\vb*{\hat{n}_3}$, and $\vb*{\hat{n}_4}$ are the LOS of each $\Delta_g$ in 4PCF and $\vb*{k_{ab\dots c}}=\vb*{k_a}+\vb*{k_b}+\dots+\vb*{k_c}$. The introduction of each LOS is crucial for the wide-angle treatment as opposed to the standard plane-parallel approximation, where a single LOS is assumed for every point in 4PCF. As we will see later, the correction to the single LOS approximation introduces a term like $\vb*{\hat{s}}\cdot\vb*{\hat{r}_\sigma}$ which is parity-odd upon fixing the reference LOS $\vb*{\hat{s}}$. Since the kernels $\mathcal{S}_{n,r}$ are linear combinations of parity-odd and even terms, $\mathcal{T}_{1113}$ and $\mathcal{T}_{1122}$ inherit the same structure multiplied by the linear matter power spectra. To assess the parity symmetry of $\hat{\zeta}$, we focus on the directionally dependent parts of these expressions, setting aside the redshift-dependent prefactors and the mode magnitude dependence. The key question is then whether the directional integral $ \int d\Omega_s$ eliminates the parity-odd contributions to $\hat{\zeta}$.

The equal-redshift approximation implies that the redshift $z$ in $\mathcal{Z}_{n,r}$ is independent of LOS direction. This allows $\int d\Omega_s$ to bypass them all and assess only the PV in the mode-mixing integral, which ultimately depends on $\vb*{s}$, $\vb*{r_1}$, $\vb*{r_2}$, $\vb*{r_3}$ and angles between them. However, we note that, while the separation of $\mathcal{Z}_{n,r}$ from the integral is exact and genuine, the redshift of each secondary point in 4PCF generally depends on the LOS direction. They correspond to their exact comoving distance, $\norm{\vb*{s}+\vb*{r_\sigma}}=\sqrt{s^2+r_\sigma^2+2\vb*{\hat{s}}\cdot\vb*{\hat{r}_\sigma}}$. Thus, they depend on $\vb*{\hat{s}}\cdot\vb*{\hat{r}_\sigma}$ and may also introduce PV into the mode-mixing integral. Nevertheless, as mentioned in the last paragraph, the deviation of the predicted 4PCF from the equal-redshift approximation is less than $5\%$, and we leave the assessment for future work.

Following \cite{Leonard:2024xvy}, we first consider the estimator under the plane-parallel approximation, where every LOS is the same, $\vb*{\hat{n}_1}=\vb*{\hat{s}}\approx\vb*{\hat{n}_2}\approx\vb*{\hat{n}_3}\approx\vb*{\hat{n}_4}$. In this case, the estimator depends on only one LOS; $\mathcal{T}_{1113}$ and $\mathcal{T}_{1122}$ are polynomials of $\mu_{\vb*{k_\sigma s}}=\vb*{\hat{k}_\sigma}\cdot\vb*{\hat{s}}$, which are the only geometrical factors that can be parity-odd. In general, the polynomials contain four cosine factors, $\mu_{\vb*{k_1 s}}$, $\mu_{\vb*{k_2 s}}$, $\mu_{\vb*{k_3 s}}$ , and $\mu_{\vb*{k_4 s}}$, corresponding to each mode vector. However, the homogeneity of the isotropic 4PCF estimator enforces
\begin{equation}
    \vb*{k_{1234}}=\vb*{k_1}+\vb*{k_2}+\vb*{k_3}+\vb*{k_4}=0
\end{equation}
through the Dirac delta in equations~\eqref{eqn: summary 1} and \eqref{eqn: summary 2}. Therefore, the cosine factors are also constrained as 
\begin{equation}
    \mu_{\vb*{k_4 s}}= -k_{123}^{-1}(k_1\mu_{\vb*{k_1 s}}+k_2\mu_{\vb*{k_2 s}}+k_3\mu_{\vb*{k_3 s}}).
\end{equation}
In other words, we can always trade one cosine factor, e.g., $\mu_{\vb*{k_4 s}}$, with the other three. So, we end up with polynomials of three cosine factors instead of all four. Then, these factors can be expanded using the spherical harmonic addition theorem,
\begin{equation}
    \mu_{\vb*{k_\sigma s}}^{\alpha_\sigma} = \sum_{j_\sigma m_{j_\sigma}}^{\alpha_\sigma} c^{(\alpha_\sigma)}_{j_\sigma} \frac{4\pi}{2j_\sigma+1} Y_{j_\sigma,m_{j_\sigma}}(\vb*{\hat{k}_\sigma})Y^*_{j_\sigma,m_{j_\sigma}}(\vb*{\hat{s}}),
\end{equation}
where $Y_{j_\sigma,m_{j_\sigma}}$ is the spherical harmonic function and $c^{(\alpha_\sigma)}_{j_\sigma}$ is some real constant.
Integrating the product of spherical harmonics over all LOS gives us the so-called \emph{Gaunt integral},
\begin{equation}
    \begin{split}
        \mathcal{G}^{j_1,j_2,j_3}_{m_{j_1},m_{j_2},m_{j_3}} &\equiv \int d\Omega_s Y_{j_1,m_{j_1}}(\hat{\vb*{s}})Y_{j_2,m_{j_2}}(\hat{\vb*{s}})Y_{j_3,m_{j_3}}(\hat{\vb*{s}}) \\
        &= \sqrt{\frac{(2j_1+1)(2j_2+1)(2j_3+1)}{4\pi}}
        \begin{pmatrix}
            j_1 & j_2 & j_3 \\
            0 & 0 & 0
        \end{pmatrix}
        \begin{pmatrix}
            j_1 & j_2 & j_3 \\
            m_{j_1} & m_{j_2} & m_{j_3}
        \end{pmatrix},
    \end{split}
\end{equation}
where the bracketed terms are \emph{Wigner 3-j symbols}. Ultimately, the product of $\mu_{\vb*{k_\sigma s}}$ after averaging over LOS can be written as a projection onto the isotropic basis functions \cite{Cahn:2020axu_isotropicbasis},
\begin{equation}
    \int d\Omega_s \, \mu_{\vb*{k_1 s}}^{\alpha_1} \mu_{\vb*{k_2 s}}^{\alpha_2} \mu_{\vb*{k_3 s}}^{\alpha_3} = (4\pi)^3 \sum_{j_1,j_2,j_3} \mathcal{C}^{\alpha_2,\alpha_1,\alpha_3}_{j_1,j_2,j_3} \mathcal{P}_{j_1,j_2,j_3}(\vb*{\Hat{k}_1},\vb*{\Hat{k}_2},\vb*{\Hat{k}_3}),
\end{equation}
where the isotropic basis function is defined as
\begin{equation}
    \begin{split}
        \mathcal{P}_{j_1,j_2,j_3}(\vb*{\Hat{k}_1},\vb*{\Hat{k}_2},\vb*{\Hat{k}_3}) =& (-1)^{j_1+j_2+j_3}  \sum_{m_{j_1},m_{j_2},m_{j_3}}
        \begin{pmatrix}
            j_1 & j_2 & j_3 \\
            m_{j_1} & m_{j_2} & m_{j_3}
        \end{pmatrix} \\
        &Y_{j_1,m_{j_1}}(\vb*{\Hat{k}_1})Y_{j_2,m_{j_2}}(\vb*{\Hat{k}_2})Y_{j_3,m_{j_3}}(\vb*{\Hat{k}_3}),
    \end{split}
\end{equation}
and the coefficients are
\begin{equation}
    \mathcal{C}^{\alpha_2,\alpha_1,\alpha_3}_{j_1,j_2,j_3} = \frac{c^{(\alpha_1)}_{j_1}c^{(\alpha_2)}_{j_2}c^{(\alpha_3)}_{j_3}}{\sqrt{4\pi(2j_1+1)(2j_2+1)(2j_3+1)}}
    \begin{pmatrix}
        j_1 & j_2 & j_3 \\
        0 & 0 & 0
    \end{pmatrix}.
\end{equation}
The benefit of projecting the estimator with the isotropic basis after averaging over LOS is that all components are separated into the fully parity-even and parity-odd contributions, determined by the sum of indices $j_1 +j_2 +j_3$. For example, the basis $\mathcal{P}_{100}$ is parity-odd since the sum of j-indices is odd, while the basis $\mathcal{P}_{110}$ is parity-even as the sum of j-indices is even. The crucial point here is that, thanks to the Gaunt integral, the coefficient $\mathcal{C}^{\alpha_2,\alpha_1,\alpha_3}_{j_1,j_2,j_3}$ carries an angular momentum selection rule: $j_1 +j_2 +j_3=\text{even}$ for a non-vanishing component. This is because the Wigner 3-j symbol obeys a parity transformation,
\begin{equation}
    \begin{pmatrix}
        j_1 & j_2 & j_3 \\
        -m_{j_1} & -m_{j_2} & -m_{j_3}
    \end{pmatrix} = (-1)^{j_1 + j_2 + j_3}
    \begin{pmatrix}
        j_1 & j_2 & j_3 \\
        m_{j_1} & m_{j_2} & m_{j_3}
    \end{pmatrix}.
\end{equation}
Thus, for $m_{j_1}=m_{j_2}=m_{j_3}=0$, the symbol is non-vanishing only if $j_1 + j_2 + j_3=\text{even}$. Therefore, the isotropic 4PCF estimator retains only a linear combination of parity-even contributions. This is consistent with the result obtained in \cite{Leonard:2024xvy}, but we extend it to the GR correction on the RSD effect up to the order $\mathcal{H}/k$. The same conclusion holds crucially because the GR correction introduces a contribution proportional to the power of the same geometrical factor,  $\mu_{\vb*{k_\sigma s}}$, as the Newtonian RSD effect. Thus, the PV associated with the LOS is removed by the averaging over LOS in the same manner.

The plane-parallel approximation ignores the size of the 4PCF configuration and assumes that every galaxy is along the same LOS. Therefore, the RSD effect in this approximation misses the internal geometry of the 4PCF entirely. As the PV of the 4PCF is examined by flipping the internal structure, $\vb*{r_\sigma}\rightarrow-\vb*{r_\sigma}$, it is therefore interesting to investigate the potential PV when the approximation is relaxed. This is particularly important for modern surveys, where the observed sky area has grown, allowing us to probe the 4PCF at large angular separations between galaxies; $\vb*{r_\sigma}$ is not much smaller than the comoving distance from the observer to galaxies. We then go beyond the plane-parallel approximation by using the local LOS,
\begin{equation}
    \vb*{\hat{n}_1}=\vb*{\hat{s}}, \quad \vb*{\hat{n}_\sigma}=\frac{\vb*{s}+\vb*{r_\sigma}}{\norm{\vb*{s}+\vb*{r_\sigma}}}, \quad \sigma \in \{1,2,3\}.
\end{equation}
This choice of definition is often referred to as the end-point LOS, where one galaxy is chosen as the reference LOS, and the others are the extensions. It is important to note that this choice of definition breaks the permutation symmetry of LOS as it selects one LOS to be the reference. This choice has been shown to artificially induce dipole and octupole in the correlation function, which do not carry any additional information about the statistics \cite{Beutler2019JCAP...03..040B}. In the same way, our final expression of 4PCF before averaging over all LOS carries similar artificial odd multipoles. However, the fact that all parity-odd contributions are removed by the LOS averaging, which is the main result of this paper, holds for the centroid definition of LOS as well. This is because the spurious parity-odd mode induced by changing the LOS definition is also removed by the averaging over all LOS (see Appendix~\ref{apx: Alternative LOS definition for Wide Angle Effect}). For mathematical simplicity, we then use the end-point LOS definition to demonstrate the removal of PV induced by the wide-angle effect in this paper.

We expand each local LOS using the generating function of Legendre polynomials to express the cosine factor $\mu_{\rho\sigma}=\vb*{\hat{k}_\rho}\cdot\vb*{\hat{n}_\sigma}$ with polynomials of the following three cosine factors,
\begin{equation}
    \mu_{\vb*{k_\rho s}}=\vb*{\hat{k}_\rho}\cdot\vb*{\hat{s}}, \quad \nu_{\rho\sigma}=\vb*{\hat{k}_\rho}\cdot\vb*{\hat{r}_\sigma}, \quad \xi_\sigma=\vb*{\hat{s}}\cdot\vb*{\hat{r}_\sigma}, \quad \rho,\sigma \in \{1,2,3\}.
\end{equation}
Together with the $\kappa_{\rho\sigma}=\vb*{\hat{k}_\rho}\cdot\vb*{\hat{k}_\sigma}$ factors, $\mathcal{T}_{1113}$ and $\mathcal{T}_{1122}$ can be fully expressed as polynomials in these three cosine factors. While other projection factors arise from the RSD and wide-angle effects, $\kappa_{\rho\sigma}$ represents the mode-mixing behaviour arising from the nonlinear dynamics of LSS, which appears in $\mathcal{T}_{1113}$ and $\mathcal{T}_{1122}$ through the kernel $S_{n,r}$. Then, the plane-wave factor and all cosine factors are expanded in terms of spherical harmonics via the addition theorem (see equations~\eqref{eqn: isotropic estimator of 4PCF with los average integral wide angle}{-}\eqref{eqn: 4PCF estimator final proportion}).

Finally, the directional integrations $\int d\Omega_s$ and $\int d\Omega_{k_\sigma}$ invoke angular momentum selection rules on the spherical harmonic functions of LOS $\vb*{\hat{s}}$ and $\vb*{\hat{k}_\sigma}$, respectively, forcing the sum of angular momentum indices to be even (see equations~\eqref{eqn: s constraint} and \eqref{eqn: k constraints}). The selection rule follows from the parity property of the spherical harmonics when subjected to integration over a symmetric domain (see Appendix~\ref{apx: Integrals of Spherical Harmonics}). This condition is true for the product of an arbitrary number of spherical harmonic functions. For instance, it corresponds to the triangular conditions on the Wigner 3-j symbols for the product of three spherical harmonics (Gaunt integral) in the plane-parallel case (see equation~\eqref{eqn: Gaunt integral}). In the wide-angle case, we instead integrate over a product of more than three spherical harmonics. Combining the two constraints gives the final selection rule on the residual spherical harmonic functions of $\vb*{\hat{r}_\sigma}$ that allow only the even sum of angular momentum indices (denoted as the $S$ index; see equation~\eqref{eqn: S parameter}), and force all parity-odd contributions to vanish (see equation~\eqref{eqn: final selection rule}). Therefore, the 4PCF isotropic estimator contains only parity-even contributions, also in the presence of the wide-angle correction.

The assessment can be generalised; the parity-odd contribution in $\zeta$ that comes in polynomials of the projection $\vb*{\hat{k}_\sigma}$ against LOS is always washed away from $\hat{\zeta}$ by averaging all LOS. This extends the conclusion beyond the leading contribution to $\zeta$ listed in Appendix~\ref{apx: Leading Contribution of the Connected 4PCF}. First, the higher-order solution of matter density in EPT will only introduce a more complex mode-mixing kernel. Nonetheless, it can be written as a polynomial of $\kappa_{\rho\sigma}=\vb*{\hat{k}_\rho}\cdot\vb*{\hat{k}_\sigma}$. Second, the expression~\eqref{eqn: full relativistic Number Density} for the observed galaxy number density introduces different orders of LOS derivatives $\partial_r$ on the galaxy number density, $\delta_g$, peculiar velocity $v_g$, gravitational potential $\Psi$, and their products. Through EPT, $v_g$ and $\Psi$ at every order can always be written in terms of $\delta_g^{(1)}$ integrated against the EPT mode-mixing kernel. Therefore, the mapping from $\delta_g$ to $\Delta_g$ can introduce parity-odd mode only through the polynomials of $\kappa_{\rho\sigma}$ and the projection of mode against LOS $\mu_{\vb*{k_\sigma}}=\vb*{\hat{k}_\sigma}\cdot\vb*{\hat{n}}$. This aligns with the discussion in Section~\ref{sec: Kernel Solutions of the Observed Galaxy Number Density}. Therefore, higher-order perturbations $\Delta^{(n)}_g$ will not reintroduce parity-odd contributions that survive the integration over all LOS. 

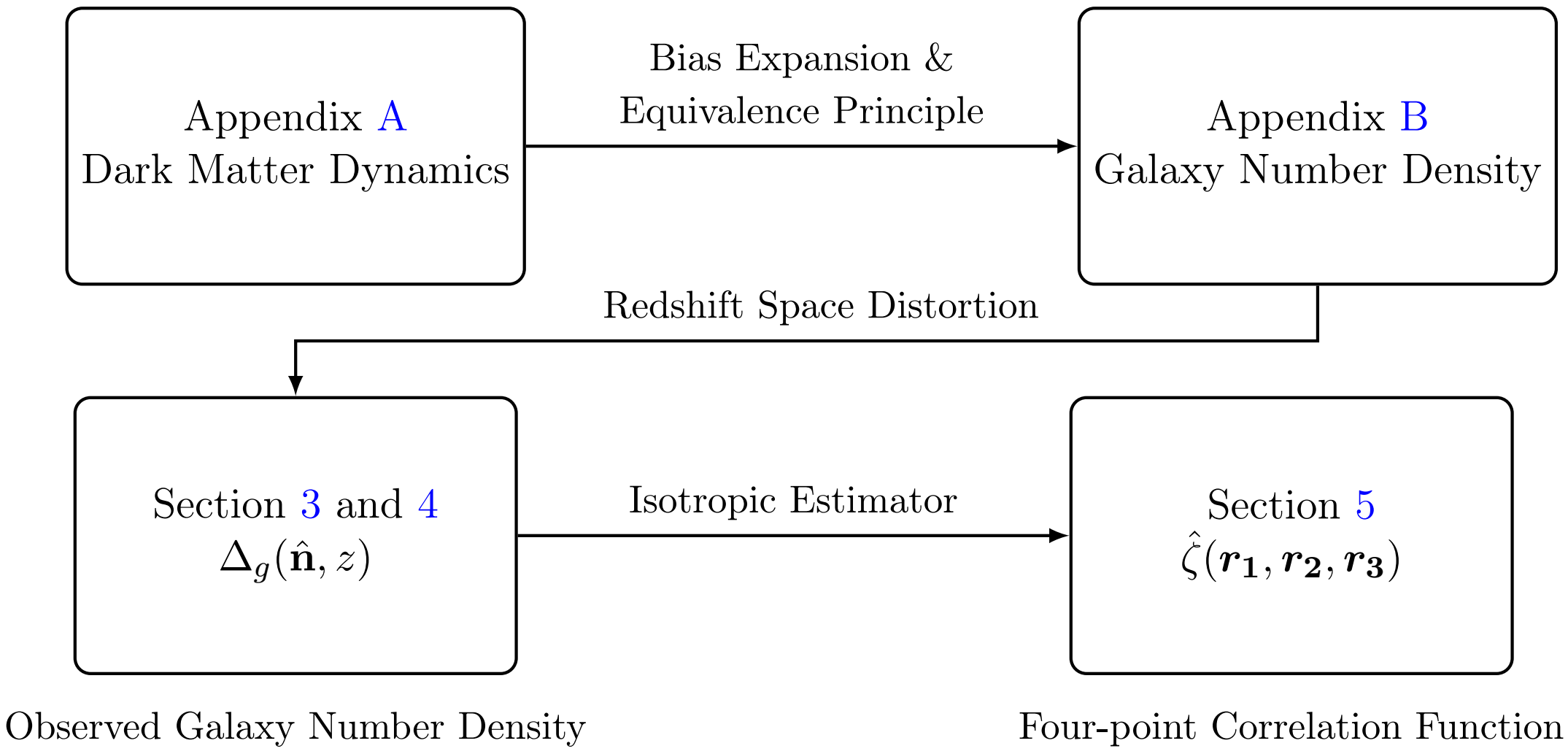
\begin{figure}[tp]
    \centering
    \begin{tikzpicture}[
    >=latex,
    every node/.style={align=center},
    box/.style={
        draw,
        rounded corners,
        thick,
        minimum width=4cm,
        minimum height=2.5cm
    },
    labelstyle/.style={font=\small, above}
    ]
    
    \node[box] (dm) {Appendix~\ref{sec: Perturbative Solution of the Large-scale Structure} \\Dark Matter Dynamics};
    
    \node[box, right=5cm of dm] (bias) { Appendix~\ref{sec: The Bias Expansion}\\ Galaxy Number Density};
    
    \node[box, below=1cm of dm] (rs) {Section~\ref{sec: The Relativistic Galaxy Number Density: Redshift Space Distortion} and \ref{sec: Kernel Solutions of the Observed Galaxy Number Density}\\ $\Delta_g(\hat{\mathbf{n}},z)$};
    \node[below=0.2cm of rs] {\small Observed Galaxy Number Density};
    
    \node[box, right=5cm of rs] (obs) {Section~\ref{sec: Probing Parity Violation With 4PCF}\\$\hat{\zeta}(\vb*{r_1},\vb*{r_2},\vb*{r_3})$};
    \node[below=0.2cm of obs] {\small Four-point Correlation Function};
    
    \draw[->, thick] (dm) -- node[above]{\small Bias Expansion \& \\ \small Equivalence Principle} (bias);
    \draw[->, thick] (bias.south) -- ++(0,-5mm)
        node[above, xshift=-4.5cm]{\small Redshift Space Distortion} -| (rs.north);
    \draw[->, thick] (rs) -- node[above]{\small Isotropic Estimator} (obs);
    \end{tikzpicture}
    \caption{Computational procedure summary. The first two stages are computed in Fourier space, while the last two are carried out in real space. Working in Fourier space is advantageous for solving the dark matter dynamics, as the time evolution can be factored out and the mode-dependent part expressed as a convolution of first-order solutions against a kernel. This structure propagates throughout the calculation, and the observed galaxy number density contrast $\Delta_g$ inherits the same form.}
    \label{fig: Summary of the computation}
\end{figure}

In conclusion, we have shown that the RSD and wide-angle effects, including the leading-order GR corrections, cannot systematically induce parity-odd modes in the 4PCF isotropic estimator. The calculation pipeline of this work is summarised schematically in Figure~\ref{fig: Summary of the computation}.

The real observation cases sit well within our framework. \cite{Hou:2022wfj, Philcox:2022hkh, Slepian:2025kbb} measure the 4PCF with point separation from the reference galaxy restricted to $20h^{-1}{-}164h^{-1}$~Mpc at the redshift ranges $0.2<z<0.7$ \cite{Hou:2022wfj, Philcox:2022hkh} and $0.4<z<1.1$ \cite{Slepian:2025kbb}. This corresponds to probing the perturbation modes $k\sim0.003h{-}0.05h$~Mpc$^{-1}$, as the smallest and largest separations between galaxies in the 4PCF are $20h^{-1}$ and $328h^{-1}$~Mpc, respectively. First, the scale at which the matter perturbation is linear (above the nonlinear scale) is $k<0.1h$~Mpc$^{-1}$ at $z\approx0$, and the $k$ upper bound increases at higher redshift \cite{Bernardeau2002PhR...367....1B}. Therefore, the length scale of interest is larger above the nonlinear scale at $z\geq0$, and the pressureless CDM is in the single-stream regime (see Appendix~\ref{sec: Perturbative Solution of the Large-scale Structure}). Second, the late-time growing comoving Hubble radius ($\mathcal{H}^{-1}$) is $\sim5$~Gpc at $z=0$ and  $\sim2$~Gpc at $z\sim1$, assuming the $\Lambda$CDM model with cosmological parameters from \cite{Planck:2018vyg_CosmoParam}. Therefore, the relevant matter perturbation modes are well within the horizon, where the GR correction is perturbative, ordered by powers of $\mathcal{H}/k\sim0.004{-}0.2$. Our conclusions therefore directly apply to the scales and redshifts probed by \cite{Hou:2022wfj,Philcox:2022hkh,Slepian:2025kbb}. 

Some extensions of the analysis presented here could introduce non-vanishing parity-odd contributions (see Section~\ref{sec: Discussion and Potential Sources of Violation}). First, the vorticity, $\vb*{\nabla}\times\vb*{v}$, in the LSS dynamics can seed a pseudo-scalar term, e.g., $\vb*{v}\cdot\vb*{\nabla}\times\vb*{v}$, whether in the bias expansion or in the EPT solutions. However, vorticity on large scales, as a transverse vector perturbation, is negligible. In the standard inflationary paradigm, the primordial vector perturbation exponentially decays, leaving no initial condition for the vorticity. In the single-stream regime, the dynamics of the pressureless CDM is governed by the Euler, continuity, and Poisson equations (see Appendix~\ref{sec: Perturbative Solution of the Large-scale Structure}). In this case, the irrotationality of the CDM fluid is preserved order by order \cite{Pueblas2009PhRvD..80d3504P}. The rotation can be generated at shell-crossing beyond the nonlinear scale, $k>0.1h$~Mpc$^{-1}$, which is beyond the scale of 4PCF probed in \cite{Hou:2022wfj,Philcox:2022hkh,Slepian:2025kbb}. Second, this work considers the GR correction up to $\mathcal{O}(\mathcal{H}/k)$. Consequently, we confine our analysis to the GR correction to the RSD effect only. GR corrections enter the CDM dynamics from the quadratic order $(\mathcal{H}/k)^2$, where the equations of motion are derived from the Einstein field equations. The question of whether the higher-order GR term can reintroduce PV into the estimator is nontrivial. One way is that it introduces a gauge-invariant pseudo-scalar, independent of the LOS, into the galaxy number density. For example, the vorticity, as a transverse vector perturbation, can be sourced on large scales through the scalar-scalar mode coupling in the second-order GR dynamics \cite{Matarrese:1993zf,Bruni:1996im,Matarrese:1997ay, Lu2009JCAP...02..023L}. However, it would be suppressed by $(\mathcal{H}/k)^2$ relative to the Newtonian squared scalar term and would be a negligible contribution to the PV detected in \cite{Hou:2022wfj,Philcox:2022hkh,Slepian:2025kbb}.

Since the various analyses of parity-violating 4PCF signal spurred a lot of interest in investigating parity violation in large-scale structures, we refer the reader to Section~\ref{sec: Discussion and Potential Sources of Violation}, where we discuss potential physical sources of PV beyond the scope of the RSD mechanism examined here. In particular, while RSD up to the leading-order GR corrections have been shown not to induce parity-odd modes, the analysis in this paper does not include other systematics such as survey geometry, selection effects, or higher-order galaxy bias.
\section{The relativistic galaxy number density: redshift space distortion}\label{sec: The Relativistic Galaxy Number Density: Redshift Space Distortion}

In this section, we review how relativistic effects distort the observed galaxy number density from the true density, following the analysis of \cite{DiDio2019JCAP...04..050D,DiDio2020JCAP...09..058D}. We begin by clarifying what we mean by the relativistic correction to the number density. Generally speaking, GR becomes relevant when the gravitational potential is large. However, the galaxy distribution is in the weak-field regime because the matter density is low; gravitational interactions between galaxies at large separations are weak. Hence, GR effects become important on large scales. In particular, the first-order correction to the matter density starts at
\begin{equation}
    \epsilon_\mathcal{H} \equiv \left( \frac{\mathcal{H}}{k} \right)^2,
\end{equation}
where $\mathcal{H}\equiv a'/a$ is the comoving Hubble parameter, $a$ is the scale factor, and the prime denotes the conformal time derivative, $'\equiv\frac{\partial}{\partial\tau}$. The first-order correction from the GR dynamics is quadratic in the ratio $\mathcal{H}/k$, as the relativistic corrections come from terms in the perturbation equations that are second order in spatial gradients, and the quantity that sets their size is the ratio of the Hubble scale to the comoving scale squared, because the Poisson equation and the Newtonian limit are quadratic in derivatives. Indeed, these contributions become important as the perturbation scale, $k^{-1}$, approaches the Hubble horizon scale, $\mathcal{H}^{-1}$. However, they do not induce PV by themselves, as the Einstein field equation, which governs the exact GR dynamics, is parity conserving.

However, as we will see shortly, the GR correction can also enter the observed density perturbation in the form of projection effects associated with the peculiar velocity and the LOS, and these start at the linear order of $\mathcal{H}/k=\epsilon_\mathcal{H} ^{1/2}$. In this work, we consider the leading correction up to the order of $\epsilon_\mathcal{H}^{1/2}$ and neglect the higher orders. (see Section~\ref{sec: Discussion and Potential Sources of Violation} for more details). Mathematically, the observed galaxy number density $\Delta_g$ can be divided into two contributions as
\begin{equation}
    \Delta_g = \Delta_N+\Delta_R \quad \text{where} \quad \mathcal{O}(\Delta_R) \sim \mathcal{O}(\epsilon_\mathcal{H}^{1/2}\Delta_N)
\end{equation}
where the relativistic contribution, $\Delta_R$, is of the order $\epsilon_\mathcal{H}^{1/2}=\mathcal{H}/k$ smaller than the Newtonian contribution, $\Delta_N$. With this framework, we will proceed to derive the observed galaxy number density subjected to the relativistic effects. A subtle point is that the calculation of \cite{DiDio2019JCAP...04..050D,DiDio2020JCAP...09..058D} is in Poisson gauge, while the calculations in the previous works \cite{Yoo2009PhRvD..80h3514Y,Yoo2010PhRvD..82h3508Y,Bouvin2011PhRvD..84f3505B,Jeong2012PhRvD..85b3504J,Bertacca:2012tp_GRcorrect,Bertacca2014JCAP...09..037B, Gervani:2025wmh} are in synchronous-comoving gauge. Hence, the galaxy number density contrast $\delta_g$ is inherently in a different gauge between those results. However, as mentioned in \cite{DiDio2019JCAP...04..050D, DiDio2020JCAP...09..058D}, the expression is interchangeable up to the order $\epsilon_\mathcal{H}^{1/2}$ since the transformation between the two gauges alters the form of the correction to the density perturbation through the GR dynamical correction term, which enters at the order $\epsilon_\mathcal{H}$ as mentioned before. Therefore, the results are consistent with those in \cite{Yoo2009PhRvD..80h3514Y,Yoo2010PhRvD..82h3508Y,Bouvin2011PhRvD..84f3505B,Jeong2012PhRvD..85b3504J,Bertacca:2012tp_GRcorrect,Bertacca2014JCAP...09..037B, Gervani:2025wmh} in this case.

To understand the difference between observational and theoretical quantities, let us define the observed galaxy number overdensity,
\begin{equation}\label{def: observed number overdensity}
    \Delta_g(\vb*{\hat{n}}_{\text{obs}},z_{\text{obs}})= \frac{N(\vb*{\hat{n}}_{\text{obs}},z_{\text{obs}})-\bar{N}(z_{\text{obs}})}{\bar{N}(z_{\text{obs}})},
\end{equation}
where $\vb*{\hat{n}}$ is the LOS unit vector, $z$ is the redshift.
The subscript 'obs' indicates quantities estimated by the observer. $N(\vb*{\hat{n}},z)$ is the number of galaxies in the observed volume bin in the $\vb*{\hat{n}}$-direction at redshift $z$ while $\bar{N}(z)$ is the number of galaxies averaged over all LOS at redshift $z$. Indeed, galaxy surveys estimate the galaxy number overdensity by counting the number of galaxies within each survey volume bin and then comparing them with the average number per bin over the LOS. Since $\Delta_g$ is constructed from observational counts, it is a gauge-invariant quantity. From the theoretical side, we can define the galaxy number overdensity field as
\begin{equation}\label{def: theoretical number overdensity}
    \delta_g(\vb*{\hat{n}},z)= \frac{\rho_g(\vb*{\hat{n}},z)-\bar{\rho}_g(z)}{\bar{\rho}_g(z)},
\end{equation}
where the number density field $\rho_g(\vb*{\hat{n}},z)$ is used in place of the number count. The observed number density is the number of galaxies per unit volume, $\rho_g(\vb*{\hat{n}},z)=N(\vb*{\hat{n}},z)/V(\vb*{\hat{n}},z)$ and $\bar{\rho}_g(z)=\bar{N}(z)/\bar{V}(z)$. $\bar{V}(z)$ is the volume computed from the unperturbed Friedmann–-Lema\^itre--Robertson–-Walker (FLRW) metric. $\bar{\rho}_g(z)$ is the background number density, corresponding to the number density in the unperturbed FLRW universe. This quantity describes the inhomogeneity of the galaxy distribution compared to the unperturbed background evolution. Hence, unlike $\Delta_g$, $\delta_g$ is a gauge-dependent quantity. We follow the convention of \cite{DiDio2019JCAP...04..050D,DiDio2020JCAP...09..058D} where they choose the Poisson gauge.

\begin{figure}[tp]
    \centering
    \includegraphics[width=0.8\linewidth]{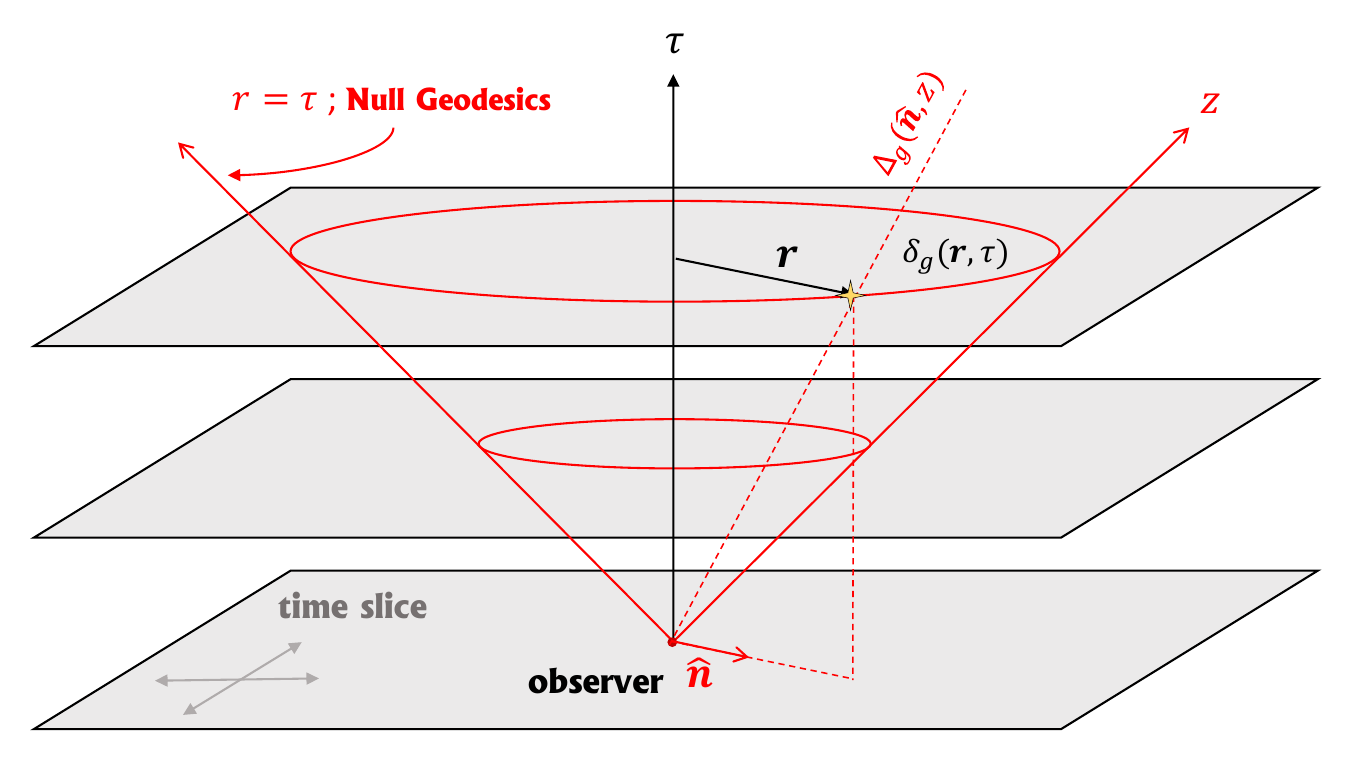} 
    \caption{A Pictorial Explanation of Observed and Theoretical Galaxy Number Overdensity: The observer's coordinates are in red while the space-time coordinates are in black. The observer's coordinate is defined along the photon path (or null geodesics) where space and time coordinates degenerate, $r=\tau$. The observed galaxy number overdensity, $\Delta_g$, is defined on the observer's coordinates, labelled with redshift and the LOS direction, $(\vb*{\hat{n}},z)$. On the other hand, the theoretical galaxy number overdensity field, $\delta_g$, is defined on the space-time coordinates, labelled with spatial and time coordinates, $(\vb*{r},\tau)$.}
    \label{fig: Lightcone}
\end{figure}

$\delta_g$ is defined on spatial comoving coordinates and conformal time, $(\vb*{r},\tau)$, which are independent of each other. By contrast, $\Delta_g$ is defined on the observer's coordinates $(\vb*{\hat{n}}, z)$. These two coordinate systems are related by,
\begin{equation}
    \vb*{r}=r\vb*{\hat{n}}\; , \quad r(z)=\int^z_0\frac{dz'}{\mathcal{H}(z')(1+z')}=\tau(z),
\end{equation}
where $r$ is the comoving radial distance. In the observer's coordinates, $r$ is degenerate with $\tau$ since the galaxy is observed by light. Hence, the observer's perspective is confined to the null geodesics (see Figure~\ref{fig: Lightcone}).

$\Delta_g$ and $\delta_g$ are related but distinct quantities. They would coincide if 1) the observed volume bin matched the one defined on the unperturbed FLRW metric, $V(\vb*{\hat{n}},z)=\bar{V}(z)$; and  2) the observed redshift and LOS direction accurately reflected the true position of the galaxy, $\vb*{\hat{n}}_{\text{obs}}=\vb*{\hat{n}}$ and $z_{\text{obs}}=z$. However, neither condition holds in practice, and the observed density thus acquires the corrections starting from the order $\epsilon_\mathcal{H}^{1/2}$. This is particularly relevant to the observational analyses of \cite{Hou:2022wfj, Philcox:2022hkh, Slepian:2025kbb} as they are estimating the 4PCF at $\mathcal{H}/k$ up to 0.2 (see discussion at the end of Section~\ref{sec: Synopsis: the Observational Implication and its Validity}). The first is violated simply because we live in a \emph{perturbed} FLRW universe rather than an unperturbed one (see \cite{Bouvin2011PhRvD..84f3505B} for details). The second is violated because the observed position, $\vb*{\hat{n}}_{\text{obs}}$, and $z_{\text{obs}}$, deviate from the true position due to various effects along the photon path. The misestimation of redshift is called the RSD effect, which is the focus of this work. That is, the observed redshift differs from the galaxy's true redshift due to relativistic effects on the photon travelling from the galaxy to the observer. In addition, gravitational lensing can deflect the LOS via the lensing effect (\cite{Hui2007PhRvD..76j3502H} and references therein). However, \cite{Bouvin2014PhRvD..89h3535B,Lepori2018JCAP...05..043L} show that the parity-odd contribution from lensing is several orders of magnitude smaller than that of other relativistic effects in the galaxy N-point correlation function; the lensing dipole contribution is suppressed by a factor of $\mathcal{H}r\sim\mathcal{H}/k$ relative to other relativistic dipole contributions and only becomes comparable at $z=2$ and separation $>200h^{-1}$~Mpc (see Figure 8 in \cite{Bouvin2014PhRvD..89h3535B} and Figure 3 in \cite{Lepori2018JCAP...05..043L}). Therefore, we disregard it throughout this work, as our primary interest is PV in the 4PCF.

\begin{figure}[tp]
    \centering
    \includegraphics[width=0.7\linewidth]{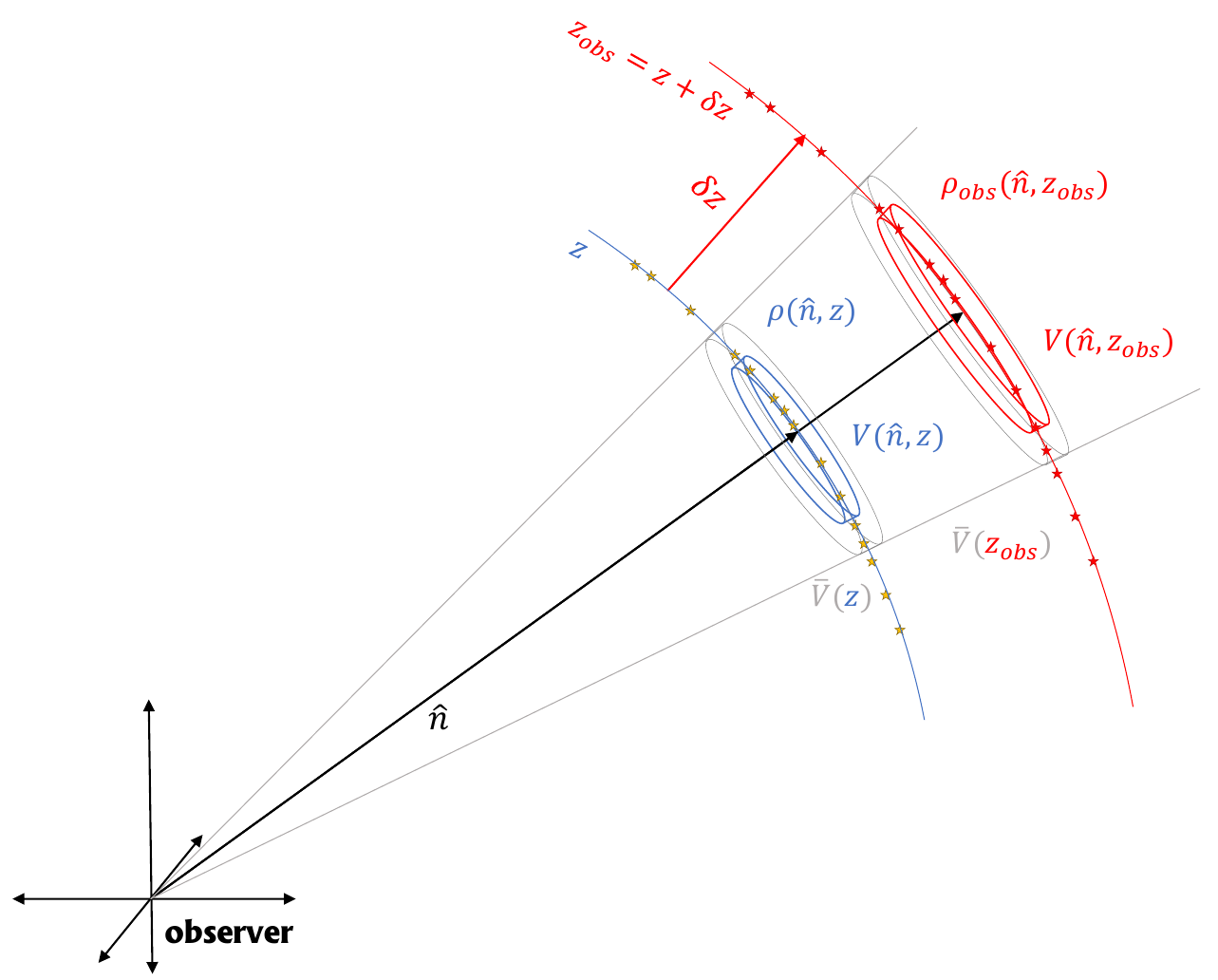}
    \caption{A Pictorial Explanation of Redshift Space Distortion Effect: The true position of the galaxies is at redshift $z$ and direction $\vb*{\hat{n}}$. Due to relativistic effects, the redshift estimation of the galaxies deviates from the true redshift by $\delta z$. So, the observer sees the redshift $z$ population at redshift $z_{\text{obs}}=z+\delta z$ instead. Furthermore, the volume element changes with the redshift. The distortions in both observed density and volume cause the misestimate in the observed galaxy number density. The LOS deviation is ignored in this work, and the volume element size is exaggerated for visualisation.}
    \label{fig:A Pictorial Explanation of Redshift Space Distortion}
\end{figure}

Let us relate the observed overdensity to the theoretical overdensity at redshift $z_\text{obs}$ step by step.  Suppose we are observing galaxies with estimated redshift $z_\text{obs}$. Rather than seeing the true galaxy population at redshift $z_\text{obs}$, the RSD effect projects the galaxies at redshift $z=z_\text{obs}-\delta z$ onto the observed slice. Thus, the observer counts the number of galaxies projected from redshift $z$ rather than $z_\text{obs}$ (see Figure~\ref{fig:A Pictorial Explanation of Redshift Space Distortion}). Consequently, the observed sky distribution at $z_\text{obs}$ is equal to the true sky distribution at redshift $z$, that is $\rho_{g,\text{obs}}(\vb*{\hat{n}},z_{\text{obs}})=\rho_g(\vb*{\hat{n}},z)$, where we introduce
\begin{equation}\label{def: observed theoretical number density}
    \rho_{g,\text{obs}}(\vb*{\hat{n}},z_\text{obs})=\bar{\rho}_{g}(z_\text{obs})[1+\delta_{g,\text{obs}}(\vb*{\hat{n}},z_\text{obs})],
\end{equation}
which is the theoretical galaxy number density evaluated on the observer coordinates. This quantity is not the same as $\rho_g$, which is the density evaluated on the comoving coordinates. Note that it is defined as the number of galaxies per unit volume, 
\begin{equation}
    \rho_{g,\text{obs}}(\vb*{\hat{n}},z_\text{obs})=\frac{N(\vb*{\hat{n}},z_\text{obs})}{V(\vb*{\hat{n}},z_\text{obs})}\;,\quad \bar{\rho}_g(z_\text{obs})=\frac{\bar{N}(z_\text{obs})}{\bar{V}(z_\text{obs})}
\end{equation}
This is equivalent to stating that $\rho_g$ transforms as a scalar field defined on the coordinates $(\vb*{\hat{n}},z)$, which maps to $(\vb*{\hat{n}},z_\text{obs})$ under the RSD effect. Using this fact together with definitions~\eqref{def: observed number overdensity} and \eqref{def: theoretical number overdensity}, we can write the relation between the observed and theoretical number overdensities as
\begin{equation}\label{eqn: RSD observed number overdensity}
    \Delta_g(\vb*{\hat{n}},z_{\text{obs}}) = (1+\delta_{g,\text{obs}}(\vb*{\hat{n}},z_\text{obs}))\left(1+\frac{\delta V(\vb*{\hat{n}},z_\text{obs})}{V(\vb*{\hat{n}},z_\text{obs})} \right),
\end{equation}
where $\delta V = V-\bar{V}$ is the volume perturbation from the FLRW background. This volume perturbation enters particularly through the Jacobian of the transformation from comoving coordinates to the observer's coordinates with the metric perturbation (see computation shortly after). The observed overdensity breaks down into two contributions: the overdensity field and the observed volume perturbation. Notice that the density contrast is the observed one. However, we would like to express this in terms of the true one. Since $\rho_g$ transforms as a scalar field, we can write $\rho_{g,\text{obs}}$ as a Taylor expansion of $\rho_g$,
\begin{equation}
    \rho_{g,\text{obs}}(\vb*{\hat{n}},z_\text{obs})=\rho_g(\vb*{\hat{n}},z)=\rho_g(\vb*{\hat{n}},z_\text{obs}-\delta z) = \sum^{\infty}_{n=0}\frac{1}{n!}\frac{d^n \rho_g}{d z^n}\Bigg|_{z_\text{obs}}(-\delta z)^n.
\end{equation}
We can rewrite $\rho_g=\bar{\rho}_g(1+\delta_g)$. So, we have the relation,
\begin{equation}\label{eqn: density perturbation expansion with redshift perturbation}
    1+\delta_{g,\text{obs}}(\vb*{\hat{n}},z_\text{obs})=\frac{1}{\bar{\rho}_g(z_\text{obs})} \sum^{\infty}_{n=0}\frac{1}{n!}\frac{d^n}{d z^n}[\bar{\rho}_g(1+\delta_g)]\Bigg|_{z_\text{obs}}(-\delta z)^n.
\end{equation}

\subsection{Redshift perturbation}\label{sec: Redshift Perturbation}
Before proceeding with the perturbative expansion of \eqref{eqn: RSD observed number overdensity}, let us identify the origin of the redshift perturbation. The misestimation of redshift arises from perturbations in the photon geodesics: specifically, the relativistic Doppler effect and the gravitational redshift contribute on top of the cosmological redshift. We assume the weak gravitational field regime and work in Poisson gauge, so that the space-time metric takes the form
\begin{equation}\label{eqn: Perturbed metric in Poisson gauge}
    ds^2 = g_{\mu\nu}dx^\mu dx^\nu=a^2(\tau)\left[ -(1+2\Psi)d\tau^2+(1-2\Phi)\delta_{ij}dx^idx^j \right]
\end{equation}
where the metric tensor $g_{\mu\nu}$ is the FLRW metric perturbed with scalar fields, $\Psi$ and $\Phi$. The choice of gauge is different in \cite{Yoo2009PhRvD..80h3514Y,Yoo2010PhRvD..82h3508Y,Bouvin2011PhRvD..84f3505B,Jeong2012PhRvD..85b3504J,Bertacca:2012tp_GRcorrect,Bertacca2014JCAP...09..037B, Gervani:2025wmh}, where they consider the synchronous-comoving gauge. Later, we will incorporate the bias expansion of the galaxy number density, which is valid in the synchronous-comoving gauge. However, we consider the general relativistic corrections up to $\mathcal{O}(\epsilon_\mathcal{H}^{1/2})$ only. Therefore, the choice of gauge here does not affect the final expression of $\Delta_g$ since the transformation between the two gauges is equal to or smaller than $\mathcal{O}(\epsilon_\mathcal{H})$ \cite{DiDio2019JCAP...04..050D,DiDio2020JCAP...09..058D}.

Suppose we have photons emitted from a source with 4-velocity
\begin{equation}\label{eqn: 4-velocity of the source}
    u^\mu_{(g)}=\frac{1}{a(\tau)}
    \begin{pmatrix}
        u^0_g \\
        v^i_g
    \end{pmatrix}.
\end{equation}
The photon then follows the null geodesic to the observer with 4-velocity of $u_{(o)}^\mu=\frac{1}{a} (u^0_o,v^i_o)^T$. The index $i\in\{1,2,3\}$ is the spatial index of the 4-vector, while index $0$ is the time index. The frequency of the photon in the frame of the source or observer is obtained by the dot product between their 4-velocity and the photon's wave 4-vector,
\begin{equation}
    k_\mu=(\omega,\; k_i)
\end{equation}
where $\omega$ and $k_i$ are the constant frequency and wave covector of the photon. The observed redshift factor is then computed from
\begin{equation}
    1+z_\text{obs}=\frac{u^{(g)}_\mu k^\mu}{u^{(o)}_\mu k^\mu}
\end{equation}
By exploiting the on-shell condition of a massive observer and a massless photon, $u^\mu u_\mu = -1$ and $k^\mu k_\mu = 0$, the measured redshift factor is
\begin{equation}
    1+z_\text{obs}=\underbrace{\left(\frac{a(\tau_g)}{a(\tau_o)}\right)}_{\text{cosmological}}\underbrace{\left(\sqrt{\frac{1+\Psi|_g}{1+\Psi|_o}}\right)}_{\text{gravitational}}\underbrace{\left(\frac{\sqrt{1+(1-2\Phi|_g)v^2_g}-\sqrt{1-2\Phi|_g}v_{g\|}}{\sqrt{1+(1-2\Phi|_o)v^2_o}-\sqrt{1-2\Phi|_o}v_{o\|}}\right)}_{\text{Doppler}},
\end{equation}
which has three contributing factors. The first parenthesis denotes the cosmological redshift due to the expansion of space, encoded in the scale factor ratio. This term represents the real redshift of the galaxy with respect to the observer,
\begin{equation}
    \frac{a(\tau_g)}{a(\tau_o)} = 1+z.
\end{equation}
The second encodes the gravitational redshift in the weak-field regime, which depends on the gravitational potential at both the source and observer positions. The last parenthesis represents the Doppler redshift, which decomposes into transverse and longitudinal (along LOS) terms. The integrated Sachs-Wolfe (ISW) contribution is neglected, as it appears at linear order in $\epsilon_{\mathcal{H}}$ in the final number density perturbation. This is because the ISW term involves an integral along the LOS of $\Psi'+\Phi'$, which is suppressed by an additional factor of $\epsilon_\mathcal{H}^{1/2}=\mathcal{H}/k$ compared to the local terms \cite{Bertacca2014JCAP...09..037B, Gervani:2025wmh}.

Hence, the full redshift deviation equation reads
\begin{equation}\label{eqn: full redshift deviation}
    1+z_\text{obs}=\left(1+z\right)\left(\sqrt{\frac{1+\Psi|_g}{1+\Psi|_o}}\right)\left(\frac{\sqrt{1+(1-2\Phi|_g)v^2_g}-\sqrt{1-2\Phi|_g}v_{g\|}}{\sqrt{1+(1-2\Phi|_o)v^2_o}-\sqrt{1-2\Phi|_o}v_{o\|}}\right),
\end{equation}
In this equation, the redshift deviation is sourced by the metric perturbations, $\Psi$ and $\Phi$, and the peculiar velocity, $v_g$. These quantities are of the order
\begin{equation}
    \mathcal{O}(\Psi) \sim \mathcal{O}(\Phi) \sim \mathcal{O}(\epsilon_\mathcal{H}\delta) \quad \text{and} \quad \mathcal{O}(v_g) \sim \mathcal{O}(\epsilon_\mathcal{H}^{1/2}\delta),
\end{equation}
which can be obtained by performing order estimation on the Poisson equation~\eqref{eqn: Poisson equation} and Euler equations~\eqref{eqn: dark matter fluid equation}.\footnote{Another ingredient of this estimation is $\mathcal{O}(v_g)\sim\mathcal{O}(v)$ presented by the no-velocity bias condition in equation~\eqref{eqn: no velocity bias}.} We are interested only in the GR correction to the number density up to $\mathcal{O}(\epsilon_\mathcal{H}^{1/2})$. For completeness, let us keep the term up to $\mathcal{O}(\epsilon_\mathcal{H})$ for now. The reason will become clear shortly. Therefore, we can expand the equation~\eqref{eqn: full redshift deviation} to the first order of $\epsilon_\mathcal{H}$,
\begin{equation}\label{eqn: approximated redshift deviation}
    1+z+\delta z = (1+z)\left[ 1+ \left( -v_{\|}-\Psi+\frac{v^2}{2} \right)\Bigg|^g_o\right]+\mathcal{O}(\epsilon_\mathcal{H}^{3/2}),
\end{equation}
We also substitute $z_\text{obs}=z+\delta z$. The source of redshift perturbation $\delta z$ is the difference between the terms in the parentheses evaluated at the source and the observer. Hence, we have the expression of redshift perturbation in quantities evaluated at the source's redshift $z$ and quantities at the observer. For an expression at the observed redshift, $z_\text{obs}$, we Taylor expand the expression around $z=z_\text{obs}-\delta z$,
\begin{equation}\label{eqn: redshift perturbation}
    \delta z = \sum_{n=0}^\infty\frac{1}{n!}(-\delta z)^n\frac{d^n}{dz^n}\left[ (1+z_\text{obs})\left( \frac{v_g^2}{2}-v_{g\|}-\Psi-\frac{v_o^2}{2}+v_{o\|}+\Psi_o \right) \right]+\mathcal{O}(\epsilon_\mathcal{H}^{3/2}),
\end{equation}
where the quantities with no subscript are evaluated at the observed redshift $z_\text{obs}$. Let us review the order of perturbation in the framework of $\epsilon_\mathcal{H}$ mentioned at the beginning of the section. Equation~\eqref{eqn: approximated redshift deviation} implies that the leading order of $\delta z$ is $\mathcal{O}(\epsilon_\mathcal{H}^{1/2})$, which implies that $\frac{d}{dz}\sim \mathcal{O}(\epsilon_\mathcal{H}^{-1/2})$. Since we are interested in the leading GR correction of order $\mathcal{O}(\epsilon_\mathcal{H}^{1/2})$ to the number density, we only neglect terms in equation~\eqref{eqn: approximated redshift deviation} if they are smaller than $\mathcal{O}(\epsilon_\mathcal{H})$. By doing so, the Taylor's expansion around $z=z_\text{obs}-\delta z$ in equation~\eqref{eqn: density perturbation expansion with redshift perturbation} captures every correction at order $\mathcal{O}(\epsilon_\mathcal{H}^{1/2})$.

The redshift perturbation can be solved from equation~\eqref{eqn: redshift perturbation} perturbatively. Given an ansatz $\delta z=\sum_i\delta z^{(i)}$ while $\delta z^{(i)}\sim\mathcal{O}(\epsilon_\mathcal{H}^{i/2})$, we can iteratively solve the perturbation order by order, by substituting the lower-order solution back into the equation~\eqref{eqn: redshift perturbation}. For example, the first-order solution is
\begin{equation}
    \delta z^{(1)}=(1+z_\text{obs})\left( \frac{v_g^2}{2}-v_{g\|}-\Psi-\frac{v_o^2}{2}+v_{o\|}+\Psi_o \right)+\mathcal{O}(\epsilon_\mathcal{H}^{3/2}).
\end{equation}
Notice that the redshift perturbation has two contributions of different orders in $\epsilon_\mathcal{H}$,
\begin{equation}
    \begin{split}
        \delta z^{(1)}&=\delta z_N^{(1)}+\delta z_R^{(1)}\;; \\
        \delta z_N^{(1)}&=-(1+z_\text{obs})(v_{g\|}-v_{o\|}), \\
        \delta z_R^{(1)}&=-(1+z_\text{obs})\left( \Psi-\Psi_o-\frac{v^2_g}{2}+\frac{v_o^2}{2} \right).
    \end{split}
\end{equation}
They represent the Newtonian and relativistic parts of the redshift distortion, respectively, where $\delta z_R\sim\mathcal{O}(\epsilon_\mathcal{H}^{1/2}\delta z_N)$. The Newtonian part corresponds to the linear Doppler effect, while the relativistic part encapsulates the transverse Doppler and gravitational redshift contributions. Solving the higher-order solutions and resumming them gives the final expression of
\begin{equation}\label{eqn: redshift perturbation final}
    \begin{split}
        \delta z =& -\sum_{i=0}^\infty\frac{1}{(i+1)!}\frac{d^i}{dz^i}\left[ (1+z_\text{obs})\left( \frac{v_g^2}{2}-v_{g\|}-\Psi-\frac{v_o^2}{2}+v_{o\|}+\Psi_o \right) \right]^{i+1}\Bigg|_{\mathcal{O}(\epsilon_\mathcal{H})} \\
        &+\mathcal{O}(\epsilon_\mathcal{H}^{3/2}).
    \end{split}
\end{equation}
One can check that this expression also satisfies equation~\eqref{eqn: redshift perturbation}. The leading contribution to $\delta z$ is at $\mathcal{O}(\epsilon_\mathcal{H}^{1/2})$ where the infinite series converges as the higher-order $\delta z^{(i)}$ shrinks incrementally with $\epsilon_\mathcal{H}^{1/2}$. Here, we keep the contributions up to $\mathcal{O}(\epsilon_\mathcal{H})$. However, we remind the reader that the field variables $v$ and $\Psi$ are non-perturbative, as we do not write them in perturbative orders. Effectively, every order of field perturbations in contributions of order lower than $\mathcal{O}(\epsilon_\mathcal{H}^{3/2})$ is resummed in this expression.

\subsection{Density perturbation}
Now that we have a formal expression for the redshift perturbation including GR corrections, we substitute expression \eqref{eqn: redshift perturbation} into equation~\eqref{eqn: density perturbation expansion with redshift perturbation} to obtain the density perturbation. To be consistent within our framework, we keep only the contribution at $\mathcal{O}(\epsilon_\mathcal{H}^{1/2})$ and neglect others from $\mathcal{O}(\epsilon_\mathcal{H})$. The approximate result is \cite{DiDio2020JCAP...09..058D} 
\begin{equation}\label{eqn: relativistic tracer density perturbation}
    \begin{split}
        &\delta_{g,\text{obs}}(\vb*{\hat{n}},z_\text{obs}) = \\
        &\sum_{n=0}^\infty\frac{1}{n!}\left(\frac{\delta z}{\mathcal{H}(1+z_\text{obs})}\right)^n\left[ \partial_r^n+\mathcal{H}n\left( \frac{1-n}{2} \left( 1-\frac{\mathcal{H}'}{\mathcal{H}^2} \right)+3-b_e\right)\partial_r^{n-1} - n\partial_r^{n-1}\partial_\tau \right]\delta_g \Bigg|_{z_\text{obs}} \\
        &+\mathcal{O}(\epsilon_\mathcal{H}),
    \end{split}
\end{equation}
where prime indicates the time derivative, $\partial_\tau$, $\partial_r=\vb*{\hat{n}}\cdot \vb*{\nabla}$ is the spatial derivative projected along LOS, and the term at $n=0$ is exclusively $\delta_g$. We remind the reader that this theoretical density field $\delta_g$ is defined on spatial (comoving) coordinates rather than the observer's redshift and LOS coordinates. Meanwhile, this expression is exactly the transformation of $\delta_g$ to the one defined on the observer's coordinates $\delta_{g,\text{obs}}$. All of the derivative operators in the bracket measure the rate of change of $\delta_g$ with redshift, which can be broken down into three contributions. The first operator accounts for the comoving distance gradient along the LOS. The second operator accounts for the time evolution of the background, i.e., the Hubble parameter $\mathcal{H}$ and the background number density $\bar{\rho}_g$. The latter is captured by the so-called \emph{evolution bias parameter} $b_e$.  The last operator accounts for the time evolution of $\delta_g$ itself. The higher-order time derivatives of $\delta_g$ are neglected as they are suppressed below $\mathcal{O}(\epsilon_\mathcal{H}^{1/2})$. In particular, after a binomial expansion of the $n$th-order redshift derivatives, the term $\delta z^n\partial_r^{n-m}\partial_\tau^m\sim\mathcal{O}(\epsilon_\mathcal{H}^{m/2})$ as $\delta z^n\sim\epsilon_\mathcal{H}^{n/2}$, $\partial_r^{n}\sim\epsilon_\mathcal{H}^{-n/2}$, and $\partial_\tau^{n}\sim\epsilon_\mathcal{H}^0$.

In obtaining the approximate formula above, we exchange the derivative with respect to the redshift for the time derivative and the derivative with respect to the comoving distance via the chain rule,
\begin{equation}
    \frac{d}{dz} = \frac{dr}{dz}\partial_r + \frac{d\tau}{dz}\partial_\tau = \frac{1}{\mathcal{H}(z)(1+z)}\left( \partial_r - \partial_\tau\right),
\end{equation}
where the definition of the comoving distance
\begin{equation}
    dr = -d\tau = -\frac{da}{a'}=a\frac{dz}{\mathcal{H}(z)}=\frac{dz}{\mathcal{H}(z)(1+z)}
\end{equation}
is exploited. $\mathcal{H}=a'/a$ is the comoving Hubble parameter, and prime denotes the conformal time derivative. In this expression, the evolution bias parameter
\begin{equation}
    b_e = 3-(1+z)\frac{d\ln{\bar{\rho}_g}}{dz},
\end{equation}
accounts for the deviation of the tracer's redshift evolution from that of the background matter density, which decays as $(1+z)^3$. If the galaxy number density and the matter density evolve identically with redshift, then $b_e=0$. For a typical low-redshift galaxy survey, e.g., the sample used in  \cite{Hou:2022wfj, Philcox:2022hkh, Slepian:2025kbb}, simulations in \cite{Beutler2020JCAP...07..048B} predict that $b_e$ typically lies between 0 and 3 at $z<0.45$ and grows slowly with $z$ (see Figure 2 in that reference). However, the result of this paper is not sensitive to the value of $b_e$, which we treat as a free parameter. The redshift derivatives of $b_e$ appear in the higher-order time derivative terms, so they contribute at an order higher than $\mathcal{O}(\epsilon_\mathcal{H}^{1/2})$ and thus are neglected.

\subsection{Volume perturbation}

Next, we consider the volume perturbation in the context of the RSD effect. We define an infinitesimal volume element around a source,
\begin{equation}\label{eqn: volume element}
    dV=\sqrt{-g}\epsilon_{\mu\nu\rho\sigma}u^\mu_{\text{(s)}}dx^\nu dx^\rho dx^\sigma = v(z,\theta,\varphi)dz d\theta d\varphi,    
\end{equation}
where $u^\mu_{\text{(s)}}$ is the 4-velocity of the source as defined in \eqref{eqn: 4-velocity of the source} and $v$ is the volume density defined on the observer coordinates. Therefore, the ratio in the second parenthesis of \eqref{eqn: RSD observed number overdensity} is computed through
\begin{equation}
    \frac{\delta V}{V} = \frac{dV_\text{obs}-dV}{dV}=\frac{v_\text{obs}-\bar{v}}{\bar{v}},
\end{equation}
where $dV$ and $dV_\text{obs}$ are the infinitesimal volume elements computed on the unperturbed background and perturbed background, respectively. This ratio is particularly important for \eqref{eqn: RSD observed number overdensity} as the observed density is sensitive to the volume, which appears distorted to the observer by the metric perturbation and RSD effect. Hence, we need to compute the volume density in two cases: a stationary source on an unperturbed background, and a moving source on a perturbed background. From equation~\eqref{eqn: volume element}, we can write the volume density as
\begin{equation}\label{eqn: volume density}
    v(z,\theta,\varphi) = \sqrt{-g}\epsilon_{\mu\nu\rho\sigma}u^\mu_{\text{(s)}}\frac{\partial x^\nu}{\partial z}\frac{\partial x^\rho}{\partial \theta} \frac{\partial x^\sigma}{\partial \varphi}.
\end{equation}
For the unperturbed case, we have  $u^\mu_{\text{(s)}}=\frac{1}{a}(1,\vb*{0})^T$ and the flat FLRW metric. In this case, we have $\sqrt{-g}=a^4r^2\sin\theta$ in spherical coordinates, and the derivative of the comoving distance with respect to the redshift gives $\partial r/\partial z =a\mathcal{H}^{-1}$. So, the unperturbed volume density can be written as
\begin{equation}
    \bar{v}(z,\theta,\varphi)= \sqrt{-g}\frac{\partial r}{\partial z}\frac{\partial \theta}{\partial \theta} \frac{\partial \varphi}{\partial \varphi}=\frac{a^4(z)r^2(z)\sin{\theta}}{\mathcal{H}(z)}.
\end{equation}
For the perturbed case, the 4-velocity \eqref{eqn: 4-velocity of the source} is used instead,
\begin{equation}
    v_\text{obs}(z_\text{obs},\theta_\text{obs},\varphi_\text{obs})=\frac{\sqrt{-g}}{a}\left(u^0_g\frac{\partial r}{\partial z_\text{obs}}+u^r_g\frac{\partial \tau}{\partial z_\text{obs}}\right)\frac{\partial \theta}{\partial \theta_\text{obs}} \frac{\partial \varphi}{\partial \varphi_\text{obs}}.
\end{equation}
Meanwhile, the perturbed metric in Poisson gauge \eqref{eqn: Perturbed metric in Poisson gauge} yields
\begin{equation}
    \sqrt{-g}=a^4r^2\sin\theta\sqrt{(1+2\Psi)(1-2\Phi)^3}=a\mathcal{H}^{-1}\bar{v}+\mathcal{O}(\epsilon_\mathcal{H}),
\end{equation}
as $\Psi, \Phi\sim\epsilon_\mathcal{H}\delta$. Therefore, perturbing the redshift $z=z_\text{obs}-\delta z$ of this metric determinant gives
\begin{equation}
    \sqrt{-g(z_\text{obs})}=a\mathcal{H}^{-1}\left(1-\frac{1}{\bar{v}}\frac{d\bar{v}}{dz}\delta z\right)+\mathcal{O}(\epsilon_\mathcal{H}).
\end{equation}
The derivatives with respect to observed redshift can be written as
\begin{equation}
    \frac{\partial \tau}{\partial z_\text{obs}}=-\frac{\partial r}{\partial z_\text{obs}}=-\frac{\partial r}{\partial z}\frac{d z}{d z_\text{obs}}=-\mathcal{H}^{-1}\left(1+\frac{d \delta z}{d z}\right)^{-1}=-\mathcal{H}^{-1}\left(1-\frac{d \delta z}{d z}\right)+\mathcal{O}(\epsilon_\mathcal{H})
\end{equation}
Combining the time and radial components of the source's 4-velocity gives
\begin{equation}
    u^0_g-u^r_g=1+\frac{v_g^2}{2}-\Psi-v_{g\|}+\mathcal{O}(\epsilon_\mathcal{H}^{2})=1+\frac{\delta z}{1+z_\text{obs}}-v_{o\|}+\mathcal{O}(\epsilon_\mathcal{H}),
\end{equation}
where we use the redshift perturbation expression \eqref{eqn: redshift perturbation final} up to the first order. Therefore, the final expression is
\begin{equation}
    v_\text{obs}(z_\text{obs},\theta_\text{obs},\varphi_\text{obs})=\bar{v}\left(1-\frac{1}{\bar{v}}\frac{d\bar{v}}{dz}\delta z\right)\left( 1-\frac{d\delta z}{dz} \right) \left ( 1+\frac{\delta z}{1+z_\text{obs}}-v_{o\|} \right)+\mathcal{O}(\epsilon_\mathcal{H}).
\end{equation}
Note that we assume no angular deviation, $\frac{\partial\theta}{\partial\theta_\text{obs}} = \frac{\partial\varphi}{\partial\varphi_\text{obs}} = 1$. This angular part of the Jacobian corresponds to the lensing contribution, which we neglect throughout this work. This is consistent with the contribution truncation at $\mathcal{O}(\epsilon_\mathcal{H}^{1/2})$ in the parity-odd sector of the 4PCF as \cite{Bouvin2014PhRvD..89h3535B,Lepori2018JCAP...05..043L} show that the lensing dipole contribution is suppressed by an additional factor of $\mathcal{H}r\sim\mathcal{H}/k=\epsilon_\mathcal{H}^{1/2}$ relative to other relativistic contributions, which already start at $\epsilon_\mathcal{H}^{1/2}$ (see discussion in Section~\ref{sec: Discussion and Potential Sources of Violation}). Now, we have the volume perturbation,
\begin{equation}\label{eqn: relativistic volume perturbation}
    \frac{\delta V(\vb*{\hat{n}},z_\text{obs})}{V(\vb*{\hat{n}},z_\text{obs})}=\left(1-\frac{1}{\bar{v}}\frac{d\bar{v}}{dz}\delta z\right)\left( 1-\frac{d\delta z}{dz} \right) \left ( 1+\frac{\delta z}{1+z_\text{obs}}-v_{o\|} \right)-1+\mathcal{O}(\epsilon_\mathcal{H}).
\end{equation}

\subsection{Relativistic number density in perturbation theory}

Substituting the perturbed density \eqref{eqn: relativistic tracer density perturbation} and volume \eqref{eqn: relativistic volume perturbation} into the equation~\eqref{eqn: RSD observed number overdensity} yields
\begin{equation}
    \begin{split}\label{eqn: full relativistic Number Density}
        &\Delta_g(\vb*{\hat{n}},z_\text{obs})= \\
        &\sum_{n=0}^\infty\frac{1}{n!}\left(\frac{\delta z}{\mathcal{H}(1+z_\text{obs})}\right)^n\left[ \partial_r^n+\mathcal{H}n\left( \frac{1-n}{2} \left( 1-\frac{\mathcal{H}'}{\mathcal{H}^2} \right)+3-b_e\right)\partial_r^{n-1} - n\partial_r^{n-1}\partial_\tau \right](1+\delta_g) \Bigg|_{z_\text{obs}} \\
        &\times\left(1-\frac{1}{\bar{v}}\frac{d\bar{v}}{dz}\delta z\right)\left( 1-\frac{d\delta z}{dz} \right) \left ( 1+\frac{\delta z}{1+z_\text{obs}}-v_{o\|} \right)-1+\mathcal{O}(\epsilon_\mathcal{H}).
    \end{split}
\end{equation}
We then substitute the redshift perturbation \eqref{eqn: redshift perturbation final} to express the observed number density purely in terms of physical fields. Although we retain only the leading relativistic correction, the result is non-perturbative in $\delta_g$, $v_g$, and $\Psi$, as we argued at the end of Section~\ref{sec: Redshift Perturbation}. This is possible as the expression of redshift perturbation \eqref{eqn: redshift perturbation final} does not require us to write those fields in perturbative order. The order of truncation is rather controlled by the relativistic perturbation parameter $\epsilon_\mathcal{H}$, and the fields in this expression remain non-perturbative. 

However, as we are considering the cosmic structure on a large scale, the fields will henceforth be treated as perturbations. Within the framework of LSS perturbation theory, the physical fields are expanded into
\begin{equation}\label{eqn: perturbative expansion of galaxy number density, velocity, and potential}
    \delta_g=\sum_{n=0}^\infty\delta_g^{(n)} \quad \text{,} \quad v_g=\sum_{n=0}^\infty v_g^{(n)} \quad \text{,} \quad \Psi=\sum_{n=0}^\infty \Psi^{(n)},
\end{equation}
where $n$ is the perturbative order of the fields. We can substitute these expansions into the master equation~\eqref{eqn: full relativistic Number Density} and recover the expression of the observed number density at each order of the field perturbation. The results up to third order are reproduced from \cite{DiDio2019JCAP...04..050D,DiDio2020JCAP...09..058D}, where the Newtonian part and the relativistic part are separated following the introduced framework. The first order is
\begin{equation}\label{eqn: first-order observed galaxy number density contrast}
    \Delta_g^{(1)}(\vb*{\hat{n}}, z_{\text{obs}})= \delta_g+\mathcal{H}^{-1} \partial_r v_{g\|}+v_{g\|}\left(\frac{\mathcal{H}'}{\mathcal{H}^2}+\frac{2}{\mathcal{H} r}-b_e\right)+\mathcal{O}\left(\epsilon_{\mathcal{H}}\right),
\end{equation}
where the superscript $(1)$ for the first-order field perturbations is omitted for brevity. In this expression, we see the intrinsic number density contrast $\delta_g$, and the rest are the leading contributions from the observational effect. The second term $\mathcal{H}^{-1}\partial_r v_{g\|}$ is the standard linear RSD term. The third term proportional to $v_{g\|}$ collects the relativistic corrections, including the evolution bias $b_e$ coming from the density perturbation, the geometrical term $2/(\mathcal{H}r)$ coming from the volume perturbation, and the Hubble rate variation $\mathcal{H}'/\mathcal{H}^2$ coming from both. 
The second order is
\begin{equation}\label{eqn: second-order observed galaxy number density contrast}
    \begin{split}
    \Delta_g^{(2)}(\vb*{\hat{n}}, z_{\text{obs}})= & \Delta_N^{(2)}(\vb*{\hat{n}}, z_{\text{obs}})+\Delta_R^{(2)}(\vb*{\hat{n}}, z_{\text{obs}})+\mathcal{O}\left(\epsilon_{\mathcal{H}}\right), \\
    \Delta_N^{(2)}(\vb*{\hat{n}}, z_{\text{obs}})= & \delta_g^{(2)}+\mathcal{H}^{-1} \partial_r v_{g\|}^{(2)}+\mathcal{H}^{-1} \partial_r\left(v_{g\|} \delta_g\right)+\mathcal{H}^{-2} \partial_r\left(v_{g\|} \partial_r v_{g\|}\right), \\
    \Delta_R^{(2)}(\vb*{\hat{n}}, z_{\text{obs}})= & \left(-1+\frac{\mathcal{H}'}{\mathcal{H}^2}+\frac{2}{\mathcal{H} r}-b_e\right)\left(v_{g\|}^{(2)}+v_{g\|} \delta_g\right)-\mathcal{H}^{-1} \left(v_{g\|}^{(2)}\right)'-2 \mathcal{H}^{-2} \partial_r v_{g\|} v'_{g\|} \\
    & +\left(-2+3 \frac{\mathcal{H}'}{\mathcal{H}^2}+\frac{4}{\mathcal{H} r}-2 b_e\right) \mathcal{H}^{-1} v_{g\|} \partial_r v_{g\|}-2 \mathcal{H}^{-2} v_{g\|} \partial_r v'_{g\|} \\
    & -\mathcal{H}^{-1} v'_{g\|} \delta_g-\mathcal{H}^{-1} v_{g\|} \delta_g '+\mathcal{H}^{-1} v^i \partial_i v_{g\|}+\mathcal{H}^{-2} \Psi \partial_r^2 v_{g\|}+\mathcal{H}^{-1} \Psi \partial_r \delta_g \\
    & +\mathcal{H}^{-1} \partial_r \Psi^{(2)}+2 \mathcal{H}^{-2} \partial_r v_{g\|} \partial_r \Psi+\mathcal{H}^{-1} \delta_g \partial_r \Psi+\mathcal{H}^{-2} v_{g\|} \partial_r^2 \Psi . \\
    \end{split}
\end{equation}
We can organise the terms, as promised at the beginning, into the Newtonian contribution $\Delta_N$ and the relativistic contribution $\Delta_R$, where $\Delta_R\sim\epsilon_\mathcal{H}^{1/2}\Delta_N$. $\Delta_N$ consists of the intrinsic $\delta_g$ and the standard RSD terms, where the couplings between $\delta_g$ and $v_{g\|}$ are the nonlinear RSD terms, in addition to the linear structure that already appeared at the first order. $\Delta_R$ contains all terms at $\mathcal{O}(\epsilon_{\mathcal{H}}^{1/2})$ arising from redshift perturbations through number-density and volume perturbations, involving time derivatives and transverse velocities. Those terms represent relativistic corrections to these nonlinear RSD couplings. The third order is
\begin{equation}\label{eqn: third-order observed galaxy number density contrast}
    \begin{split}
    \Delta_g^{(3)}(\vb*{\hat{n}}, z_{\text{obs}})= & \Delta_N^{(3)}(\vb*{\hat{n}}, z_{\text{obs}})+\Delta_R^{(3)}(\vb*{\hat{n}}, z_{\text{obs}})+\mathcal{O}\left(\epsilon_{\mathcal{H}}\right), \\
    \Delta_N^{(3)}(\vb*{\hat{n}}, z_{\text{obs}})= & \delta_g^{(3)}+\frac{\partial_r v_{g\|}^{(3)}}{\mathcal{H}}+\left[\mathcal{H}^{-1} \partial_r\left(v_{g\|} \delta_g\right)\right]^{(3)}+\left[\mathcal{H}^{-2} \partial_r\left(v_{g\|} \partial_r v_{g\|}\right)\right] \\
    & +\frac{1}{6} \mathcal{H}^{-3} \partial_r^3 v_{g\|}^3+\frac{1}{2} \mathcal{H}^{-2} \partial_r^2\left(\delta_g v_{g\|}^2\right), \\
    \Delta_R^{(3)}(\vb*{\hat{n}}, z_{\text{obs}})= & \left\{\left(-1+\frac{\mathcal{H}'}{\mathcal{H}^2}+\frac{2}{\mathcal{H} r}-b_e\right)\left(v_{g\|}^{(3)}+\left[v_{g\|} \delta_g\right]^{(3)}\right)-\mathcal{H}^{-1} \left(v_{g\|}^{(3)}\right)'\right. \\
    & -2 \mathcal{H}^{-2}\left[\partial_r v_{g\|} v'_{g\|}\right]^{(3)}+\left(-2+3 \frac{\mathcal{H}'}{\mathcal{H}^2}+\frac{4}{\mathcal{H} r}-2 b_e\right)\mathcal{H}^{-1}\left[ v_{g\|} \partial_r v_{g\|}\right]^{(3)} \\
    & -2 \mathcal{H}^{-2}\left[v_{g\|} \partial_r v'_{g\|}\right]^{(3)}-\mathcal{H}^{-1}\left[v'_{g\|} \delta_g\right]^{(3)}-\mathcal{H}^{-1}\left[v_{g\|} \delta_g '\right]^{(3)}+\mathcal{H}^{-1}\left[v_g^i \partial_i v_{g\|}\right]^{(3)} \\
    & +\mathcal{H}^{-2}\left[\Psi \partial_r^2 v_{g\|}\right]^{(3)}+\mathcal{H}^{-1}\left[\Psi \partial_r \delta_g\right]^{(3)}+\mathcal{H}^{-1} \partial_r \Psi^{(3)}+2 \mathcal{H}^{-2}\left[\partial_r v_{g\|} \partial_r \Psi\right]^{(3)} \\
    & \left.+\mathcal{H}^{-1}\left[\delta_g \partial_r \Psi\right]^{(3)}+\mathcal{H}^{-2}\left[v_{g\|} \partial_r^2 \Psi\right]^{(3)}\right\} \\
    & +\frac{1}{2 \mathcal{H}^3} \partial_r^3\left(v_{g\|}^2 \Psi\right)-\frac{1}{2 \mathcal{H}^3} \partial_\tau \partial_r^2 v_{g\|}^3+\frac{1}{\mathcal{H}^2} \partial_r^2\left(\Psi v_{g\|} \delta_g\right)-\frac{1}{2 \mathcal{H}^2} \partial_r^2\left(v_{g\|} v_g^2\right) \\
    & -\frac{1}{\mathcal{H}^2} \partial_\tau \partial_r\left(\delta_g v_{g\|}^2\right)-\frac{1}{2 \mathcal{H}^2} \partial_r\left(\delta_g v_g^2\right)+\frac{1}{\mathcal{H}^2} \partial_r\left(\partial_r v_{g\|} v_{g\|}^2\right)\left(3 \frac{\mathcal{H}'}{\mathcal{H}^2}+\frac{3}{\mathcal{H} r}-\frac{3}{2} b_e\right) \\
    & +\frac{1}{\mathcal{H}} \partial_r\left(\delta_g v_{g\|}^2\right)\left(-\frac{1}{2}+\frac{3}{2} \frac{\mathcal{H}'}{\mathcal{H}^2}+\frac{2}{\mathcal{H} r}-b_e\right) .
    \end{split}
\end{equation}
where the square brackets denote all possible combinations of quantities that yield the indicated perturbation order. In particular, the bracket $[AB]^{(3)}$ in the expression above means first-order A multiplied by second-order B plus vice versa. The term does not include the third-order fields. For example, $\left[\left(v_{g\|} \delta_g\right)\right]^{(3)}=v_{g\|}^{(2)}\delta_g+v_{g\|}\delta_g^{(2)}$, and does not include $v_{g\|}^{(3)}$ or $\delta_g^{(3)}$. All quantities are evaluated at $z_{\text{obs}}$. For instance, $\delta_g=\delta_g(\vb*{r}(z_{\text{obs}}),\tau(z_{\text{obs}}))$. We will henceforth drop the subscript "obs" for brevity. In obtaining these expressions, the first-order Euler equation,
\begin{equation}
    {v^{(1)}_{g\|}}'+\mathcal{H}v^{(1)}_{g\|}-\partial_r\Psi^{(1)}=0,
\end{equation}
was used to eliminate certain terms, along with the repeatedly used relation,
\begin{equation}
    \frac{1+z}{\bar{v}}\frac{d\bar{v}}{dv}=\frac{\mathcal{H}'(z)}{\mathcal{H}^2(z)}+\frac{2}{\mathcal{H}(z)r(z)}-4.
\end{equation}

In this work, all quantities evaluated at the observer position are set to zero, $v_o\equiv \Psi_o\equiv 0$. These can be reinstated if needed, as done for $v_o$ in the N-body simulation of \cite{Beutler2020JCAP...07..048B}. Neither of these quantities can introduce the parity-odd mode into the 4PCF at the monopole level. $\Psi_o$ is a true scalar quantity and does not carry any preferred direction, which guarantees invariance under parity transformation. $\vb*{v}_{o}$ is a constant vector with a preferred direction. It enters the observed galaxy number density through the redshift perturbation, which is introduced only in the form of $v_{o\|}$ and $v_o^2$. Any derivative of this quantity is zero, as it is a constant vector. Like the galaxy peculiar velocity $v_{g\|}$, $v_{o\|}$ is odd in the LOS. Therefore, the only potential Parity Violation in $\Delta_g$ from $v_{o\|}$ is the term arising from the replacement $v_{g\|}\longrightarrow v_{g\|}-v_{o\|}$, which are some terms replacing $v_{g\|}$ with constant $v_{o\|}$. What matters here is that the power of $\vb*{\hat{n}}$ stays the same after the replacement, but $v_{o\|}$ is a constant term with no fluctuation, unlike $v_{g\|}$. The term linear in $v_{o\|}$ is the global dipole contribution described in, e.g., \cite{Bouvin2014PhRvD..89h3535B}. As we will conclude in this paper, every odd-in-LOS contribution to 4PCF from these terms is washed away by averaging over LOS. The LOS-related parity-odd contribution carried in $v_{o\|}$ will be removed identically. However, this work does not consider the effect of LOS aberration introduced by the perpendicular term $\vb*{v}_{o}$ \cite{Bertacca2014JCAP...09..037B, Gervani:2025wmh}, as we are assuming the LOS is unperturbed. Therefore, the full consideration of potential PV from $\vb*{v}_{o}$ remains outside the scope of this work.

All contributions integrated along the LOS, including the lensing, ISW, and Shapiro time delay terms, are neglected. The ISW and Shapiro time delay terms contribute at order $\mathcal{O}(\epsilon_\mathcal{H})$ and are therefore subdominant in this framework. The lensing term, while comparable to the Newtonian contribution in magnitude \cite{Hui2007PhRvD..76j3502H, Hui2008PhRvD..77f3526H}, contributes to the parity-odd correlation function only at a level suppressed by an additional factor of $\mathcal{H}r\sim\mathcal{H}/k=\epsilon_\mathcal{H}^{1/2}$ below the Doppler contribution \cite{Bouvin2014PhRvD..89h3535B, Lepori2018JCAP...05..043L} (see discussion in Section~\ref{sec: Discussion and Potential Sources of Violation}). It is therefore neglected throughout this work. 
\section{Kernel solutions of the observed galaxy number density}\label{sec: Kernel Solutions of the Observed Galaxy Number Density}
The Fourier transform of the observed galaxy number density contrast is given by
\begin{equation}
    \Delta_{g}^{(n)}(\vb*{r}) = \int_{\vb*{k}}\Delta_{g,\vb*{k}}^{(n)}~e^{i\vb*{k}\cdot\vb*{\vb*{\hat{n}}}r},
\end{equation}
where $\vb*{r}=r(z)\vb*{\hat{n}}$ is the position in the redshift space, while $r=r(z)$ is the comoving distance. Since our goal is to compute the observed 4PCF in redshift space, we do not require the exact form of $\Delta^{(n)}_{g,\vb*{k}}$. Therefore, we do not Fourier transform the full expressions in equations \eqref{eqn: first-order observed galaxy number density contrast}, \eqref{eqn: second-order observed galaxy number density contrast}, and \eqref{eqn: third-order observed galaxy number density contrast}. Instead, we factor out the redshift-dependent quantities, such as $r(z)$ and $\mathcal{H}(z)$, and apply the inverse Fourier transform only to the fields $v_g$, $v_{g\|}$, $\delta_g$, $\Psi$, and their derivatives. Accordingly, we write
\begin{equation}\label{eqn: local term of delta_g}
    \Delta_{g}^{(n)}(\vb*{\hat{n}},z) = \sum^{t_n}_{r=1} Z_{r}(z)\mathcal{A}_r(\vb*{r},\tau ),
\end{equation}
where $t_n$ is the number of terms in the expression of $\Delta_{g}^{(n)}$ and $\mathcal{A}_r$'s denote the inverse Fourier transforms of those terms, which are products of $\delta^{(n)}_g$, $v_{g\|}^{(n)}$, $\vb*{v_g}^{(n)}$, $\Psi^{(n)}$, and their derivatives. $Z_r(z)$'s collect the purely redshift-dependent factors, involving $\mathcal{H}(z)$ and the comoving distance $r(z)$. For example, the first term in $\Delta_g^{(3)}$ consists of the redshift factor, $\left( -1+\frac{\mathcal{H}'}{\mathcal{H}^2}+\frac{2}{\mathcal{H}r}-b_e \right)$ multiplied by the field $v^{(3)}_{g\|}(\vb*{r},\tau)$. This decomposition follows a similar approach to that of \cite{Raccanelli:2013dza_GRpot, Bertacca:2012tp_GRcorrect, Raccanelli:2016avd_wideangleandGR}. This section demonstrates that the observed galaxy number density contrast up to third order can be written in the integral form,
\begin{equation}\label{eqn: Fourier observed galaxy number density constrast}
    \Delta_g^{(n)}(\vb*{\hat{n}},z)=\sum^{R_n}_{r=1}\mathcal{Z}_{n,r}(z)\int_{\vb*{k}}e^{-i\vb*{k}\cdot\vb*{\vb*{\hat{n}}}r}\int_{\vb*{k_1}} \dots \int_{\vb*{k_n}}\delta^{(3)}_D\left(\vb*{k}-\sum^n_{\sigma=1}\vb*{k_\sigma}\right)\mathcal{S}_{n,r}(\vb*{k_1},\dots ,\vb*{k_n};\vb*{\hat{n}})\prod^n_{\sigma=1}\delta^{(1)}_{\vb*{k_\sigma}},
\end{equation}
where $R_n$ is the number of linearly separated kernels, which is $t_n$ multiplied by the number of subkernels in each term of $\Delta_g^{(n)}$. The discrepancy between the numbers of summation terms in real space (equation~\ref{eqn: local term of delta_g}) and Fourier space (equation~\ref{eqn: Fourier observed galaxy number density constrast}), i.e., $R_n$ and $t_n$, comes from the fact that some term in $\Delta_g^{(n)}$ may be decomposed into multiple subkernels $\mathcal{S}_{n,r}$ in Fourier space. For example, the term $\mathcal{H}^{-1}\partial_r(v_{g\|}\delta_{g\|})$ from $\Delta^{(2)}_N$ in Fourier space is the combination of two subkernels: from $v_{g\|}\partial_r\delta_{g\|}$ and $\delta_{g\|}\partial_rv_{g\|}$. Let us start with some remarks on equations \eqref{eqn: first-order observed galaxy number density contrast}, \eqref{eqn: second-order observed galaxy number density contrast}, and \eqref{eqn: third-order observed galaxy number density contrast}.
\begin{enumerate}
    \item Ignoring time derivatives, $\mathcal{A}_r$'s take the general form,
    \begin{equation}\label{eqn: general form of relativistic correction terms}
    \partial_r^a\left[\prod_i\prod_j\big(\partial_r^iv^j_{g\|}\big)^{b_{ij}}\big(\partial_r^i\delta^j_{g}\big)^{c_{ij}}\big(\partial_r^i\Psi^j\big)^{d_{ij}}\right]^{(n)}\;; \quad a,b_{ij},c_{ij},d_{ij} \in \mathbb{Z},
    \end{equation}
    except for the terms $[v_g^i\partial_i v_{g\|}]^{(n)}$ and $\partial^2_r(v_{g\|}v_g^2)$ which will be discussed separately. $n$ is the order of perturbation of the expression, either 1, 2, or 3 in this case.
    \item From Standard Eulerian Perturbation Theory (EPT) of LSS \cite{Bernardeau2002PhR...367....1B, Fasiello2016PhRvD..94f3516F, Takahashi2008PThPh.120..549T} and the bias expansion \cite{McDonald2009JCAP...08..020M} (see Appendices~\ref{sec: Perturbative Solution of the Large-scale Structure} and \ref{sec: The Bias Expansion}), $\delta^{(n)}_g$, $v_g^{(n)}$, and $\Psi^{(n)}$ in Fourier space are written in the integral form with factorisable kernels, as combinations of the EPT kernels and bias operators. Therefore, all time-dependent factors can also be factored out from expression \eqref{eqn: general form of relativistic correction terms}. In Fourier space, $\vb*{v_{g,k}}^{(n)}=-i\frac{\vb*{k}}{k^2}\theta^{(n)}_{g,\vb*{k}}$ where $\theta^{(n)}_{g,\vb*{k}}$ is the galaxy peculiar velocity divergence. Meanwhile, $\Psi^{(n)}_{\vb*{k}}=\frac{3\Omega_m(\tau)\mathcal{H}^2(\tau)}{2k^2}\delta^{(n)}_{\vb*{k}}$ in the Newtonian approximation.\footnote{In fact, the EPT solutions are calculated entirely in the Newtonian approximation. This is valid for the correction up to order $\epsilon_{\mathcal{H}}^{1/2}$ since the GR correction to the LSS equation of motion comes in order $\epsilon_{\mathcal{H}}\delta$ or smaller \cite{Bardeen:1980kt, Mukhanov1992PhR...215..203M}.}
    \item A time derivative acts only on the time evolution factor and leaves the mode-dependent part unchanged. Therefore, the time-dependent parts can still be factored out in this case, nonetheless.
\end{enumerate}
This list is complete since every term in $\Delta_g^{(3)}$ is local (we ignore integration along the LOS; see Section \ref{sec: The Relativistic Galaxy Number Density: Redshift Space Distortion}) and the perturbations on redshift, number density, and volume always come in the form of an $n$th-order gradient of the fields along the LOS (we assume no perturbation on $\vb*{\hat{n}}$). In particular, all contributions are constructed through equation~\eqref{eqn: full relativistic Number Density} and the redshift perturbation \eqref{eqn: redshift perturbation final}. Other than the exception $[v_g^i\partial_i v_{g\|}]^{(n)}$ and $\partial^2_r(v_{g\|}v_g^2)$, which come from the derivative along the LOS on $v^2_g$ in the redshift perturbation \eqref{eqn: redshift perturbation final}, every term must be only derivatives of $v_{g\|}^{(n)}$, $\delta_g^{(n)}$, and $\Psi^{(n)}$.

We will demonstrate that the Fourier transform of a term in the form \eqref{eqn: general form of relativistic correction terms} can always be expressed in the kernel integral form \eqref{eqn: Fourier observed galaxy number density constrast}. Firstly, we use the general Leibniz rule,
\begin{equation}
    \partial_r^n\left(\prod^k_{i=1}f_i(\vb*{r},\tau)\right)=\sum_{\{j_i\}}
    \begin{pmatrix}
        n \\
        j_1,j_2,\dots,j_k
    \end{pmatrix}
    \prod^k_{i=1}\partial^{j_i}_rf_i(\vb*{r},\tau),
\end{equation}
to distribute the LOS derivative to the individual $v_{g\|}$, $\delta_g$, and $\Psi$. We remind the reader again that the partial derivative directed along a fixed LOS $\vb*{\hat{n}}$ acts only on the part of the field that depends on $r$ explicitly. Although the time variable $\tau$ depends on $r$ implicitly through the lightcone, the fields $v_g$, $\delta_g$, and $\Psi$ are defined on comoving coordinates, where $r$ and $\tau$ are independent variables, not on the lightcone as observer's coordinates. Therefore, the Leibniz rule distributes the derivative nicely with $\partial_r\tau=0$. Then, the expression \eqref{eqn: general form of relativistic correction terms} can be written as a sum of
\begin{equation}\label{eqn: general form of relativistic correction terms 2}
    \propto\left[\prod_j\left(\partial_r^j v_{g\|}\right)^{a_j}\left(\partial_r^j \delta_{g}\right)^{b_j}\left(\partial_r^j \Psi\right)^{c_j}\right]^{(n)}\;; \quad j,a_j,b_j,c_j \in \mathbb{Z},
\end{equation}
ignoring the constants introduced by the general Leibniz rule. Following the convolution theorem,
\begin{equation}\label{eqn: convolution theorem}
    \mathcal{F}_{\vb*{k}}\left[\prod^N_{i=1}f_i(\vb*{r})\right]=\int_{\vb*{q_1}} \dots \int_{\vb*{q_N}}\delta^{(3)}_D\left(\vb*{k}-\sum^N_{i=1}\vb*{q_i}\right)\prod^N_{i=1}\mathcal{F}_{\vb*{q_i}}\left[f_i(\vb*{r})\right],
\end{equation}
The Fourier transform of \eqref{eqn: general form of relativistic correction terms 2} can be broken down into a convolution of all terms in the bracket. According to EPT and the bias expansion, the perturbative solutions of the galaxy number density contrast and velocity divergence in Fourier space are written as
\begin{equation}
    \begin{split}
        \delta_{g,\vb*{k}}^{(n)}(\tau)&=\int_{\vb*{k_1}} \dots \int_{\vb*{k_n}}\delta^{(3)}_D\left(\vb*{k}-\sum^n_{\sigma=1}\vb*{k_\sigma}\right)\Delta^S_{g,n}(\vb*{k_1},\dots ,\vb*{k_n};\tau)\prod^n_{\sigma=1}\delta^{(1)}_{\vb*{k_\sigma}}, \\
        \theta_{g,\vb*{k}}^{(n)}(\tau)&=\int_{\vb*{k_1}} \dots \int_{\vb*{k_n}}\delta^{(3)}_D\left(\vb*{k}-\sum^n_{\sigma=1}\vb*{k_\sigma}\right)\Theta^S_{g,n}(\vb*{k_1},\dots ,\vb*{k_n};\tau)\prod^n_{\sigma=1}\delta^{(1)}_{\vb*{k_\sigma}},
    \end{split}
\end{equation}
where the time-dependent kernels can always be separated,
\begin{equation}
    \begin{split}
        \Delta^S_{g,n}(\vb*{k_1},\dots ,\vb*{k_n};\tau)&=\sum^{N_n}_{\lambda=1}D_{g,n,\lambda}(\tau)\Delta^S_{g,n,\lambda}(\vb*{k_1},\dots ,\vb*{k_n}), \\
        \Theta^S_{g,n}(\vb*{k_1},\dots ,\vb*{k_n};\tau)&=\sum_{\{X_n\}}E_{X_n}(\tau)\Delta^S_{X_n}(\vb*{k_1},\dots ,\vb*{k_n}).
    \end{split}
\end{equation}
The derivation of these results is in Appendices~\ref{sec: Perturbative Solution of the Large-scale Structure} and \ref{sec: The Bias Expansion}. $N_n$ is the number of kernels separated in $\Delta^S_{g,n}$, and the set $\{X_n\}$ contains the label of every possible kernel $\Delta^S_{X_n}$ for the EPT solution at $n$th order. Then, the Fourier transform of each term in \eqref{eqn: general form of relativistic correction terms 2} can be written as
\begin{equation}\label{eqn: Fourier transform of each terms in general correction 1}
    \begin{split}
        \mathcal{F}_{\vb*{q}}&\left[\partial^j_rv_{g\|}^{(m)}(\vb*{r},\tau)\right]=(i\mu_{\vb*{q}}q)^j\left(-i\mu_{\vb*{q}}\frac{\theta^{(m)}_{\vb*{q}}(\tau)}{q}\right) \\
        =&\int_{\vb*{k_1}}\dots\int_{\vb*{k_m}}\delta^{(3)}_D\left(\vb*{q}-\sum_{\sigma=1}^m\vb*{k_\sigma}\right)
        \underbrace{\sum_{\{X_m\}}E_{X_m}(\tau)\frac{-\left(i\sum_{\sigma=1}^m \mu_{\vb*{k_\sigma}}k_\sigma\right)^{j+1}}{\norm{\sum_{\sigma=1}^m \vb*{k_\sigma}}^2}\Delta^S_{X_m}(\vb*{k_1},\dots,\vb*{k_m})}_{\mathcal{V}_j^{(m)}(\vb*{k_1}\dots,\vb*{k_m};\vb*{\hat{n}},\tau)}
        \prod_{\sigma=1}^m\delta^{(1)}_{\vb*{k_\sigma}}, \\
        \mathcal{F}_{\vb*{q}}&\left[\partial^j_r\delta _{g}^{(l)}(\vb*{r},\tau)\right]=(i\mu_{\vb*{q}}q)^j\delta^{(l)}_{g,\vb*{q}}(\tau) \\
        =&\int_{\vb*{k_1}}\dots\int_{\vb*{k_l}}\delta^{(3)}_D\left(\vb*{q}-\sum_{\sigma=1}^l\vb*{k_\sigma}\right)
        \underbrace{\sum_{\lambda=1}^{N_l}D_{g,l,\lambda}(\tau)\left(i\sum_{\sigma=1}^l \mu_{\vb*{k_\sigma}}k_\sigma\right)^j \Delta^S_{g,l,\lambda}(\vb*{k_1},\dots,\vb*{k_l})}_{\mathcal{D}_j^{(l)}(\vb*{k_1}\dots,\vb*{k_l};\vb*{\hat{n}},\tau)}
        \prod_{\sigma=1}^l\delta^{(1)}_{\vb*{k_\sigma}}, \\
         \mathcal{F}_{\vb*{q}}&\left[\partial^j_r\Psi^{(p)}(\vb*{r},\tau)\right]=(i\mu_{\vb*{q}}q)^j\frac{3\Omega_m(\tau)\mathcal{H}^2(\tau)}{2q^2}\delta^{(p)}_{\vb*{q}}(\tau) \\
        =&\int_{\vb*{k_1}}\dots\int_{\vb*{k_p}}\delta^{(3)}_D\left(\vb*{q}-\sum_{\sigma=1}^p\vb*{k_\sigma}\right) \\
        &\underbrace{\sum_{\{X_p\}}\frac{3}{2}\Omega_m(\tau)\mathcal{H}^2(\tau)D_{X_p}(\tau)\frac{\left(i\sum_{\sigma=1}^p \mu_{\vb*{k_\sigma}}k_\sigma\right)^j}{\norm{\sum_{\sigma=1}^p \vb*{k_\sigma}}^2} \Delta^S_{X_p}(\vb*{k_1},\dots,\vb*{k_p})}_{\mathcal{I}_j^{(p)}(\vb*{k_1}\dots,\vb*{k_p};\vb*{\hat{n}},\tau)}
        \prod_{\sigma=1}^p\delta^{(1)}_{\vb*{k_\sigma}}, \\
    \end{split}
\end{equation}
where we use the relation $\partial_r=i\mu_{\vb*{k}}k$, $v_{g,\vb*{k}\|}=\mu_{\vb*{k}}v_{g,\vb*{k}}$, and $\mu_{\vb*{k}}=\vb*{\hat{k}}\cdot\vb*{\hat{n}}$. The Dirac delta also imposes a constraint $\vb*{q}=\sum_{\sigma=1}^m\vb*{k_\sigma}$, which implies the relation between the cosine factors
\begin{equation}
    \begin{split}
        \vb*{q}\cdot\vb*{\hat{n}}&=\left(\sum_{\sigma=1}^m\vb*{k_\sigma}\right)\cdot\vb*{\hat{n}} \\
        q\vb*{\hat{q}}\cdot\vb*{\hat{n}}&=\sum_{\sigma=1}^mk_\sigma\vb*{\hat{k}_\sigma}\cdot\vb*{\hat{n}} \\
        \mu_{\vb*{q}}q&=\sum_{\sigma=1}^m \mu_{\vb*{k_\sigma}}k_{\sigma}.
    \end{split}
\end{equation}
We use this relation to trade the external mode vector $\vb*{q}$ with the internal ones $\vb*{k_\sigma}$ and allow ourselves to construct the unified Fourier kernels $\mathcal{V}^{(m)}_j$, $\mathcal{D}^{(l)}_j$, and $\mathcal{I}^{(p)}_j$ defined in equations~\eqref{eqn: Fourier transform of each terms in general correction 1}. The key point is that the time-dependent functions in these kernels are completely factorised from the mode-dependent part. Therefore, any time derivative acting on these terms affects only the time-evolution factors of the kernels, leaving the mode-mixing integral unchanged. For example, $\partial_rv^{(2)'}_{g\|}$ yields the same mode dependence as $\partial_rv^{(2)}_{g\|}$, namely a convolution of $\delta^{\text{(1)}}$'s against the kernel $\mathcal{V}^{(2)}_1$. 

The advantage of this form is that the convolution of these terms has a kernel which is just a product of the three kernels, $\mathcal{V}^{(m)}_j$, $\mathcal{D}^{(l)}_j$, and $\mathcal{I}^{(p)}_j$. For example,
\begin{equation}
    \begin{split}
        \mathcal{F}_{\vb*{k}}&\left[v_{g\|}^{(1)}(\vb*{r},\tau)\partial_rv_{g\|}^{(2)}(\vb*{r},\tau)\right]=\int_{\vb*{q_1}}\int_{\vb*{q_2}} \delta^{(3)}_D\left(\vb*{k}-\vb*{q_1}-\vb*{q_2}\right)\mathcal{F}_{\vb*{q_1}}\left[v_{g\|}^{(1)}(\vb*{r},\tau)\right]\mathcal{F}_{\vb*{q_2}}\left[\partial_rv_{g\|}^{(2)}(\vb*{r},\tau)\right] \\
        &=\int_{\vb*{k_1}}\int_{\vb*{k_2}}\int_{\vb*{k_3}}\delta^{(3)}_D\left(\vb*{k}-\vb*{k_1}-\vb*{k_2}-\vb*{k_3}\right)\left[\mathcal{V}^{(1)}_0\mathcal{V}^{(2)}_1\right]^S(\vb*{k_1},\vb*{k_2},\vb*{k_3};\vb*{\hat{n}},\tau)\delta^{(1)}_{\vb*{k_1}}\delta^{(1)}_{\vb*{k_2}}\delta^{(1)}_{\vb*{k_3}},
    \end{split}
\end{equation}
where $\left[\mathcal{V}^{(1)}_0\mathcal{V}^{(2)}_1\right]^S(\vb*{k_1},\vb*{k_2},\vb*{k_3};\vb*{\hat{n}},\tau)$ is the symmetrised $\mathcal{V}^{(1)}_0(\vb*{k_1};\vb*{\hat{n}},\tau)\mathcal{V}^{(2)}_1(\vb*{k_2},\vb*{k_3};\vb*{\hat{n}},\tau)$ with respect to $\vb*{k_\sigma}$'s in the arguments. Since both $\mathcal{V}^{(m)}_j$'s are factorisable, the product of them is also factorisable. Meanwhile, the number of the mixed modes $\vb*{k_\sigma}$ is equal to the order of perturbation. Hence, the convolution, which constitutes $n$th order of perturbation, mixes $n$ modes of $\delta^{(1)}$. For example, the kernel $\left[\mathcal{V}^{(1)}_0\mathcal{V}^{(2)}_1\right]^S$ depends on 3 mode vectors.

Let us now consider the remaining terms in item 1, starting with $\left[v_g^i\partial_iv_{g\|}\right]^{(n)}$. As before, let us write the Fourier transform of each term,
\begin{equation}
    \begin{split}
        &\mathcal{F}_{\vb*{q}}\left[\left(v_g^{(m)}\right)^\lambda(\vb*{r},\tau)\right]=\left(-iq^\lambda\frac{\theta^{(m)}_{\vb*{q}}(\tau)}{q^2}\right) \\
        &=\int_{\vb*{k_1}}\dots\int_{\vb*{k_m}}\delta^{(3)}_D\left(\vb*{q}-\sum_{\sigma=1}^m\vb*{k_\sigma}\right)
        \sum_{\{X_m\}}E_{X_m}(\tau)\frac{-i\sum_{\sigma=1}^m \left(k_\sigma\right)^\lambda}{\norm{\sum_{\sigma=1}^m \vb*{k_\sigma}}^2}\Delta^S_{X_m}(\vb*{k_1},\dots,\vb*{k_m})
        \prod_{\sigma=1}^m\delta^{(1)}_{\vb*{k_\sigma}},
    \end{split}
\end{equation}
\begin{equation}
    \begin{split}
        &\mathcal{F}_{\vb*{q}}\left[\partial_\lambda v_{g\|}^{(l)}(\vb*{r},\tau)\right]=i\mu_{\vb*{q}}q_\lambda\left(-i\mu_{\vb*{q}}\frac{\theta^{(l)}_{\vb*{q}}(\tau)}{q}\right) \\
        &=\int_{\vb*{k_1}}\dots\int_{\vb*{k_l}}\delta^{(3)}_D\left(\vb*{q}-\sum_{\sigma=1}^l\vb*{k_\sigma}\right)\sum_{\{X_l\}}E_{X_l}(\tau)\frac{\left[\sum_{\sigma=1}^l \left(k_\sigma\right)_\lambda\right]\left[\sum_{\sigma=1}^l \mu_{\vb*{k_\sigma}}k_\sigma\right]^2}{\norm{\sum_{\sigma=1}^l \vb*{k_\sigma}}^3}\Delta^S_{X_l}(\vb*{k_1},\dots,\vb*{k_l})\prod_{\sigma=1}^l\delta^{(1)}_{\vb*{k_\sigma}},
    \end{split}
\end{equation}
Note that $\lambda$ in these expressions is a spatial index in the Einstein summation convention rather than an exponent. Hence, the convolution of these two terms yields
\begin{equation}\label{eqn: Fourier transform of each terms in general correction 2}
    \begin{split}
        \mathcal{F}_{\vb*{k}}&\left[\left(v_g^{(m)}\right)^\lambda \partial_\lambda v_{g\|}^{(l)}\right]
        =\int_{\vb*{k_1}}\dots\int_{\vb*{k_{m+l}}}\delta^{(3)}_D\left(\vb*{k}-\sum_{\sigma=1}^{m+l}\vb*{k_\sigma}\right)\left(\prod_{\sigma=1}^{m+l}\delta^{(1)}_{\vb*{k_\sigma}}\right)\sum_{\{X_l\}}\sum_{\{X_m\}}E_{X_l}(\tau)E_{X_m}(\tau) \\
        &
        \underbrace{\frac{\left[-i\sum_{\rho=1}^m \sum_{\sigma=m+1}^{m+l} \vb*{k_\rho}\cdot \vb*{k_\sigma}\right]\left[\sum_{\sigma=m+1}^{m+l} \mu_{\vb*{k_\sigma}}k_\sigma\right]^2}{\norm{\sum_{\sigma=1}^m \vb*{k_\sigma}}^2\norm{\sum_{\sigma=m+1}^{m+l} \vb*{k_\sigma}}^3}
        \Delta^S_{X_m}(\vb*{k_1},\dots,\vb*{k_m})\Delta^S_{X_l}(\vb*{k_{m+1}},\dots,\vb*{k_{m+l}})}_{\mathcal{C}^{(m,l)}(\vb*{k_1},\dots,\vb*{k_{m+l}};\vb*{\hat{n}},\tau)}. \\
    \end{split}
\end{equation}
Once again, the kernels are factorisable as in equations \eqref{eqn: Fourier transform of each terms in general correction 1}. As before, the final kernel will always be $\vb*{k_\sigma}$-symmetrised by the product $\prod_{\sigma=1}^{m+l}\delta^{(1)}_{\vb*{k_\sigma}}$.

Finally, we consider the term $\partial^2_r(v_{g\|}v_g^2)$. The Fourier transform of each term is
\begin{equation}
    \begin{split}
        \mathcal{F}_{\vb*{q}}&\left[v_{g\|}^{(1)}(\vb*{r},\tau)\right]=\int_{\vb*{k_1}}\delta^{(3)}_D\left(\vb*{q}-\vb*{k_1}\right)\sum_{\{X_1\}}E_{X_1}(\tau)\frac{-i\mu_{\vb*{k_1}}k_1}{k_1^2}\Delta^S_{X_1}(\vb*{k_1})\delta^{(1)}_{\vb*{k_1}}=-E_1(\tau)\frac{i\mu_{\vb*{q}}q}{q^2}\delta^{(1)}_{\vb*{q}}, \\
        \mathcal{F}_{\vb*{q}}&\left[\left(v_{g}^{(1)}\right)^\lambda\left(v_{g}^{(1)}\right)_\lambda\right] = \int_{\vb*{k_1}}\int_{\vb*{k_2}}\delta^{(3)}_D\left(\vb*{q}-\vb*{k_1}-\vb*{k_2}\right)\mathcal{F}_{\vb*{k_1}}\left[\left(v_{g}^{(1)}\right)^i\right]\mathcal{F}_{\vb*{k_2}}\left[\left(v_{g}^{(1)}\right)_i\right] \\
        &= \int_{\vb*{k_1}}\int_{\vb*{k_2}}\delta^{(3)}_D\left(\vb*{q}-\vb*{k_1}-\vb*{k_2}\right)E_1^2(\tau)\frac{-\vb*{k_1}\cdot\vb*{k_2}}{k_1^2k_2^2}\delta^{(1)}_{\vb*{k_1}}\delta^{(1)}_{\vb*{k_2}},
    \end{split}
\end{equation}
where the first-order EPT kernel is unity, $\Delta^S_{X_1}(\vb*{k})=1$ (see Appendix~\ref{sec: Perturbative Solution of the Large-scale Structure}). Therefore, we have
\begin{equation}
    \begin{split}
        \mathcal{F}_{\vb*{k}}&\left[\partial^2_r(v_{g\|}v_g^2)\right]=-(\mu_{\vb*{k}}k)^2\int_{\vb*{q_1}}\int_{\vb*{q_2}}\delta^{(3)}_D\left(\vb*{k}-\vb*{q_1}-\vb*{q_2}\right)\mathcal{F}_{\vb*{q_1}}\left[v_{g\|}^{(1)}\right]\mathcal{F}_{\vb*{q_2}}\left[\left(v_{g}^{(1)}\right)^\lambda\left(v_{g}^{(1)}\right)_\lambda\right] \\
        &=\int_{\vb*{k_1}}\int_{\vb*{k_2}}\int_{\vb*{k_3}}\delta^{(3)}_D\left(\vb*{k}-\vb*{k_1}-\vb*{k_2}-\vb*{k_3}\right)\underbrace{E_1^3(\tau)\frac{i\mu_{\vb*{k_1}}k_1(\vb*{k_2}\cdot\vb*{k_3})}{k_1^2k_2^2k_3^2}}_{\mathcal{J}(\vb*{k_1},\vb*{k_2},\vb*{k_3};\vb*{\hat{n}},\tau)}\delta^{(1)}_{\vb*{k_1}}\delta^{(1)}_{\vb*{k_2}}\delta^{(1)}_{\vb*{k_3}}.
    \end{split}
\end{equation}
The kernel must also be symmetrised as usual through
\begin{equation}
    \begin{split}
        \mathcal{J}^S(\vb*{k_1},\vb*{k_2},\vb*{k_3};\vb*{\hat{n}},\tau)&=\frac{1}{3!}\sum_{\vb*{k_1},\vb*{k_2},\vb*{k_3}} \mathcal{J}(\vb*{k_1},\vb*{k_2},\vb*{k_3};\vb*{\hat{n}},\tau) \\
        &=\frac{E_1^3(\tau)}{3}\frac{i\left[\mu_{\vb*{k_1}}k_1(\vb*{k_2}\cdot\vb*{k_3})+\mu_{\vb*{k_2}}k_2(\vb*{k_1}\cdot\vb*{k_3})+\mu_{\vb*{k_3}}k_3(\vb*{k_1}\cdot\vb*{k_2})\right]}{k_1^2k_2^2k_3^2},
    \end{split}
\end{equation}
where we use the symmetry $\vb*{k_i}\cdot\vb*{k_j} = \vb*{k_j}\cdot\vb*{k_i}$ to reduce 6 terms to 3 terms. We can see that the time-dependent functions factor nicely out of the mode-dependent part, again as in equation~\eqref{eqn: Fourier transform of each terms in general correction 1}.

The results suggest that the convolutions of the terms in equations \eqref{eqn: Fourier transform of each terms in general correction 1} and \eqref{eqn: Fourier transform of each terms in general correction 2} are of the form \eqref{eqn: Fourier observed galaxy number density constrast}. The examples of kernels we considered, $\left[\mathcal{V}^{(1)}_0\mathcal{V}^{(2)}_1\right]^S$, and the symmetrised $\mathcal{C}^{(m,l)}$ and $\mathcal{J}$, are factorisable. With respect to the expression \eqref{eqn: Fourier observed galaxy number density constrast}, their mode-mixing parts are stored in $\mathcal{S}_{n,r}(\vb*{k_1},\dots ,\vb*{k_n};\vb*{\hat{n}})$ where the time-dependent factors are separated to $\mathcal{Z}_{n,r}(z)$. This is possible since the Fourier integral ignores the time evolution part of the solution.

Finally, let us discuss the general structure of the mode-dependent part. As a product of the sub-kernels introduced above, it must be expressible as a sum of products of the following terms,
\begin{equation}\label{eqn: relevant terms in the kernel}
    \left\{ i\sum_{\{\sigma\}} \mu_{\vb*{k_\sigma}}k_\sigma\; , \; \frac{1}{\norm{\sum_{\{\sigma\}} \vb*{k_\sigma}}^n}\; , \; \sum_{\{\rho\}} \sum_{\{\sigma\}} \vb*{k_\rho}\cdot \vb*{k_\sigma} \right\}\, ; \quad n \in \mathbb{Z}.
\end{equation}
where $\{\rho\},\{\sigma\} \subset \{1,2,3\}$. These expressions can be traced back to the structure of $\mathcal{V}^{(m)}_j$, $\mathcal{D}^{(l)}_j$, $\mathcal{I}^{(p)}_j$, and $\mathcal{C}^{(m,l)}$, defined in equations \eqref{eqn: Fourier transform of each terms in general correction 1} and \eqref{eqn: Fourier transform of each terms in general correction 2}, which are built from the tidal combinations $S$ and $F$ defined in equations \eqref{eqn: tidal combinations in Fourier space}, and the EPT source kernels $\alpha$ and $\beta$ defined in equation~\eqref{eqn: EPT source kernels} (see Appendices \ref{sec: Perturbative Solution of the Large-scale Structure} and \ref{sec: The Bias Expansion} for details). All of these sub-kernels involve only three types of mode expression: projection along the LOS, (inverse) norm, and dot products between mode vectors. These combinations are scalars, as required for an integral kernel of the scalar field $\Delta_g$. Any kernel of this type can be expressed exactly as a polynomial in $\mu_{\vb*{k_\sigma}}$, the norm of the mode vector, $k_\sigma$, and the directional dot product, $\vb*{\hat{k}_\rho}\cdot \vb*{\hat{k}_\sigma}$. Let us first point out that the inverse norm can always be expanded via the generating function of the \emph{Legendre polynomials},
\begin{equation}\label{eqn: generating function of legendre polynomials}
    \frac{1}{\norm{\vb*{k_a}+\vb*{k_b}}}=\frac{1}{k_a}\sum^\infty_{n=0}\left(-\frac{k_b}{k_a}\right)^n\mathcal{P}_n(\vb*{\hat{k}_a} \cdot \vb*{\hat{k}_b}),
\end{equation}
where $\mathcal{P}_n$ is the $n$th Legendre polynomial and the series converges only if $k_b<k_a$. In practice, we can also use the generating function for $k_b>k_a$ by just inverting the ratio $\frac{k_b}{k_a}$ and splitting the Fourier integral domain into two: where $k_b<k_a$ and $k_a<k_b$. Crucially, the angular part of the kernel is the same, and the domain of the solid angle $\Omega _ {k_\sigma}$ remains intact.

$\mathcal{P}_n(\vb*{\hat{k}_a} \cdot \vb*{\hat{k}_b})$ is simply a polynomial in $\vb*{\hat{k}_a} \cdot \vb*{\hat{k}_b}$. This relation can be applied multiple times if more than two vectors are summed. In the end, it will be expressed as a polynomial of the dot products. Similarly, the third expression in the set \eqref{eqn: relevant terms in the kernel} is a sum of dot products weighted by their norms, $\sum_{\{\rho\}} \sum_{\{\sigma\}} k_\rho k_\sigma(\vb*{\hat{k}_\rho}\cdot \vb*{\hat{k}_\sigma})$. Finally, it follows that exponents or products between members in the set \eqref{eqn: relevant terms in the kernel} can always be expressed as polynomials of $\mu_{\vb*{k_\sigma}}$, $k_\sigma$, and $\kappa_{\rho\sigma}=\vb*{\hat{k}_\rho}\cdot \vb*{\hat{k}_\sigma}$. Therefore, the kernel can always be rewritten as
\begin{equation}\label{eqn: polynomial form of redshift kernel}
    S_{n,r}(\vb*{k_1},\dots,\vb*{k_n};\vb*{\hat{n}})=\sum_{\{\alpha_n\},\{\beta_n\},\{\gamma_{n\neq n}\}} C_{\{\alpha_n\},\{\beta_n\},\{\gamma_{n\neq n}\}} \prod_{\sigma=1}^n(i\mu_{\vb*{k_\sigma}})^{\alpha_\sigma} k_\sigma^{\beta_\sigma}\prod_{\rho\neq\sigma}^n\kappa_{\rho\sigma}^{\gamma_{\rho\sigma}},
\end{equation}
where the set of exponents $\{\alpha_n\},\{\beta_n\},\{\gamma_{n\neq n}\}$ contains integers and $C_{\{\alpha_n\},\{\beta_n\},\{\gamma_{n\neq n}\}}$ is a real constant assigned to each monomial. The exponents $\alpha_\sigma$ and $\gamma_{\rho\sigma}$ are always non-negative, as the kernels are always the convolution of subkernels that contain only positive powers of $\mu_{\vb*{k_\sigma}}$ and $\kappa_{\rho\sigma}$. However, $\beta_\sigma$ can be zero, negative, or positive, as some subkernels contain the inverse norm of the mode vector. The index $n\neq n$ indicates that the set of indices is traceless, i.e., there is no $\gamma_{\sigma\sigma}$ in the set. This is because the dot product $\kappa_{\sigma\sigma}=\vb*{\hat{k}_\sigma}\cdot \vb*{\hat{k}_\sigma}=1$. We emphasise that no approximation is involved, and the polynomial expression of the kernel is exact. This is because we keep the whole infinite expansion of the inverse norm via the Legendre polynomial generating function. The truncation at finite order would introduce an approximation.

A kernel is parity-violating when $\sum_{\sigma=1}^n\alpha_\sigma=\text{odd}$, since flipping the mode vector introduces a minus sign to $\mu_{\vb*{k_\sigma}}$. Thus, the local N-point galaxy statistics, including the 4PCF, can in principle contain non-zero parity-odd modes, which are purely observational in origin and unrelated to the initial conditions or the structure formation process. It is therefore important to understand this effect, which represents a potential systematic bias in measurements of intrinsic PV. This has been extensively studied in \cite{Paul:2024uim, Paul:2024xff,Paul2025arXiv250803249P}. In Section \ref{sec: Probing Parity Violation With 4PCF}, however, we show how the isotropic estimator used in \cite{Hou:2022wfj,Philcox:2022hkh, Slepian:2025kbb} suppresses these modes by averaging over all LOS. In particular, we will show how these kernels can be projected onto an isotropic basis functions in the plane-parallel approximation, which imposes a selection rule on the angular momentum indices of the 4PCF. This particular selection rule is what constrains all parity-odd contributions, i.e., those with angular momentum indices that sum to an odd number, to zero. Furthermore, we will relax the plane-parallel approximation and introduce the local LOS to express the wide-angle effect. Then, we will show that the selection rule can still be derived by expanding the kernels with spherical harmonic functions. Therefore, the parity-odd contributions vanish identically, which is crucial for wide galaxy surveys used in \cite{Hou:2022wfj,Philcox:2022hkh, Slepian:2025kbb}.
\section{Galaxy four-point correlation function and parity violation}\label{sec: Probing Parity Violation With 4PCF}

Recent studies \cite{Hou:2022wfj,Philcox:2022hkh,Slepian:2025kbb} have reported the detection of PV through the detection of a non-zero parity-odd mode in the galaxy 4PCF (see however also other later studies cited in the Introduction).  Assuming ergodicity, the 4PCF is measured via the isotropic estimator \cite{Philcox2022MNRAS.509.2457P},
\begin{equation}\label{eqn: 4PCF isotropic estimator}
    \hat{\zeta}(\vb*{r_1},\vb*{r_2},\vb*{r_3}) = \int \frac{d^3s}{V_s}  \Delta_{g}(\vb*{s})\Delta_g(\vb*{s}+\vb*{r_1})\Delta_g(\vb*{s}+\vb*{r_2})\Delta_g(\vb*{s}+\vb*{r_3}),
\end{equation}
where the statistical average is estimated by the spatial average. Here, $\vb*{s}$ is the reference galaxy, which defines the anchor line of sight of the 4PCF, and $\vb*{r_1}$, $\vb*{r_2}$, and $\vb*{r_3}$ are the three positions relative to the reference galaxy. We also combine the LOS direction and redshift into the galaxy position, $\vb*{s}(\vb*{\hat{n}},z)=s(z)\vb*{\hat{n}}$. If the estimator is unbiased under this assumption, then taking the ensemble average gives
\begin{equation}
    \hat{\zeta}(\vb*{r_1},\vb*{r_2},\vb*{r_3}) = \int \frac{d^3s}{V_s}  \zeta(\vb*{s};\vb*{r_1},\vb*{r_2},\vb*{r_3}),
\end{equation}
where 
\begin{equation}
    \zeta(\vb*{s};\vb*{r_1},\vb*{r_2},\vb*{r_3}) = \expval{\Delta_g(\vb*{s})\Delta_g(\vb*{s}+\vb*{r_1})\Delta_g(\vb*{s}+\vb*{r_2})\Delta_g(\vb*{s}+\vb*{r_3})}
\end{equation}
is the theoretical galaxy 4PCF. While $\zeta$ can be anisotropic due to its LOS dependence, this dependence is removed in $\hat{\zeta}$ because the estimator averages over the survey volume and hence over all LOS directions. The dependence on $\vb*{r_1}$, $\vb*{r_2}$, and $\vb*{r_3}$ remains and describes the internal geometry of the tetrahedron formed by the quadruplet of galaxies. The estimator is isotropic in the LOS; it does not depend on the direction of this tetrahedron from the observer. However, different orientations of the tetrahedron can give different values of $\zeta$. As we will show, this averaging suppresses the parity-odd contributions introduced by $\mu_{\vb*{k}}$. To compute $\zeta$, we use the perturbative solutions of the observed galaxy number overdensity, $\Delta_g=\Delta_g^{(1)}+\Delta_g^{(2)}+\Delta_g^{(3)}$, in the integral form \eqref{eqn: Fourier observed galaxy number density constrast} derived in Section~\ref{sec: Kernel Solutions of the Observed Galaxy Number Density}. The two leading contributions to the connected part are

\begin{equation}\label{eqn: observed 4PCF leading contribution}
    \begin{split}
        \zeta(\vb*{s};\vb*{r_1},\vb*{r_2},\vb*{r_3})& =\expval{\Delta_g^{(1)}(\vb*{s})\Delta_g^{(1)}(\vb*{s}+\vb*{r_1})\Delta_g^{(1)}(\vb*{s}+\vb*{r_2})\Delta_g^{(3)}(\vb*{s}+\vb*{r_3})}+3\;\text{perm.} \\
        &+\expval{\Delta_g^{(1)}(\vb*{s})\Delta_g^{(1)}(\vb*{s}+\vb*{r_1})\Delta_g^{(2)}(\vb*{s}+\vb*{r_2})\Delta_g^{(2)}(\vb*{s}+\vb*{r_3})}+5\;\text{perm.},
    \end{split}
\end{equation}
where "3 perm." and "5 perm." denote the combinatorial terms arising from distributing $\Delta_g=\Delta_g^{(1)}+\Delta_g^{(2)}+\Delta_g^{(3)}$, obtained by permuting the position arguments of $\Delta_g^{(3)}$ in the first line and the two $\Delta_g^{(2)}$ factors in the second (see Appendix~\ref{apx: Leading Contribution of the Connected 4PCF}). The first contribution is
\begin{equation}\label{eqn: first observed 4PCF leading contribution}
    \begin{split}
        &\expval{\Delta_g^{(1)}(\vb*{s})\Delta_g^{(1)}(\vb*{s}+\vb*{r_1})\Delta_g^{(1)}(\vb*{s}+\vb*{r_2})\Delta_g^{(3)}(\vb*{s}+\vb*{r_3})}=\sum_{r_1=1}^{R_1}\sum_{r_2=1}^{R_1}\sum_{r_3=1}^{R_1}\sum_{r_4=1}^{R_3} \\
        &\mathcal{Z}_{1,r_1}(z_1)\mathcal{Z}_{1,r_2}(z_2)\mathcal{Z}_{1,r_3}(z_3)\mathcal{Z}_{3,r_4}(z_4)\int_{\vb*{k_1}}\int_{\vb*{k_2}}\int_{\vb*{k_3}}\int_{\vb*{k_4}} \exp\left(i\sum_{\sigma=1}^3\vb*{k_\sigma}\cdot\vb*{r_\sigma}\right)\delta_D^{(3)}\left(\vb*{k_{1234}}\right) \\
        &\underbrace{\mathcal{S}_{1,r_1}(\vb*{k_1};\vb*{\hat{n}_1})\mathcal{S}_{1,r_2}(\vb*{k_2};\vb*{\hat{n}_2})\mathcal{S}_{1,r_3}(\vb*{k_3};\vb*{\hat{n}_3})\mathcal{S}_{3,r_4}(\vb*{k_1},\vb*{k_2},\vb*{k_3};\vb*{\hat{n}_4})P_m(k_1)P_m(k_2)P_m(k_3)}_{\mathcal{T}_{1113}(\vb*{k_1},\vb*{k_2},\vb*{k_3};\vb*{k_4};\vb*{\hat{n}_1},\vb*{\hat{n}_2},\vb*{\hat{n}_3};\vb*{\hat{n}_4})}.
    \end{split}
\end{equation}
The second contribution is
\begin{equation}\label{eqn: second observed 4PCF leading contribution}
    \begin{split}
        &\expval{\Delta_g^{(1)}(\vb*{s})\Delta_g^{(1)}(\vb*{s}+\vb*{r_1})\Delta_g^{(2)}(\vb*{s}+\vb*{r_2})\Delta_g^{(2)}(\vb*{s}+\vb*{r_3})}=\sum_{r_1=1}^{R_1}\sum_{r_2=1}^{R_1}\sum_{r_3=1}^{R_2}\sum_{r_4=1}^{R_2} \\
        &\mathcal{Z}_{1,r_1}(z_1)\mathcal{Z}_{1,r_2}(z_2)\mathcal{Z}_{2,r_3}(z_3)\mathcal{Z}_{2,r_4}(z_4)\int_{\vb*{k_1}}\int_{\vb*{k_2}}\int_{\vb*{k_3}}\int_{\vb*{k_4}} \exp\left(i\sum_{\sigma=1}^3\vb*{k_\sigma}\cdot\vb*{r_\sigma}\right)\delta_D^{(3)}\left(\vb*{k_{1234}}\right) \\
        &\mathcal{S}_{1,r_1}(\vb*{k_1};\vb*{\hat{n}_1})\mathcal{S}_{1,r_2}(\vb*{k_2};\vb*{\hat{n}_2})\Big[\mathcal{S}_{2,r_3}(\vb*{k_1},\vb*{k_{24}};\vb*{\hat{n}_3})\mathcal{S}_{2,r_4}(\vb*{k_2},\vb*{k_{13}};\vb*{\hat{n}_4})P_m(k_{13}) \\
        &\underbrace{+\mathcal{S}_{2,r_3}(\vb*{k_1},\vb*{k_{23}};\vb*{\hat{n}_3})\mathcal{S}_{2,r_4}(\vb*{k_2},\vb*{k_{14}};\vb*{\hat{n}_4})P_m(k_{14})\Big]P_m(k_1)P_m(k_2),\quad \quad \quad \quad}_{\mathcal{T}_{1122}(\vb*{k_1},\vb*{k_2};\vb*{k_3},\vb*{k_4};\vb*{\hat{n}_1},\vb*{\hat{n}_2};\vb*{\hat{n}_3},\vb*{\hat{n}_4})}
    \end{split}
\end{equation}
where $\vb*{k_{ab\dots c}}=\vb*{k_a}+\vb*{k_b}+\dots+\vb*{k_c}$. The unit vectors $\vb*{\hat{n}_1}$, $\vb*{\hat{n}_2}$, $\vb*{\hat{n}_3}$, and $\vb*{\hat{n}_4}$ correspond to the LOS of each galaxy. Note that these contributions are repeated with permuted LOS directions and redshifts in the "3 perm." and "5 perm." terms. The redshifts $z_1$, $z_2$, $z_3$, and $z_4$ correspond to the respective comoving distances $s$, $\norm{\vb*{s}+\vb*{r_1}}$, $\norm{\vb*{s}+\vb*{r_2}}$ and $\norm{\vb*{s}+\vb*{r_3}}$, respectively. We denote the integrand of each contribution by $\mathcal{T}_{1113}$ and $\mathcal{T}_{1122}$.  The initial linear matter power spectrum $P_m(k)$ is defined through
\begin{equation}
    \expval{\delta_{\vb*{k_1}}^{(1)}\delta_{\vb*{k_2}}^{(1)}}=\delta^{(3)}_D\left(\vb*{k_1}+\vb*{k_2}\right)P_m(k_1),  
\end{equation}
as an isotropic function. As discussed in connection with equation~\eqref{eqn: polynomial form of redshift kernel}, each redshift kernel is a polynomial in $k_\sigma$, $i\mu_{\vb*{k_\sigma}}=i\vb*{\hat{k}_\sigma}\cdot \vb*{\hat{n}}$, and $\kappa_{\rho\sigma}=\vb*{\hat{k}_\rho}\cdot\vb*{\hat{k}_\sigma}$. The Dirac delta enforces $\vb*{k_4}=-\vb*{k_{123}}$. Throughout this work, we adopt the equal-redshift approximation for the 4PCF,
\begin{equation}
    \bar{z}=z_1\approx z_2\approx z_3 \approx z_4.
\end{equation}
The impact of relaxing this approximation is discussed in Section~\ref{sec: Discussion and Potential Sources of Violation}.

\subsection{Parity properties of 4PCF}\label{sec: Parity Properties of 4PCF}

Before assessing the parity properties of the 4PCF, we briefly introduce the relevant symmetry. Let $\mathbb{P}$ be the parity transformation that introduces a minus sign to any physical direction. For example, a position vector transforms as $\mathbb{P}[\vb*{x}]=-\vb*{x}$. Then, a parity-conserving quantity $O$ must satisfy
\begin{equation}
    \mathbb{P}[O(\vb*{x_1}, \vb*{x_2}, \dots, \vb*{x_n})] = O(-\vb*{x_1}, -\vb*{x_2}, \dots, -\vb*{x_n}) = O(\vb*{x_1}, \vb*{x_2}, \dots, \vb*{x_n}),
\end{equation}
where $\vb*{x_1}, \vb*{x_2}, \dots, \vb*{x_n}$ are spatial coordinates. In Fourier space, the transformation reflects the mode vector direction instead of the spatial coordinates. However, we hold the LOS direction, which is set by the observer, fixed, $\vb*{\hat{n}}\rightarrow\vb*{\hat{n}}$. Under a general parity transformation, $\vb*{\hat{n}}$ would flip to $-\vb*{\hat{n}}$. In practice, though, observations \cite{Hou:2022wfj, Philcox:2022hkh, Slepian:2025kbb} measure the 4PCF and its parity counterpart within the same patch of sky, and hence along the same LOS. We therefore leave $\vb*{\hat{n}}$ unchanged to isolate the PV induced by RSD when the statistics are measured from a single direction ($\vb*{\hat{n}}$) rather than also from the opposite LOS ($-\vb*{\hat{n}}$). For an example of a setup with $\vb*{\hat{n}}$ fixed, see \cite{Bouvin2014PhRvD..89h3535B}. This choice of LOS parity behaviour does not affect the results of this paper, however, since the isotropic 4PCF estimator averages over all LOS. Therefore, this transformation effectively introduces a minus sign to the cosine factor,
\begin{equation}
    \mathbb{P}[\mu_{\vb*{k_\sigma}}]=\mathbb{P}[\vb*{\hat{k}_\sigma}\cdot\vb*{\hat{n}}]=(-\vb*{\hat{k}_\sigma})\cdot\vb*{\hat{n}}=-\mu_{\vb*{k_\sigma}},
\end{equation}
but leaves, for example, the cosine of the angle between modes the same,
\begin{equation}
    \mathbb{P}[\kappa_{\rho\sigma}]=\mathbb{P}[\vb*{\hat{k}_\sigma}\cdot\vb*{\hat{k}_\rho}]=(-\vb*{\hat{k}_\sigma})\cdot(-\vb*{\hat{k}_\rho})=\kappa_{\rho\sigma}.
\end{equation}
The estimator \eqref{eqn: 4PCF isotropic estimator} admits a decomposition into parity-even and parity-odd modes,
\begin{equation}
    \hat{\zeta}(\vb*{r_1},\vb*{r_2},\vb*{r_3})=\hat{\zeta}_{\text{even}}(\vb*{r_1},\vb*{r_2},\vb*{r_3})+\hat{\zeta}_{\text{odd}}(\vb*{r_1},\vb*{r_2},\vb*{r_3}),
\end{equation}
whose transformation properties under parity are
\begin{equation}
    \begin{split}
        &\mathbb{P}\left[\hat{\zeta}_{\text{even}}(\vb*{r_1},\vb*{r_2},\vb*{r_3})\right]=\hat{\zeta}_{\text{even}}(-\vb*{r_1},-\vb*{r_2},-\vb*{r_3})=\hat{\zeta}_{\text{even}}(\vb*{r_1},\vb*{r_2},\vb*{r_3}), \\
        &\mathbb{P}\left[\hat{\zeta}_{\text{odd}}(\vb*{r_1},\vb*{r_2},\vb*{r_3})\right]=\hat{\zeta}_{\text{odd}}(-\vb*{r_1},-\vb*{r_2},-\vb*{r_3})=-\hat{\zeta}_{\text{odd}}(\vb*{r_1},\vb*{r_2},\vb*{r_3}).
    \end{split}
\end{equation}
The LOS-dependent part of $\hat{\zeta}$ is therefore the relevant object to analyse. In Fourier space, the cosine factor $\mu_{\vb*{k_\sigma}}$ changes sign when the direction of $\vb*{r_\sigma}$, equivalently the direction of $\vb*{k_\sigma}$, is reversed, making it the source of potential parity-odd contributions. All other factors, such as $\kappa_{\rho\sigma}$ and the magnitudes of mode vectors, are invariant under parity.

\subsection{Isotropic 4PCF in the plane-parallel limit: PV-free estimator}\label{sec: Isotropic 4PCF with Plane Parallel Limit}

We begin by assessing the parity symmetry of the 4PCF in the simplest case, where the LOS directions of all four galaxies are taken to be identical:
\begin{equation}
    \vb*{\hat{n}_1}\approx\vb*{\hat{n}_2}\approx\vb*{\hat{n}_3}\approx\vb*{\hat{n}_4}\approx \vb*{\hat{s}}.
\end{equation}
This is the \emph{plane-parallel} approximation. Under this assumption, every cosine factor $\mu_{\vb*{k_\sigma}}$ in the kernels is the dot product of the corresponding mode vector against the single global LOS direction $\vb*{\hat{s}}$. The functions $\mathcal{T}_{1113}$ and $\mathcal{T}_{1122}$ then take the polynomial form
\begin{equation}\label{eqn: Plane-parallel 4PCF integrands}
    \begin{split}
        \mathcal{T}_{1113}(\vb*{k_1},\vb*{k_2},\vb*{k_3};\vb*{k_4};\vb*{\hat{s}}) &= \sum_{\{\alpha_3\}}i^{\alpha_1+\alpha_2+\alpha_3} t_{1113,\{\alpha_3\}}(\vb*{k_1},\vb*{k_2},\vb*{k_3})\mu_{\vb*{k_1 s}}^{\alpha_1} \mu_{\vb*{k_2 s}}^{\alpha_2} \mu_{\vb*{k_3 s}}^{\alpha_3}, \\
        \mathcal{T}_{1122}(\vb*{k_1},\vb*{k_2};\vb*{k_3},\vb*{k_4};\vb*{\hat{s}}) &= \sum_{\{\alpha_3\}}i^{\alpha_1+\alpha_2+\alpha_3} t_{1122,\{\alpha_3\}}(\vb*{k_1},\vb*{k_2},\vb*{k_3})\mu_{\vb*{k_1 s}}^{\alpha_1} \mu_{\vb*{k_2 s}}^{\alpha_2} \mu_{\vb*{k_3 s}}^{\alpha_3}.
    \end{split}
\end{equation}
where $\mu_{\vb*{k_\sigma}s}=\vb*{\hat{k}_\sigma}\cdot \vb*{\hat{s}}$ is the cosine of the angle between the mode and the global LOS. The functions $t_{1113,\{\alpha_3\}}$ and $t_{1122,\{\alpha_3\}}$ encode all remaining factors in the polynomial kernel \eqref{eqn: polynomial form of redshift kernel}: the constants $C_{\{\alpha_3\},\{\beta_3\},\{\gamma_{3\neq3}\}}$, the mode magnitudes $k_\sigma^{\beta_\sigma}$, the angular dot products, $\kappa_{\rho\sigma}$, and the linear matter power spectrum, $P_m(k_\sigma)$. 

We can express the functions as a polynomial in just three cosine factors, $\mu_{\vb*{k_1 s}}$, $\mu_{\vb*{k_2 s}}$, and $\mu_{\vb*{k_3 s}}$, thanks to the following projection relation. The constraint on the sum of internal mode vectors imposed by the Dirac delta in equations~\eqref{eqn: first observed 4PCF leading contribution} and \eqref{eqn: second observed 4PCF leading contribution} is
\begin{equation}
    \vb*{k_{ab\dots c}}=\vb*{k_a}+\vb*{k_b}+\dots+\vb*{k_c}.
\end{equation}
Taking the dot product with $\vb*{\hat{s}}$ gives
\begin{equation}
    \mu_{\vb*{k_{ab\dots c}s}}k_{ab\dots c}=\mu_{\vb*{k_{a}s}}k_{a}+\mu_{\vb*{k_{b}s}}k_{b}+\dots\mu_{\vb*{k_{c}s}}k_{c},
\end{equation}
which is the projection relation used to redistribute composite cosine factors onto individual modes, such as $\mu_{\vb*{k_{14}}s}=\frac{1}{k_{14}}(\mu_{\vb*{k_{1}}s}k_1+\mu_{\vb*{k_4}s}k_4)$. All permuted terms in equation~\eqref{eqn: observed 4PCF leading contribution} admit the same polynomial form via the same procedure. Consequently, the theoretical 4PCF takes the form,
\begin{equation}
    \begin{split}
        \zeta(\vb*{s};\vb*{r_1},\vb*{r_2},\vb*{r_3})=&\sum_{r_1=1}^{R_1}\sum_{r_2=1}^{R_1}\sum_{r_3=1}^{R_1}\sum_{r_4=1}^{R_3}\mathcal{Z}_{1,r_1}(\bar{z})\mathcal{Z}_{1,r_2}(\bar{z})\mathcal{Z}_{1,r_3}(\bar{z})\mathcal{Z}_{3,r_4}(\bar{z}) \int_{\vb*{k_1}}\int_{\vb*{k_2}}\int_{\vb*{k_3}}\int_{\vb*{k_4}} \\
        &\exp\left(i\sum_{\sigma=1}^3\vb*{k_\sigma}\cdot\vb*{r_\sigma}\right)\delta_D^{(3)}\left(\vb*{k_{1234}}\right)\left[\mathcal{T}_{1113}(\vb*{k_1},\vb*{k_2},\vb*{k_3},\vb*{k_4};\vb*{\hat{s}})+3\;\text{perm.}\right] \\
        +&\sum_{r_1=1}^{R_1}\sum_{r_2=1}^{R_1}\sum_{r_3=1}^{R_2}\sum_{r_4=1}^{R_2}\mathcal{Z}_{1,r_1}(\bar{z})\mathcal{Z}_{1,r_2}(\bar{z})\mathcal{Z}_{2,r_3}(\bar{z})\mathcal{Z}_{2,r_4}(\bar{z})\int_{\vb*{k_1}}\int_{\vb*{k_2}}\int_{\vb*{k_3}}\int_{\vb*{k_4}} \\
        &\exp\left(i\sum_{\sigma=1}^3\vb*{k_\sigma}\cdot\vb*{r_\sigma}\right)\delta_D^{(3)}\left(\vb*{k_{1234}}\right)\left[\mathcal{T}_{1122}(\vb*{k_1},\vb*{k_2},\vb*{k_3},\vb*{k_4};\vb*{\hat{s}})+5\;\text{perm.}\right] \\
        \propto & \sum_{\{\alpha_3\}}i^{\alpha_1+\alpha_2+\alpha_3} t_{\{\alpha_3\}}(\vb*{k_1},\vb*{k_2},\vb*{k_3})\mu_{\vb*{k_1 s}}^{\alpha_1}\mu_{\vb*{k_2 s}}^{\alpha_2} \mu_{\vb*{k_3 s}}^{\alpha_3},
    \end{split}
\end{equation}
where $t_{\{\alpha_3\}}$ is the sum of $t_{1113,\{\alpha_3\}}$, $t_{1122,\{\alpha_3\}}$, and all their permutations. Having identified the LOS dependence, the isotropic estimator reduces to
\begin{equation}\label{eqn: isotropic estimator of 4PCF with los average integral}
    \begin{split}
        \hat{\zeta}(\vb*{r_1},\vb*{r_2},\vb*{r_3}) &= \int \frac{s^2ds}{V_s} \int d\Omega_s\,\zeta(\vb*{s};\vb*{r_1},\vb*{r_2},\vb*{r_3})  \\
        &\propto \int d\Omega_s \, \mu_{\vb*{k_1 s}}^{\alpha_1} \mu_{\vb*{k_2 s}}^{\alpha_2} \mu_{\vb*{k_3 s}}^{\alpha_3}.
    \end{split}
\end{equation}
We now show that this angular integral eliminates all parity-odd contributions. Expanding each cosine factor in Legendre polynomials,
\begin{equation}
    \begin{split}
        \mu_{\vb*{k_\sigma s}}^{\alpha_\sigma} &= (\vb*{\hat{k}_\sigma} \cdot \vb*{\hat{s}})^{\alpha_\sigma} \\
        &=\sum_{j=1}^{\alpha_\sigma} c^{(\alpha_\sigma)}_j \mathcal{P}_j(\vb*{\Hat{k}_\sigma} \cdot \hat{\vb*{s}}),
    \end{split}
\end{equation}
and applying the \emph{spherical harmonic addition theorem},
\begin{equation}\label{eqn: cosine decomposition with spherical harmonics}
    \mu_{\vb*{k_\sigma s}}^{\alpha_\sigma} = \sum_{j_\sigma m_{j_\sigma}}^{\alpha_\sigma} c^{(\alpha_\sigma)}_{j_\sigma} \frac{4\pi}{2j_\sigma+1} Y_{j_\sigma,m_{j_\sigma}}(\vb*{\hat{k}_\sigma})Y^*_{j_\sigma,m_{j_\sigma}}(\vb*{\hat{s}}),
\end{equation}
where $Y_{j,m_j}$ is the spherical harmonic function of degree $j$ and order $m_j$ and we use the shorthand
\begin{equation}
    \sum_{j_\sigma m_\sigma}^J=\sum_{j_\sigma=1}^J\sum_{m_{j_\sigma}=-j_\sigma}^{j_\sigma}.
\end{equation}
The LOS integral then takes the form
\begin{equation}
    \begin{split}
        \int d\Omega_s &\, \mu_{\vb*{k_1 s}}^{\alpha_1} \mu_{\vb*{k_1 s}}^{\alpha_2} \mu_{\vb*{k_1 s}}^{\alpha_3} = (4\pi)^3 \sum_{j_1,j_2,j_3} \sum_{m_{j_1},m_{j_2},m_{j_3}} \frac{c^{(\alpha_1)}_{j_1}c^{(\alpha_2)}_{j_2}c^{(\alpha_3)}_{j_3}}{(2j_1+1)(2j_2+1)(2j_3+1)} \\
        &Y_{j_1,m_{j_1}}(\vb*{\Hat{k}_1})Y_{j_2,m_{j_2}}(\vb*{\Hat{k}_2})Y_{j_3,m_{j_3}}(\vb*{\Hat{k}_3}) \int d\Omega_s Y^*_{j_1,m_{j_1}}(\hat{\vb*{s}})Y^*_{j_2,m_{j_2}}(\hat{\vb*{s}})Y^*_{j_3,m_{j_3}}(\hat{\vb*{s}}). \\
    \end{split}
\end{equation}
The remaining angular integral is the \emph{Gaunt integral},
\begin{equation}\label{eqn: Gaunt integral}
    \begin{split}
        \mathcal{G}^{j_1,j_2,j_3}_{m_{j_1},m_{j_2},m_{j_3}} &\equiv \int d\Omega_s Y^*_{j_1,m_{j_1}}(\hat{\vb*{s}})Y^*_{j_2,m_{j_2}}(\hat{\vb*{s}})Y^*_{j_3,m_{j_3}}(\hat{\vb*{s}}) \\
        &= \sqrt{\frac{(2j_1+1)(2j_2+1)(2j_3+1)}{4\pi}}
        \begin{pmatrix}
            j_1 & j_2 & j_3 \\
            0 & 0 & 0
        \end{pmatrix}
        \begin{pmatrix}
            j_1 & j_2 & j_3 \\
            m_{j_1} & m_{j_2} & m_{j_3}
        \end{pmatrix},
    \end{split}
\end{equation}
where the bracketed terms are \emph{Wigner 3-j symbols}. The second symbol imposes the constraints,
\begin{equation}
    \begin{split}
        &\text{1)}\; m_{j_1} + m_{j_2} + m_{j_3} = 0 \\
        &\text{2)}\; \abs{j_1 - j_2} \leq j_3 \leq j_1 + j_2 \; \text{(Triangle inequality)},
    \end{split}
\end{equation}
and transforms as
\begin{equation}\label{eqn:3-j symbol parity}
    \begin{pmatrix}
        j_1 & j_2 & j_3 \\
        -m_{j_1} & -m_{j_2} & -m_{j_3}
    \end{pmatrix} = (-1)^{j_1 + j_2 + j_3}
    \begin{pmatrix}
        j_1 & j_2 & j_3 \\
        m_{j_1} & m_{j_2} & m_{j_3}
    \end{pmatrix}.
\end{equation}
When all projection indices vanish, $m_{j_1}=m_{j_2}=m_{j_3}=0$, equation~\eqref{eqn:3-j symbol parity} requires $(-1)^{j_1 + j_2 + j_3}=1$. So, the first symbol with zero $m_j$'s enforces $j_1 + j_2 + j_3 = \text{even}$. This will become important shortly. Expressing the averaging integral in terms of the isotropic basis functions \cite{Cahn:2020axu_isotropicbasis} gives
\begin{equation}
    \int d\Omega_s \, \mu_{\vb*{k_1 s}}^{\alpha_1} \mu_{\vb*{k_2 s}}^{\alpha_2} \mu_{\vb*{k_3 s}}^{\alpha_3} = (4\pi)^3 \sum_{j_1,j_2,j_3} \mathcal{C}^{\alpha_2,\alpha_1,\alpha_3}_{j_1,j_2,j_3} \mathcal{P}_{j_1,j_2,j_3}(\vb*{\Hat{k}_1},\vb*{\Hat{k}_2},\vb*{\Hat{k}_3}),
\end{equation}
where the isotropic basis function is defined as
\begin{equation}\label{eqn: isotropic basis}
    \begin{split}
        \mathcal{P}_{j_1,j_2,j_3}(\vb*{\Hat{k}_1},\vb*{\Hat{k}_2},\vb*{\Hat{k}_3}) =& (-1)^{j_1+j_2+j_3}  \sum_{m_{j_1},m_{j_2},m_{j_3}}
        \begin{pmatrix}
            j_1 & j_2 & j_3 \\
            m_{j_1} & m_{j_2} & m_{j_3}
        \end{pmatrix} \\
        &Y_{j_1,m_{j_1}}(\vb*{\Hat{k}_1})Y_{j_2,m_{j_2}}(\vb*{\Hat{k}_2})Y_{j_3,m_{j_3}}(\vb*{\Hat{k}_3}),
    \end{split}
\end{equation}
and the coefficients are
\begin{equation}\label{eqn: plane parallel coefficient of isotropic basis}
    \mathcal{C}^{\alpha_2,\alpha_1,\alpha_3}_{j_1,j_2,j_3} = \frac{c^{(\alpha_1)}_{j_1}c^{(\alpha_2)}_{j_2}c^{(\alpha_3)}_{j_3}}{\sqrt{4\pi(2j_1+1)(2j_2+1)(2j_3+1)}}
    \begin{pmatrix}
        j_1 & j_2 & j_3 \\
        0 & 0 & 0
    \end{pmatrix}.
\end{equation}
The isotropic basis is particularly useful for identifying parity: since a spherical harmonic of odd degree $j$ is parity-odd, the basis function is parity-odd when $j_1+j_2+j_3 = \text{odd}$ and parity-even otherwise.  A non-vanishing coefficient $\mathcal{C}^{\alpha_2,\alpha_1,\alpha_3}_{j_1,j_2,j_3}$ with $j_1+j_2+j_3 = \text{odd}$ would therefore constitute direct evidence of PV in the 4PCF, and this is precisely the signal sought in \cite{Hou:2022wfj,Philcox:2022hkh, Slepian:2025kbb}.

However, the coefficient \eqref{eqn: plane parallel coefficient of isotropic basis} is constrained by the first Wigner 3-j symbol to satisfy $j_1+j_2+j_3 = \text{even}$. Consequently, all parity-odd contributions are removed from the 4PCF isotropic estimator in the plane-parallel approximation. A similar analysis of the LOS average appears in Appendix G of \cite{Leonard:2024xvy}, but without the GR correction to the RSD effect. The present result extends that conclusion to the leading-order GR correction at order $\epsilon_\mathcal{H}^{1/2}=\mathcal{H}/k$, as they enter through the same polynomial structure in $\mu_{\vb*{k_\sigma}}$. Therefore, the contributions with those corrections in 4PCF are subject to the same selection rule.

\subsection{Wide-angle effect: theoretical comeback}\label{sec: Wide-angle Effect: Theoretical Comeback}

\begin{figure}[tp]
    \centering
    \includegraphics[width=0.7\linewidth]{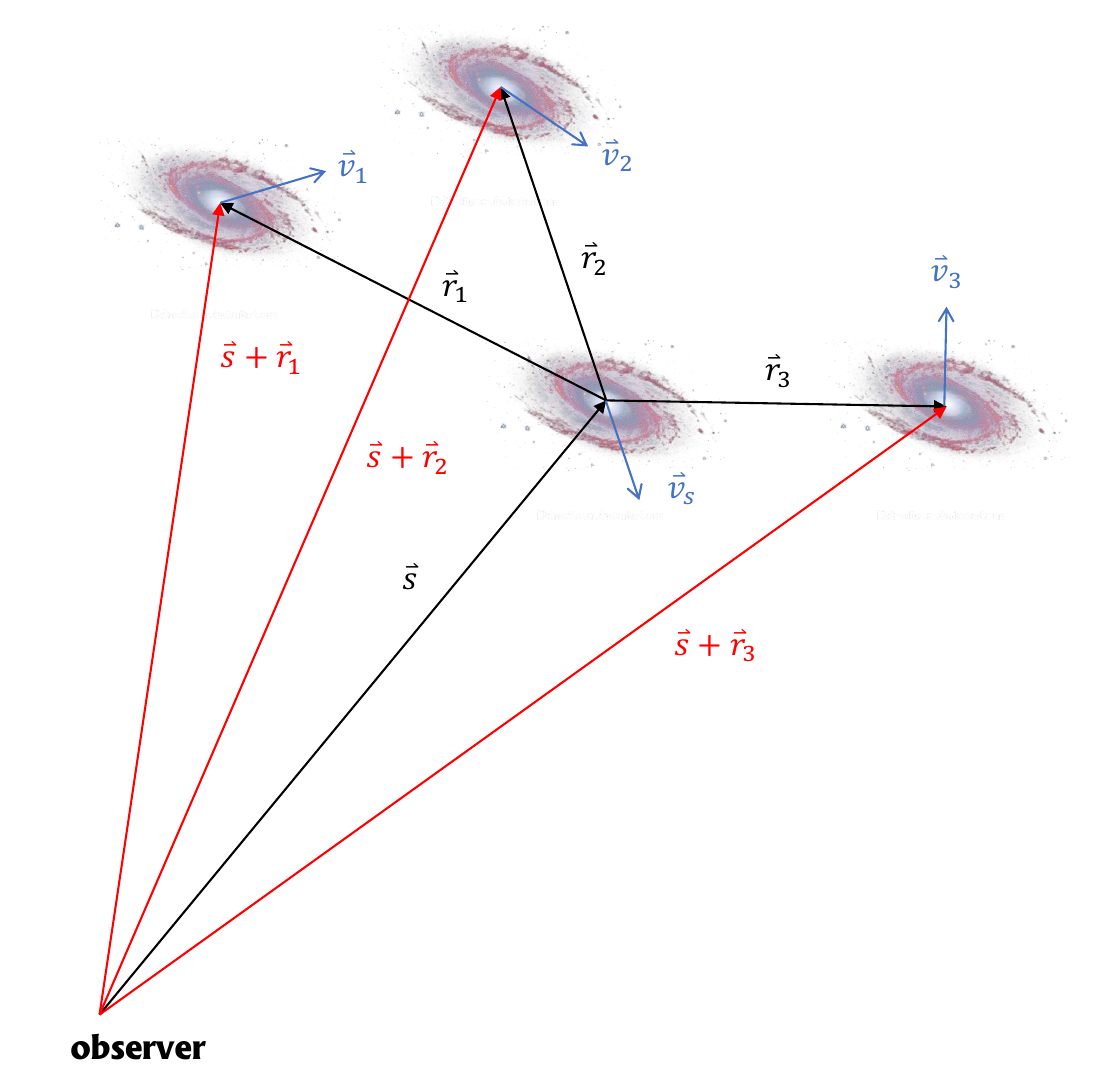}
    \caption[Geometry of local LOS for each galaxy in 4PCF]{Geometry of the local LOS for each galaxy in the 4PCF. The reference LOS is denoted by $\vb*{s}$, directed towards the reference galaxy. The arm vectors $\vb*{r_1}$, $\vb*{r_2}$, and $\vb*{r_3}$ give the positions of the three remaining galaxies relative to the reference. The studies \cite{Philcox:2022hkh, Hou:2022wfj, Slepian:2025kbb} consider arm-vector lengths in the range $20{-}164\,h^{-1}$\,Mpc.}
    \label{fig: LOS_correction}
\end{figure}

The plane-parallel approximation is overly simplistic for modern galaxy surveys \cite{BOSS:2015ewx, DESI2019AJ....157..168D, Euclid2022A&A...662A.112E}, where the 4PCF arm lengths $r_1$, $r_2$, and $r_3$ can be comparable to the radial distance from the observer, $s$. A more accurate treatment requires assigning a local LOS to each galaxy. Consider the configuration of four galaxies: the reference galaxy is at position $\vb*{s}$ and the three remaining galaxies are at positions $\vb*{r_1}$, $\vb*{r_2}$, and $\vb*{r_3}$ relative to it (see figure~\ref{fig: LOS_correction}). The LOS of the reference galaxy is $\hat{\vb*{s}}$ as usual, while the local LOS of each of the other three galaxies is (see \cite{Reimberg:2015jma_wideaglesep} and references therein),
\begin{equation}
    \vb*{\hat{n}_\sigma}=\widehat{\vb*{s}+\vb*{r_\sigma}} = \frac{\vb*{s}+\vb*{r_\sigma}}{\norm{\vb*{s}+\vb*{r_\sigma}}} \quad , \quad \sigma \in \{1,2,3\}.
\end{equation}
In this case, the polynomial structure of $\mathcal{T}_{1113}$ and $\mathcal{T}_{1122}$ becomes more complicated, since each mode is projected onto a distinct local LOS that also appears permuted among the redshift kernels $\mathcal{S}_{n,r}$ in the full 4PCF expression. The more general expansions are
\begin{equation}\label{eqn: wide-angle 4PCF integrands 1}
    \begin{split}
        &\mathcal{T}_{1113}(\vb*{k_1},\vb*{k_2},\vb*{k_3};\vb*{k_4};\vb*{\hat{n}_1},\vb*{\hat{n}_2},\vb*{\hat{n}_3};\vb*{\hat{n}_4}) \\
        &= \sum_{\{\alpha_{33}\},\{\gamma_{3\neq3}\}}i^{(\sum_{\rho\sigma}\alpha_{\rho\sigma})} t_{1113,\{\alpha_{33}\},\{\gamma_{3\neq 3}\}}(k_1,k_2,k_3)\prod^4_{\rho\sigma}\mu_{\rho\sigma}^{\alpha_{\rho\sigma}}\prod^4_{i\neq j}\kappa_{ij}^{\gamma_{ij}}, \\
        &\mathcal{T}_{1122}(\vb*{k_1},\vb*{k_2};\vb*{k_3},\vb*{k_4};\vb*{\hat{n}_1},\vb*{\hat{n}_2};\vb*{\hat{n}_3},\vb*{\hat{n}_4}) \\
        &= \sum_{\{\alpha_{33}\},\{\gamma_{3\neq3}\}}i^{(\sum_{\rho\sigma}\alpha_{\rho\sigma})} t_{1122,\{\alpha_{33}\},\{\gamma_{3\neq 3}\}}(k_1,k_2,k_3)\prod^4_{\rho\sigma}\mu_{\rho\sigma}^{\alpha_{\rho\sigma}}\prod^4_{i\neq j}\kappa_{ij}^{\gamma_{ij}}.
    \end{split}
\end{equation}
where $\mu_{\rho\sigma}=\vb*{\hat{k}_\rho}\cdot\vb*{\widehat{s+r_\sigma}}$. We also explicitly separate the mode-mode dot products $\kappa_{ij}$. Each local LOS factor can be further expanded in terms of the main LOS, $\vb*{\hat{s}}$, the arm direction $\vb*{\hat{r}_\sigma}$, and the mode directions $\vb*{k_\sigma}$ by applying the generating function of the Legendre polynomials,
\begin{equation}
        \widehat{\vb*{s}+\vb*{r_\sigma}}=\frac{\vb*{s}+\vb*{r_\sigma}}{\norm{\vb*{s}+\vb*{r_\sigma}}}=\left(\vb*{\hat{s}}+\frac{r_\sigma}{s}\vb*{\hat{r}_\sigma}\right)\left[\sum_{j_\sigma=0}^\infty \left(\frac{r_\sigma}{s}\right)^{j_\sigma}\mathcal{P}_{j_\sigma}(\vb*{\hat{s}}\cdot\vb*{\hat{r}_\sigma})\right].
\end{equation}
Projecting a mode vector onto this local LOS yields the exact cosine factor,
\begin{equation}\label{eqn: expansion of cosine factor}
    \mu_{\rho\sigma}=\vb*{\hat{k}_\rho}\cdot\widehat{\vb*{s}+\vb*{r_\sigma}}=\left(\mu_{\vb*{k_\rho s}}+\frac{r_\sigma}{s}\nu_{\rho\sigma}\right)\left[\sum_{j_\sigma=0}^\infty \left(\frac{r_\sigma}{s}\right)^{j_\sigma}\mathcal{P}_{j_\sigma}(\xi_\sigma)\right],
\end{equation}
where $r_\sigma<s$. Therefore, the new projection is expressed in terms of the three geometrical factors,
\begin{equation}\label{eqn: wide-angle projection factors}
    \mu_{\vb*{k_\rho s}}=\vb*{\hat{k}_\rho}\cdot\vb*{\hat{s}} \; , \quad \nu_{\rho\sigma}=\vb*{\hat{k}_\rho}\cdot\vb*{\hat{r}_\sigma} \; , \quad \xi_\sigma=\vb*{\hat{s}}\cdot\vb*{\hat{r}_\sigma}.
\end{equation}
Substituting expansion \eqref{eqn: expansion of cosine factor} into \eqref{eqn: wide-angle 4PCF integrands 1}, $\mathcal{T}_{1113}$ and $\mathcal{T}_{1122}$ become polynomials in the four types of projections,
\begin{equation}
    \prod^4_{\sigma=1}\mu_{\vb*{k_\sigma s}}\;,\quad \prod^3_{\sigma=1}\xi_\sigma\;,\quad \prod^4_{\rho=1}\prod^3_{\sigma=1}\nu_{\rho\sigma}\;,\quad \prod^4_{\rho\neq\sigma}\kappa_{\rho\sigma}. 
\end{equation}
Note that $\mathcal{P}_{j_\sigma}(\xi_\sigma)$ are themselves polynomials of $\xi_\sigma$. Since $\vb*{\hat{k}_4}=k_{123}^{-1}(\vb*{k_1}+\vb*{k_2}+\vb*{k_3})$, and the inverse norm $k_{123}^{-1}$ can again be expanded with the generating function of Legendre polynomials,  the dependence on $\vb*{\hat{k}_4}$ can always be traded for combinations of the three independent modes. The set of projections therefore reduces to
\begin{equation}
    \prod^3_{\sigma=1}\mu_{\vb*{k_\sigma s}}\;,\quad \prod^3_{\sigma=1}\xi_\sigma\;,\quad \prod^3_{\rho,\sigma=1}\nu_{\rho\sigma}\;,\quad \prod^3_{\rho\neq\sigma}\kappa_{\rho\sigma}. 
\end{equation}
The 4PCF integrands accordingly take the final polynomial forms,
\begin{equation}\label{eqn: wide-angle 4PCF integrands 2}
    \begin{split}
        &\mathcal{T}_{1113}(\vb*{k_1},\vb*{k_2},\vb*{k_3};\vb*{k_4};\vb*{\hat{n}_1},\vb*{\hat{n}_2},\vb*{\hat{n}_3};\vb*{\hat{n}_4}) \\
        &= \sum_{\{\alpha_{3}\},\{\beta_{3}\},\{\tau_{33}\},\{\gamma_{3\neq3}\}}i^{(\sum_{\sigma}\alpha_{\sigma})} t_{1113,\{\alpha_{3}\},\{\beta_{3}\},\{\tau_{33}\},\{\gamma_{3\neq3}\}}(k_1,k_2,k_3,s,r_1,r_2,r_3) \\
        &\prod^3_{\sigma=1}\mu^{\alpha_{\sigma}}_{\vb*{k_\sigma s}}\xi_\sigma^{\beta_{\sigma}} \prod^3_{\rho=1}\nu^{\tau_{\rho\sigma}}_{\rho\sigma} \prod^3_{i\neq j}\kappa^{\gamma_{ij}}_{ij}, \\
        &\mathcal{T}_{1122}(\vb*{k_1},\vb*{k_2};\vb*{k_3},\vb*{k_4};\vb*{\hat{n}_1},\vb*{\hat{n}_2};\vb*{\hat{n}_3},\vb*{\hat{n}_4}) \\
        &= \sum_{\{\alpha_{3}\},\{\beta_{3}\},\{\tau_{33}\},\{\gamma_{3\neq3}\}}i^{(\sum_{\sigma}\alpha_{\sigma})} t_{1122,\{\alpha_{3}\},\{\beta_{3}\},\{\tau_{33}\},\{\gamma_{3\neq3}\}}(k_1,k_2,k_3,s,r_1,r_2,r_3) \\
        &\prod^3_{\sigma=1}\mu^{\alpha_{\sigma}}_{\vb*{k_\sigma s}}\xi_\sigma^{\beta_{\sigma}} \prod^3_{\rho=1}\nu^{\tau_{\rho\sigma}}_{\rho\sigma} \prod^3_{i\neq j}\kappa^{\gamma_{ij}}_{ij}.
    \end{split}
\end{equation}
The new geometrical factors appear as potential new sources of parity-odd contributions to the 4PCF. We argue in the following subsection, however, that they do not survive the averaging over all LOS.

We emphasise that these expressions are exact: the Legendre expansion in \eqref{eqn: expansion of cosine factor} is valid for all $r_\sigma < s$ without further approximation. For the case $r_\sigma>s$, one may instead expand in $s/r_\sigma$, which yields the same set of dot products with the roles of $\mu_{\vb*{k_\rho s}}$ and $\nu_{\rho\sigma}$ interchanged.

\subsection{Isotropic wide-angle 4PCF: still PV-free}

We now extend the analysis of Section~\ref{sec: Isotropic 4PCF with Plane Parallel Limit} to the wide-angle case. In addition to the LOS integral, we must also handle the directional integrals over the mode vectors. Writing the Fourier integral as
\begin{equation}
    \int_{\vb*{k_\lambda}}e^{i\vb*{k_\lambda}\cdot\vb*{r_\lambda}}=\int dk_\lambda \, 4\pi k_\lambda^2 \sum^\infty_{l_\lambda m_\lambda}i^{l_\lambda} j_{l_\lambda}(k_\lambda r_\lambda)Y_{l_\lambda m_\lambda}(\vb*{\hat{r}_\lambda}) \int d \Omega_{k_\lambda} Y^*_{l_\lambda m_\lambda}(\vb*{\hat{k}_\lambda}),
\end{equation}
where we have applied the \emph{Rayleigh expansion},
\begin{equation}
    e^{i\vb*{k_\lambda}\cdot\vb*{r_\lambda}} = 4\pi \sum^\infty_{l_\lambda m_\lambda} i^{l_\lambda} j_{l_\lambda}(k_\lambda r_\lambda) Y_{l_\lambda,m_\lambda}(\vb*{\Hat{r}_\lambda})Y^*_{l_\lambda,m_\lambda}(\vb*{\Hat{k}_\lambda}).
\end{equation}
$j_{l_\lambda}$ is the $l_\lambda$th-\emph{spherical Bessel function}. This decomposition is necessary to resolve the mode-direction dependence in the factors $\nu_{\rho\sigma}$ and $\kappa_{\rho\sigma}$, ultimately expressing everything in terms of $\vb*{\hat{r}_\sigma}$ alone. The isotropic estimator then takes the form,
\begin{equation}\label{eqn: isotropic estimator of 4PCF with los average integral wide angle}
    \begin{split}
        \hat{\zeta}(\vb*{r_1},\vb*{r_2},\vb*{r_3}) &= \int \frac{s^2ds}{V_s} \int d\Omega_s\,\zeta(\vb*{s};\vb*{r_1},\vb*{r_2},\vb*{r_3}) \\
        &\propto \int d\Omega_s \,\sum_{\{\alpha_{3}\},\{\beta_{3}\},\{\tau_{33}\},\{\gamma_{3\neq3}\}} \sum_{\{l_3 m_3\}}^\infty \prod_{\lambda=1}^3 i^{l_\lambda} j_{l_\lambda}(k_\lambda r_\lambda)Y_{l_\lambda m_\lambda}(\vb*{\hat{r}_\lambda}) \\
        &\int d\Omega_{\vb*{k_1}} \, \int d\Omega_{\vb*{k_2}} \, \int d\Omega_{\vb*{k_3}} Y_{l_\lambda m_\lambda}(\vb*{\hat{k}_\lambda}) \, 
        \prod^3_{\sigma=1}\mu^{\alpha_{\sigma}}_{\vb*{k_\sigma s}}\xi_\sigma^{\beta_{\sigma}} \prod^3_{\rho=1}\nu^{\tau_{\rho\sigma}}_{\rho\sigma} \prod^3_{i\neq j}\kappa^{\gamma_{ij}}_{ij}.
    \end{split}
\end{equation}
Following Section~\ref{sec: Isotropic 4PCF with Plane Parallel Limit}, we expand each cosine factor in spherical harmonics. In addition to equation~\eqref{eqn: cosine decomposition with spherical harmonics}, the remaining factors decompose as
\begin{equation}
    \begin{split}\label{eqn: wide angle cosine decomposition with spherical harmonics}
         \xi_\sigma^{\beta_\sigma} &= \sum_{p_\sigma,m_{p_\sigma}}^{\beta_\sigma} c^{(\beta_\sigma)}_{p_\sigma} \frac{4\pi}{2p_\sigma+1} Y_{p_\sigma,m_{p_\sigma}}(\vb*{\hat{s}})Y^*_{p_\sigma,m_{p_\sigma}}(\vb*{\hat{r}_\sigma}), \\
         \nu_{\rho\sigma}^{\tau_{\rho\sigma}} &= \sum_{\omega_{\rho\sigma},m_{\omega_{\rho\sigma}}}^{\tau_{\rho\sigma}} c^{(\tau_{\rho\sigma})}_{\omega_{\rho\sigma}} \frac{4\pi}{2\omega_{\rho\sigma}+1} Y_{\omega_{\rho\sigma},m_{\omega_{\rho\sigma}}}(\vb*{\hat{k}_\rho})Y^*_{\omega_{\rho\sigma},m_{\omega_{\rho\sigma}}}(\vb*{\hat{r}_\sigma}), \\
         \kappa_{ij}^{\gamma_{ij}} &= \sum_{\eta_{ij},m_{\eta_{ij}}}^{\gamma_{ij}} c^{(\gamma_{ij})}_{\eta_{ij}} \frac{4\pi}{2\eta_{ij}+1} Y_{\eta_{ij},m_{\eta_{ij}}}(\vb*{\hat{k}_i})Y^*_{\eta_{ij},m_{\eta_{ij}}}(\vb*{\hat{k}_j}),
    \end{split}
\end{equation}
where we introduce new angular momentum indices, $p_\sigma$, $\omega_{\rho\sigma}$, and $\eta_{ij}$, associated with each cosine factor. Substituting these expansions into \eqref{eqn: isotropic estimator of 4PCF with los average integral wide angle} and focusing on the directional integrations, one finds
\begin{equation}\label{eqn: 4PCF estimator final proportion}
    \begin{split}
        \hat{\zeta}&(\vb*{r_1},\vb*{r_2},\vb*{r_3})\propto \\
        &\sum_{\{\alpha_{3}\},\{\beta_{3}\},\{\tau_{33}\},\{\gamma_{3\neq3}\}} \sum^\infty_{\{ l_3,m_3 \}} \sum_{\{j_3 m_{j_3}\}}^{\{\alpha_3\}}\sum_{\{p_3,m_{p_3}\}}^{\{\beta_3\}} \sum_{\{\omega_{33},m_{\omega_{33}}\}}^{\{\tau_{33}\}} \sum_{\{\eta_{3\neq3},m_{\eta_{3\neq3}}\}}^{\{\gamma_{3\neq 3}\}} \prod_{\lambda=1}^3\left[i^{l_\lambda}j_{l_\lambda}(k_\lambda r_\lambda)Y_{l_\lambda m_\lambda}(\vb*{\hat{r}_\lambda})\right]\\
        &\prod_{\sigma=1}^3 \left[\frac{c^{(\alpha_\sigma)}_{j_\sigma}}{2j_\sigma+1} \frac{c^{(\beta_\sigma)}_{p_\sigma}}{2p_\sigma+1}Y^*_{p_\sigma,m_{p_\sigma}}(\vb*{\hat{r}_\sigma})\prod_{\rho=1}^3 \frac{c^{(\tau_{\rho\sigma})}_{\omega_{\rho\sigma}}}{2\omega_{\rho\sigma}+1}Y^*_{\omega_{\rho\sigma},m_{\omega_{\rho\sigma}}}(\vb*{\hat{r}_\sigma}) \right]\prod_{i\neq j}^3\left[ \frac{c^{(\gamma_{ij})}_{\eta_{ij}}}{2\eta_{ij}+1} \right] \\
        &\int d\Omega_{\vb*{k_1}} \int d\Omega_{\vb*{k_2}} \int d\Omega_{\vb*{k_3}} \int d\Omega_s  \prod_{\lambda =1}^3 \left[Y_{l_\lambda m_\lambda}(\vb*{\hat{k}_\lambda})\right]\prod_{\sigma=1}^3 \Bigg[ Y_{j_\sigma,m_{j_\sigma}}(\vb*{\hat{k}_\sigma})Y^*_{j_\sigma,m_{j_\sigma}}(\vb*{\hat{s}})\\ 
        &Y_{p_\sigma,m_{p_\sigma}}(\vb*{\hat{s}})\prod_{\rho=1}^3 Y_{\omega_{\rho\sigma},m_{\omega_{\rho\sigma}}}(\vb*{\hat{k}_\rho}) \Bigg] \prod_{i\neq j}^3\left[ Y_{\eta_{ij},m_{\eta_{ij}}}(\vb*{\hat{k}_i})Y^*_{\eta_{ij},m_{\eta_{ij}}}(\vb*{\hat{k}_j}) \right].
    \end{split}
\end{equation}
The LOS integral will capture the spherical harmonic functions of $\vb*{\hat{s}}$ arising from the expansions of $\mu_{\vb*{k_\sigma}\vb*{s}}$ and $\xi_\sigma$. Each summand is therefore proportional to
\begin{equation}
    \int d\Omega_s \prod^3_{\sigma=1}Y^*_{j_\sigma,m_{j_\sigma}}(\hat{\vb*{s}})Y_{p_\sigma,m_{p_\sigma}}(\hat{\vb*{s}}).
\end{equation}

As shown in Appendix~\ref{apx: Integrals of Spherical Harmonics}, this integral is non-vanishing only if
\begin{equation}\label{eqn: s constraint}
    \sum^3_{\sigma=1} (j_\sigma+p_\sigma) = \text{even} \quad \text{and} \quad \sum^3_{\sigma=1} (-m_{j_\sigma}+m_{p_\sigma}) = 0.
\end{equation}
These selection rules follow from the parity properties of spherical harmonics under $\vb*{\hat{s}} \to -\vb*{\hat{s}}$. When the integration measure is isotropic or at least symmetric, only the combination of spherical harmonics which is an even function of $\vb*{\hat{s}}$ survives, and the odd function vanishes. Therefore, the selection rules that make the product of spherical harmonics an even function are enforced as a non-vanishing condition. This is equivalent to the constraint imposed by the Wigner 3-j symbols arising from the Gaunt integral in equation~\eqref{eqn: Gaunt integral}.

These conditions constitute the first set of constraints on the spherical harmonic indices from the LOS integration. The subsequent integrations over $\Omega_{k_n}$ capture the spherical harmonic functions of $\vb*{\hat{k}_\sigma}$ from $\mu_{\vb*{k_\sigma}\vb*{s}}$, $\nu_{\rho\sigma}$, $\kappa_{ij}$ and the plane-wave factors, imposing analogous constraints on the angular momentum indices:
\begin{equation}\label{eqn: k constraints}
    \begin{split}
        l_1+j_1+\sum_{\sigma=1}^3\omega_{1\sigma}+\eta_{12}+\eta_{21}+\eta_{13}+\eta_{31} = \text{even} \quad &\text{from} \quad \int d\Omega_{k_1}, \\
        l_2+j_2+\sum_{\sigma=1}^3\omega_{2\sigma}+\eta_{12}+\eta_{21}+\eta_{23}+\eta_{32} = \text{even} \quad &\text{from} \quad \int d\Omega_{k_2}, \\
        l_3+j_3+\sum_{\sigma=1}^3\omega_{3\sigma}+\eta_{13}+\eta_{31}+\eta_{23}+\eta_{32} = \text{even} \quad &\text{from} \quad \int d\Omega_{k_3}.
    \end{split}
\end{equation}
This is equivalent to the constraint imposed by the Wigner 3-j symbols arising from the Gaunt integral, but for every mode vector, $\vb*{k_1}$, $\vb*{k_2}$, and $\vb*{k_3}$. Each selection rule is obtained by collecting the spherical harmonics of the same $\vb*{k_\sigma}$ and integrating over all directions of $\vb*{k_\sigma}$, $Y_{l_\lambda m_\lambda}$, $Y_{j_\sigma m_{j_\sigma}}$, $Y_{\omega_{\rho\sigma} m_{\omega_{\rho\sigma}}}$, $Y_{\eta_{ij} m_{\eta_{ij}}}$, and $Y^*_{\eta_{ij} m_{\eta_{ij}}}$, in particular.

The integral \eqref{eqn: 4PCF estimator final proportion} is non-vanishing only when all these conditions are satisfied simultaneously. After evaluating the integrals in \eqref{eqn: isotropic estimator of 4PCF with los average integral wide angle}, the estimator depends on $\vb*{\hat{r}_1}$, $\vb*{\hat{r}_2}$, and $\vb*{\hat{r}_3}$ only through the remaining spherical harmonics. Ignoring the constants and spherical Bessel functions, this gives
\begin{equation}\label{eqn: 4PCF isotropic estimator with only spherical harmonics}
    \begin{split}
        \hat{\zeta}(\vb*{r_1},\vb*{r_2},\vb*{r_3}) \propto& \sum_{\{\alpha_{3}\},\{\beta_{3}\},\{\tau_{33}\},\{\gamma_{3\neq3}\}} \sum^\infty_{\{ l_3,m_3 \}} \sum_{\{j_3 m_{j_3}\}}^{\{\alpha_3\}}\sum_{\{p_3,m_{p_3}\}}^{\{\beta_3\}} \sum_{\{\omega_{33},m_{\omega_{33}}\}}^{\{\tau_{33}\}} \sum_{\{\eta_{3\neq3},m_{\eta_{3\neq3}}\}}^{\{\gamma_{3\neq 3}\}} \\
        & \prod_{\lambda=1}^3Y_{l_\lambda m_\lambda}(\vb*{\hat{r}_\lambda})\prod_{\sigma=1}^3 \left[Y^*_{p_\sigma,m_{p_\sigma}}(\vb*{\hat{r}_\sigma})\prod_{\rho=1}^3 Y^*_{\omega_{\rho\sigma},m_{\omega_{\rho\sigma}}}(\vb*{\hat{r}_\sigma}) \right].
    \end{split}
\end{equation}
Recall that the parity transformation is $\mathbb{P}[\hat{\zeta}(\vb*{r_1},\vb*{r_2},\vb*{r_3})] = \hat{\zeta}(-\vb*{r_1},-\vb*{r_2},-\vb*{r_3})$ and the parity transformation of spherical harmonic function is $Y_{l,m}(-\vb*{\Hat{r}}) = (-1)^lY_{l,m}(\vb*{\Hat{r}})$. The estimator is therefore parity-even if and only if the total angular momentum sum,
\begin{equation}\label{eqn: S parameter}
    S = \sum_{\lambda=1}^3 l_\sigma + \sum_{\sigma=1}^3 \left(p_\sigma+\sum_{\rho=1}^3\omega_{\rho\sigma}\right),
\end{equation}
is always even. Adding all the constraints from \eqref{eqn: s constraint} and \eqref{eqn: k constraints} together yields
\begin{equation}\label{eqn: final selection rule}
    S + 2\left(\sum_{\sigma=1}^3 j_\sigma + \sum_{i\neq j}^3\eta_{ij}\right) = \text{even} \longrightarrow S = \text{even}.
\end{equation}
Hence, the 4PCF isotropic estimator \eqref{eqn: 4PCF estimator final proportion} is non-vanishing only for configurations where $S$ is even, corresponding to a parity-even product of spherical harmonics in \eqref{eqn: 4PCF isotropic estimator with only spherical harmonics}. We have thus demonstrated explicitly that the parity-odd modes are suppressed by the LOS averaging even when the exact local LOS is used for each galaxy in the 4PCF.

Intuitively, the integration over all LOS in a symmetric domain in elevation angle removes odd multipoles of $\zeta$ as a function of the LOS orientation. We demonstrate that the parity-odd contribution from the RSD effect up to the leading GR correction always couples with LOS through the cosine factor. This is because it exploits the fact that LOS can be used to construct a scalar factor $\mu_{\vb*{k}_\sigma\vb*{s}}$ which is an odd power of $\vb*{k}$ and, therefore, exists as a parity-odd contribution in a scalar expression of $\Delta_g$. However, the LOS coupling is exactly why those parity-odd contributions get averaged out by integration over all LOS, as they simultaneously are odd powers of $\vb*{\hat{n}}$ and hence odd multipoles of the LOS orientation. In other words, the preferred handedness in the observed galaxy distribution introduced by the RSD effect, which is coupled to the LOS, is washed out by averaging over all orientations of $\vb*{\hat{n}}$.
\section{Discussion and potential sources of parity violation}\label{sec: Discussion and Potential Sources of Violation}

We now discuss the caveats of our results and identify potential physical sources of PV that lie outside the scope of the present work.
\subsection{Sky coverage}\label{sec: sky coverage}

Our analysis assumes full-sky coverage of the galaxy survey, which is crucial: the constraint on the angular momentum indices \eqref{eqn: s constraint} follows from integration over all LOS on the complete spherical sky. In practice, this condition is not satisfied, since parts of the sky are obscured by the galactic plane, dividing the observable sky into the Northern Galactic Cap (NGC) and the Southern Galactic Cap (SGC). For ground-based surveys such as SDSS (see Figure 6 in \cite{SDSS2015ApJS..219...12A}) and DESI (see Figure 1 in \cite{DESI2019AJ....157..168D}), the sky access is further limited by the horizon, and since both instruments are located in the northern hemisphere, the NGC is more accessible than the SGC.

One way to account for incomplete sky coverage is to introduce a survey window function $W_{s}(\vb*{s})$ into the estimator:
\begin{equation}\label{eqn: estimator with window function}
    \begin{split}
        \hat{\zeta}&(\vb*{r_1},\vb*{r_2},\vb*{r_3}) = \int \frac{d^3s}{V_s}  (\vb*{s})W_{s}(\vb*{\hat{s}})W_{s}(\vb*{\widehat{s+r_1}})W_{s}(\vb*{\widehat{s+r_2}})W_{s}(\vb*{\widehat{s+r_3}})\zeta(\vb*{s};\vb*{r_1},\vb*{r_2},\vb*{r_3}) \\
        &=\frac{1}{V_s} \int s^2ds\int_{\mathcal{S}} d\Omega_s\,  W_{s}(\vb*{\hat{s}})W_{s}(\vb*{\widehat{s+r_1}})W_{s}(\vb*{\widehat{s+r_2}})W_{s}(\vb*{\widehat{s+r_3}})\zeta(\vb*{s};\vb*{r_1},\vb*{r_2},\vb*{r_3}),
    \end{split}
\end{equation}
where $\mathcal{S}$ denotes the full spherical surface. The survey window function is applied to the position of each of the four $\Delta_g$ fields in $\zeta$. As discussed in Appendix~\ref{apx: Integrals of Spherical Harmonics}, the parity-odd modes are removed entirely if the LOS integration domain on sky coordinates is a full revolution in azimuthal angle and symmetric around 0 elevation angle. Therefore, the result of this paper depends on how close the observable sky, which is the full sky modulated by $W_{s}$, is to this requirement. Theoretically, the reference point is free. Practically, to avoid the bright galactic plane and maximise the sky-surveyed field, one can aim to survey the NGC and SGC regions equally to satisfy the domain condition.

An alternative viewpoint of this situation is that we can expand each $W_{s}$ in multipoles of LOS. The parity leakage happens when $W_{s}$ has odd multipoles. In particular, as PV from the RSD effect is introduced through dot products of $\vb*{\hat{s}}\cdot\vb*{\hat{r}_\sigma}$, the parity-violating terms that are odd in $\vb*{\hat{r}_\sigma}$ (hence, a parity-odd contribution to $\hat{\zeta}$) are also odd in $\vb*{\hat{s}}$. Without the window function, these terms are washed out by integrating over all LOS precisely because they are odd in $\vb*{\hat{s}}$. However, this window function can contain the term which is odd purely in $\vb*{\hat{s}}$. Upon coupling, the parity-violating terms become even in $\vb*{\hat{s}}$ and leak through the integration over all LOS. But they are still odd in $\vb*{\hat{r}_\sigma}$ and contribute as parity-odd terms in $\hat{\zeta}$. Therefore, $W_{s}$ must contain only even multipoles in $\vb*{\hat{s}}$ to preserve the constraint \eqref{eqn: s constraint}.

\begin{figure}[tp]
  \centering
  \begin{subfigure}[b]{\textwidth}
    \centering
    \includegraphics[width=0.8\textwidth,height=0.22\textheight,keepaspectratio]{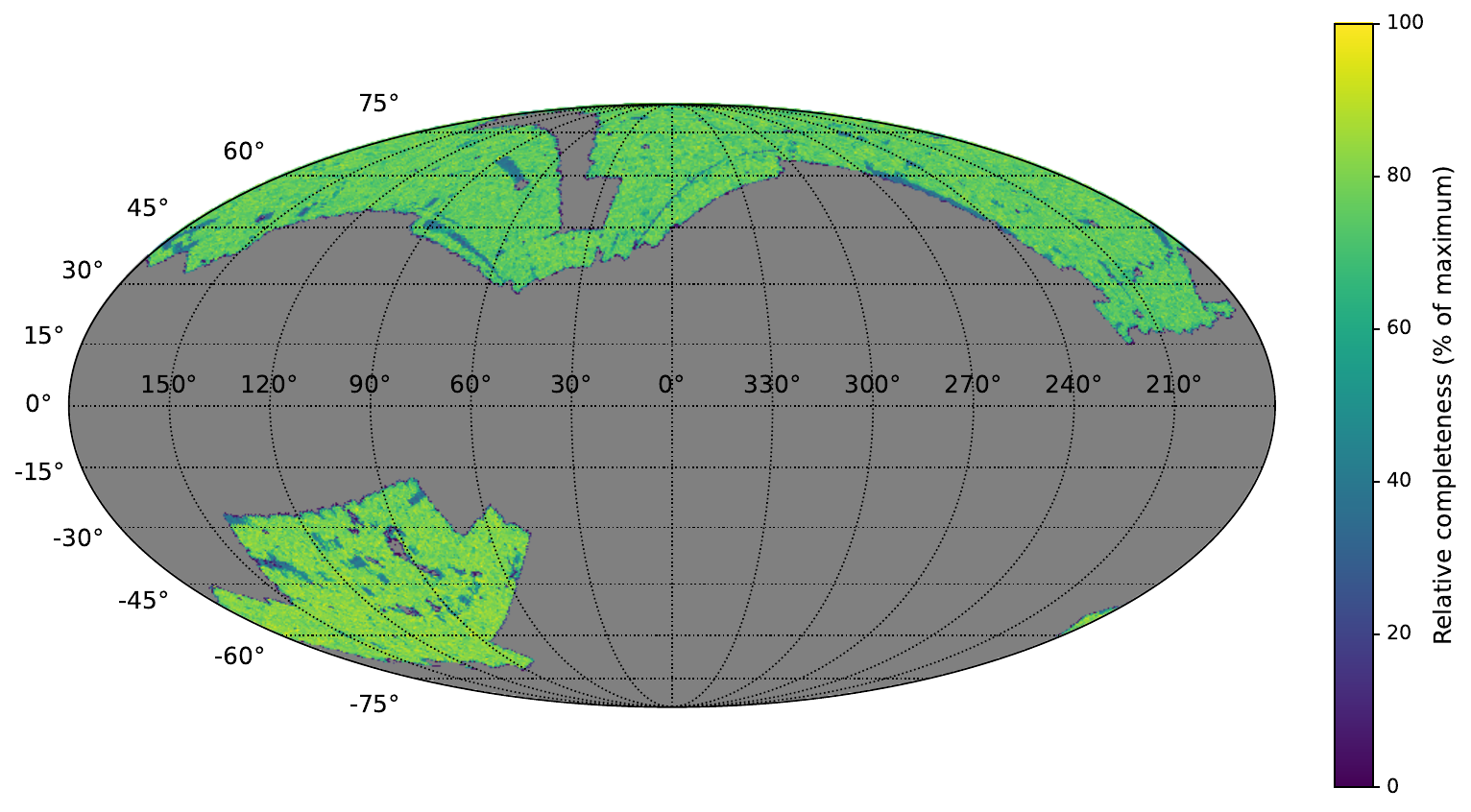}
    \caption{Relative completeness of the SDSS-BOSS DR12 survey, LOWZ catalogue at $0.15<z<0.43$.}
  \end{subfigure}
  \vspace{1em}
  \begin{subfigure}[b]{\textwidth}
    \centering
    \includegraphics[width=0.8\textwidth,height=0.22\textheight,keepaspectratio]{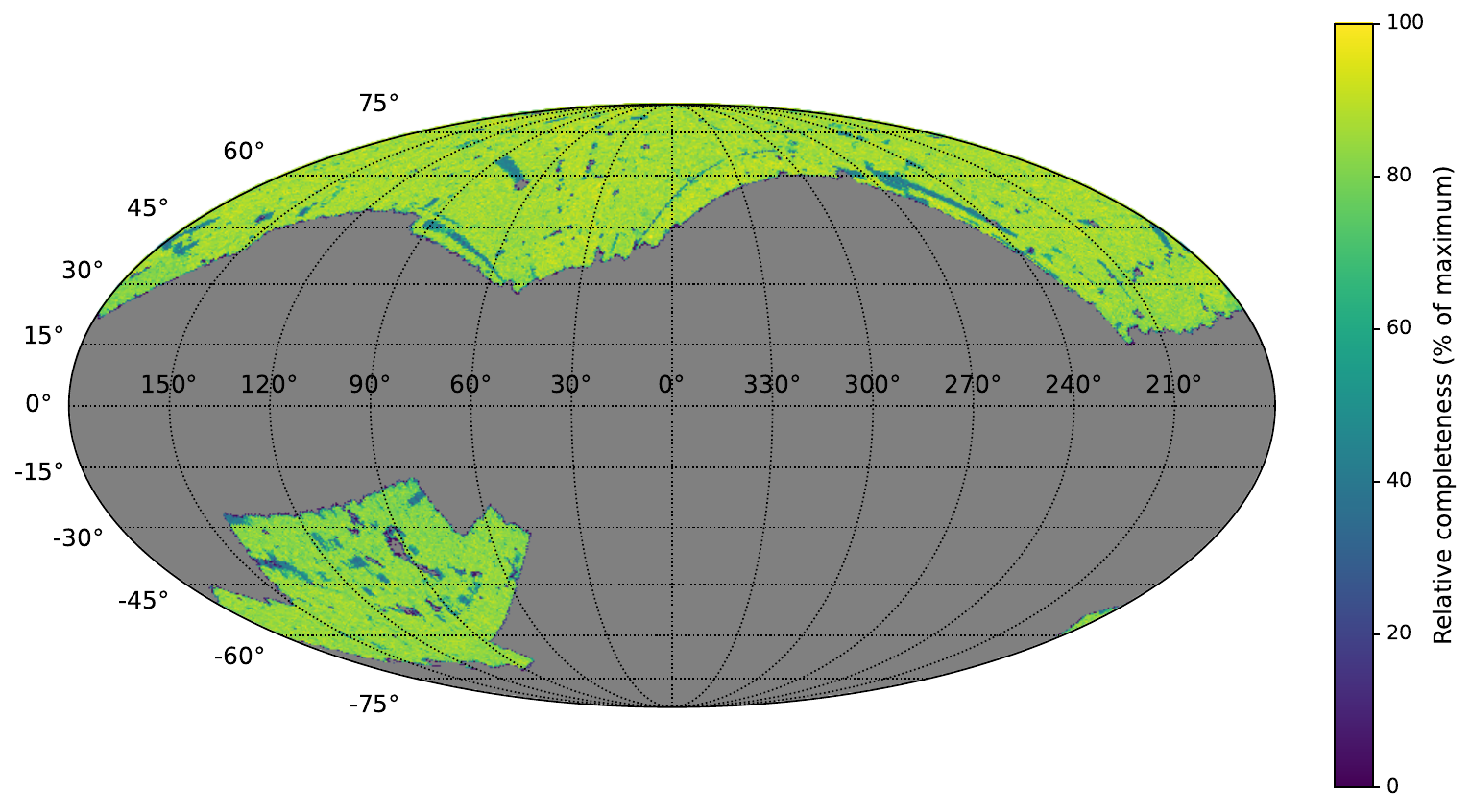}
    \caption{Relative completeness of the SDSS-BOSS DR12 survey, CMASS catalogue at $0.43<z<0.7$.}
  \end{subfigure}
  \vspace{1em}
  \begin{subfigure}[b]{\textwidth}
    \centering
    \includegraphics[width=0.8\textwidth,height=0.22\textheight,keepaspectratio]{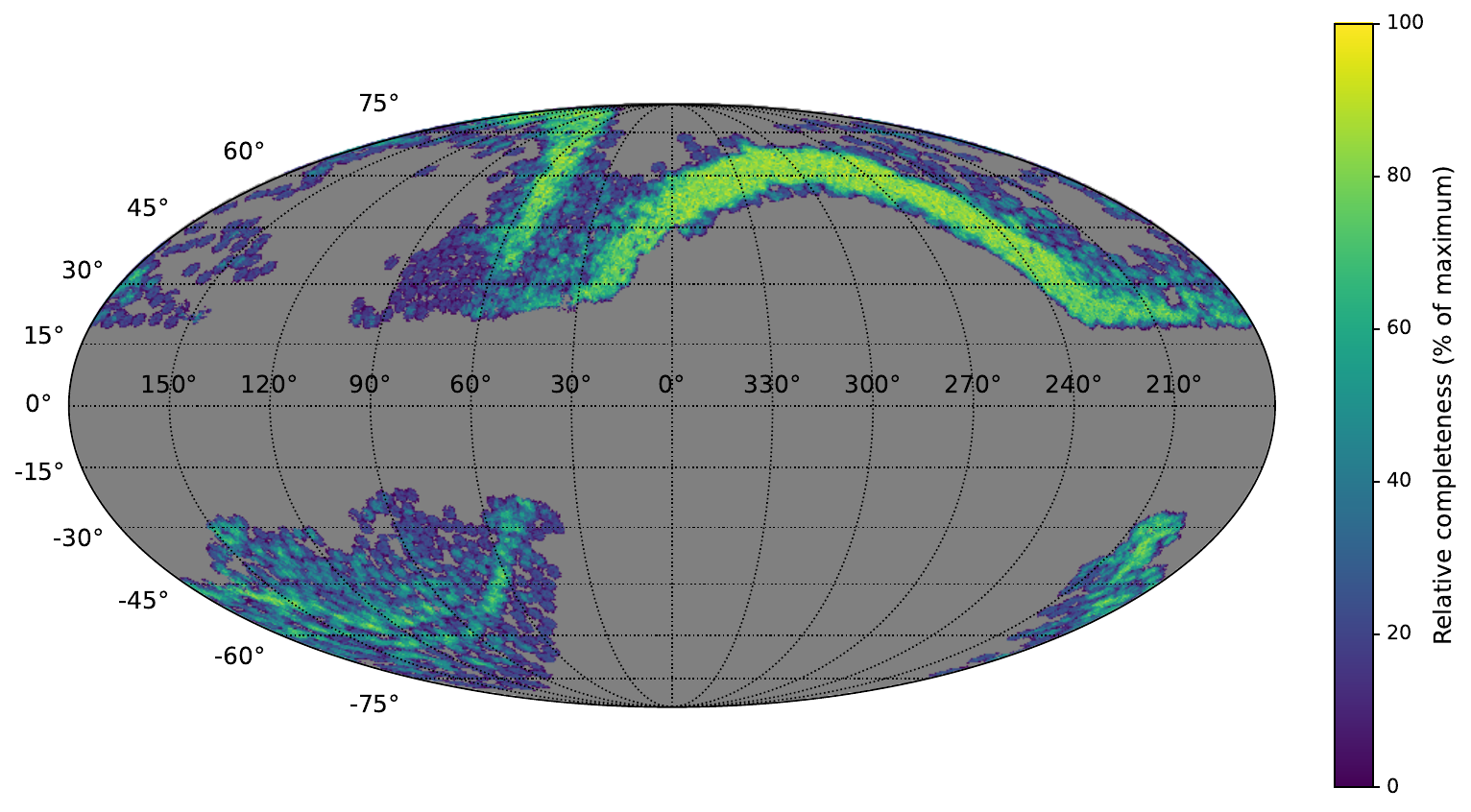}
    \caption{Relative completeness of the DESI LRG Y1 survey at $0.4<z<1.1$.}
  \end{subfigure}
  \caption{\label{fig: sky completeness} Relative completeness of the galaxy surveys. The quantity is defined from the normalised survey window function, $100\times W_{s}(\vb*{\hat{n}_p})/\max_{\vb*{\hat{n}_p}}W_{s}$. The function is projected on the galactic coordinates, where the horizontal and vertical degree axes are the galactic longitude and latitude, respectively. The galactic plane lies along the $0^\circ$ latitude line. So, the upper half of the sky is the NGC region while the lower half is the SGC region. Panels (a) and (b) show the relative completeness of the SDSS-BOSS DR12 survey, LOWZ and CMASS catalogues, respectively; panel (c) shows the relative completeness of the DESI LRG Y1 survey.}
\end{figure}

We estimate the parity-odd and parity-even parts of the weighted survey angular window function from the SDSS-BOSS DR12 catalogue \cite{BOSS:2015ewx} and the DESI Y1 catalogue \cite{DESI2025JCAP...01..125R}. In particular, we construct the angular survey window using the random catalogues associated with each survey. Random catalogues are designed to reproduce the angular and radial selection functions of the corresponding galaxy samples. We pixelise the random catalogues using \texttt{HEALPix} \cite{Healpix2005ApJ...622..759G} and estimate the weighted angular selection function in each pixel as
\begin{equation}
    W_{s}(\vb*{\hat{n}_p})=M_p\frac{\sum_{i\in p}w_i}{\left\langle\sum_{i\in p}w_i\right\rangle_{\rm footprint}}
\end{equation}
where $\vb*{\hat{n}_p}$ is the direction to the pixel, and $w_i$ is the random-catalogue weight. The binary angular footprint $M_p$ of the telescope is defined as
\begin{equation}
    M_p=
    \begin{cases}
        1 \; ;\; p\in \mathcal{F} \\
        0\; ;\; p\notin \mathcal{F}
    \end{cases}
\end{equation}
where $\mathcal{F}$ is the set of pixels within the angular field-of-view of the survey. Figure~\ref{fig: sky completeness} shows the sky relative completeness percentage, which is the normalised survey window function $100\times W_{s}(\vb*{\hat{n}_p})/\max_{\vb*{\hat{n}_p}}W_{s}$, of each survey in the galactic coordinate system.

We subsequently decompose this window into antipodally symmetric and antisymmetric components,
\begin{equation}
    W_{\text{even}}(\vb*{\hat{n}_p})=\frac{W_{s}(\vb*{\hat{n}_p})+W_{s}(-\vb*{\hat{n}_p})}{2}\quad\text{and}\quad W_{\text{odd}}(\vb*{\hat{n}_p})=\frac{W_{s}(\vb*{\hat{n}_p})-W_{s}(-\vb*{\hat{n}_p})}{2},
\end{equation}
respectively. We quantify the magnitude of the odd multipoles in these windows by computing the RMS ratio
\begin{equation}\label{eqn: window rms ratio}
    R=\frac{\sigma_{\rm odd}}{\sigma_{\rm even}}\quad \text{where}\quad \sigma_{\text{h}}^2=\sum_{p} \abs{W_{\text{h}}(\vb*{\hat{n}_p})}^2 \; ;\; \text{h} \in \{\text{even},\text{odd}\}.
\end{equation}
The results are reported in Table~\ref{tab:window_parity_rms}. A useful estimate of the odd-power leakage in $\hat{\zeta}$ by the window function of a given survey is that the leading parity-odd mode in $\zeta$ is suppressed by a factor of $\sim R$ relative to the parity-even mode. Therefore, we expect the parity-odd modes from the RSD effect to be suppressed by around 15\% and 40\% relative to the even modes, for the SDSS-BOSS DR12 catalogue (used in \cite{Hou:2022wfj, Philcox:2022hkh}) and the DESI LRG Y1 catalogue (used in \cite{Slepian:2025kbb}), respectively. We remind the reader that the parity-odd mode in $\zeta$ from the RSD effect is already generally suppressed by a factor of $\epsilon_\mathcal{H}^{1/2}=\mathcal{H}/k\sim0.004{-}0.2$ (see discussion at the end of Section~\ref{sec: Synopsis: the Observational Implication and its Validity}) relative to the parity-even mode before applying the window functions. This quantitative estimation is done by approximating $W_{s}(\vb*{\widehat{s+r_\sigma}})\approx W_{s}(\vb*{\hat{s}})$ in equation~\eqref{eqn: estimator with window function}. The correction to this approximation can introduce more parity-odd leakage to the isotropic estimator. However, the authors of \cite{Hou:2022wfj} estimate this survey geometry correction to be of the order $10^{-6}$ compared to $W_{s}(\vb*{\hat{s}})$ for the SDSS-BOSS DR12 catalogue (see section 6.3.2 in \cite{Hou:2022wfj}). Hence, the parity-odd leakage from this correction is negligible.
\begin{table}[t]
\centering
\caption{Redshift ranges and odd-to-even RMS ratios of the
angular survey windows for the SDSS-BOSS DR12 and DESI LRG Y1 samples.
The RMS ratio is defined in equation~\eqref{eqn: window rms ratio}.}
\label{tab:window_parity_rms}
\begin{tabular}{lcc}
\toprule
Sample & Redshift range & $R$ \\
\midrule
BOSS LOWZ  & $0.20 < z < 0.40$ & 0.85 \\
BOSS CMASS & $0.43 < z < 0.70$ & 0.85 \\
DESI LRG   & $0.40 < z < 1.10$ & 0.60 \\
\bottomrule
\end{tabular}
\end{table}

The suppression is expected to be stronger in forthcoming galaxy surveys such as the Euclid Wide Survey \cite{Euclid2022A&A...662A.112E}, where the planned survey footprint is much larger than that of SDSS-BOSS DR12 and DESI LRG Y1.

\subsection{Integration along LOS, observer terms, and beyond $\mathcal{O}(\mathcal{H}/k)$}

As discussed in Section~\ref{sec: The Relativistic Galaxy Number Density: Redshift Space Distortion}, we do not include contributions from integration along the LOS, namely the ISW effect, the Shapiro time delay, and gravitational lensing (\cite{Bertacca2014JCAP...09..037B,Gervani:2025wmh} and references therein).  While the lensing contribution is comparable in magnitude to the Newtonian term on large separations $r_\sigma$ \cite{Hui2007PhRvD..76j3502H, Hui2008PhRvD..77f3526H}, studies \cite{Bouvin2014PhRvD..89h3535B, Lepori2018JCAP...05..043L} show that it contributes to odd multipoles of the power spectrum at a level several orders of magnitude below that of other relativistic effects; the lensing dipole contribution is suppressed by a factor of $\mathcal{H}r\sim\mathcal{H}/k$ relative to other relativistic dipole contributions. In particular, Figure 8 in \cite{Bouvin2014PhRvD..89h3535B} and Figure 3 in \cite{Lepori2018JCAP...05..043L} show that it only becomes comparable at $z=2$ and separation $>200h^{-1}$~Mpc. Lensing is therefore a subdominant source of parity-odd contributions to the 4PCF relative to the Doppler and gravitational redshift effects.

The ISW and Shapiro time delay effects contribute to the observed galaxy number density at order $(\mathcal{H}/k)^2$ due to the integration along the LOS. Consequently, they are smaller than the corrections considered here. Whether these secondary effects can source parity-odd contributions in the 4PCF after LOS averaging or not remains an open question. These LOS-integrated effects cannot be incorporated trivially into our framework because they break the local structure of the polynomial kernel that was exploited in the main analysis (see equation~\eqref{eqn: local term of delta_g} and discussion after). Furthermore, should one wish to include corrections at order $\epsilon_\mathcal{H}\equiv(\mathcal{H}/k)^2$, the EPT solutions must also be corrected accordingly, since GR modifies the Newtonian equations of motion starting from order $\mathcal{O}(\epsilon_\mathcal{H}\delta)$ \cite{Bardeen:1980kt, Mukhanov1992PhR...215..203M}.

This work also neglects the effect of the observer terms, i.e., $\Psi_o$ and $\vb*{v}_o$ (see Section~\ref{sec: The Relativistic Galaxy Number Density: Redshift Space Distortion}). First, any derivatives of these quantities are zero, as they are constant. $\Psi_o$ is the true scalar, which does not carry any preferred direction, and thus contributes only to the monopole term of the 4PCF. Therefore, it cannot be a source of PV through the observed galaxy number density. $\vb*{v}_o$ is, however, a constant vector with a preferred direction. It appears in the redshift perturbation as $v^2_o/2$ and $v_{o\|}$ in equation~\eqref{eqn: redshift perturbation final}, which goes into the observed galaxy number density through equation~\eqref{eqn: full relativistic Number Density}. $v^2_o/2$ is a true scalar and has no preferred direction, similar to $\Psi_o$. However, the term with odd powers of $v_{o\|}$ is the global dipole term discussed in, e.g., \cite{Bouvin2014PhRvD..89h3535B}. But, similar to other RSD terms, the parity-odd component from $v_{o\|}$ is associated with the LOS. Therefore, the LOS-average removes them, identically to other terms. 

However, as this work assumes no perturbation in the LOS from $\vb*{v}_o$, the LOS aberration term \cite{Bertacca2014JCAP...09..037B,Gervani:2025wmh}, which involves the perpendicular component of $\vb*{v}_o$, remains outside the analysis of this work. So, the effect of $\vb*{v}_o$ on the PV of 4PCF requires further investigation.

\subsection{Unequal-redshift effect}

A key approximation in this work is that all galaxies in the 4PCF are assigned the same redshift. In reality, the galaxies lie on the lightcone and their redshifts differ according to their positions along the LOS. The studies \cite{Philcox:2022hkh, Hou:2022wfj,Slepian:2025kbb} restrict the arm-vector lengths to $20\,h^{-1}<r<164\,h^{-1}$ Mpc, which can introduce a redshift difference of up to $\Delta z\lesssim 0.05$. This difference affects the redshift-dependent factors $\mathcal{Z}_{n,r}(z)$, which we have assumed to always take the common value of $\bar{z}$. For example, the time evolution factor associated with the kernel of linear $\delta_g$ is $\mathcal{Z}_{1,1}(z)\propto D_1(z)\propto(1+z)^{-1}$ and receives a fractional change of $\left(1+\frac{\Delta z}{1+z}\right)^{-1}\gtrsim0.95$ for $0<z<1.1$, ignoring the time evolution of the bias parameter. Therefore, we can expect $\sim5\%$ error in the leading contribution of 4PCF. Under a parity transformation, reflecting the arm vector $\vb*{r_\sigma}$ changes the galaxy's comoving position relative to the observer, and hence its effective redshift. The parity transformation of $\hat{\zeta}(\vb*{r_1},\vb*{r_2},\vb*{r_3})$ therefore shifts the argument of $\mathcal{Z}_{n,r}(z)$ and can induce a parity-odd mode of a few percent relative to the equal-redshift estimation. 

This effect is present even in the absence of RSD, since $\mathcal{Z}_{n,r}(z)$ already encodes the time evolution of the dark matter density. It is still an observational effect and does not come from the primordial PV in the initial conditions. It comes from the fact that the observer's coordinates are embedded in the lightcone, where the comoving distance is implicitly a function of look-back time (see figure~\ref{fig: Lightcone}). The choice of background cosmology is relevant here, as it determines the growth factors $D_{X_n}(\tau)$ and $E_{X_n}(\tau)$ (see Appendix~\ref{sec: Second-, Third-, and Higher-order Solutions}).

\subsection{Eulerian perturbation theory and the bias expansion}

The prediction in this paper relies on the EPT (see Appendix~\ref{sec: (Standard) Eulerian Perturbation Theory}), which is valid above the nonlinear scale, $k<0.1\,h$~Mpc$^{-1}$ \cite{Bernardeau2002PhR...367....1B}. On such large scales, it is reasonable to assume that the vorticity of the dark matter fluid is negligible \cite{Pueblas2009PhRvD..80d3504P} and that galaxies move with the same velocity as the dark matter fluid. The generation of vorticity  \cite{Matarrese:1993zf,Bruni:1996im,Matarrese:1997ay, Lu2009JCAP...02..023L,Pichon1999A&A...343..663P} and velocity bias between galaxies and dark matter \cite{Baumann2012JCAP...07..051B, Mirbabayi:2014zca, Zheng2015PhRvD..91l3512Z} have been explored in the literature, but we do not include these subdominant dynamical effects here.

The question of PV in the perturbative bias expansion deserves attention. Since the bias expansion is a true scalar relation, all bias operators must be scalars. Constructing a parity-odd operator requires either the Levi-Civita tensor $\epsilon_{ijk}$ (to form a pseudo-scalar) or a dot product with a pseudo-vector. The vorticity field is one example, given that it is a pseudo-vector. So, including it in the bias expansion could introduce parity-violating operators.

It is not possible to build a parity-violating operator from the gravitational potential alone, as in the bias expansion of \cite{McDonald2009JCAP...08..020M} adopted here, because combinations such as $\epsilon_{ijk}\partial_i\delta\partial_j\delta\partial_k\delta$ vanish. In particular,
\begin{equation}
    \epsilon_{ijk}\partial_i\delta\partial_j\delta\partial_k\delta=\vb*{\nabla}\delta\cdot(\vb*{\nabla}\delta\times\vb*{\nabla}\delta)=0
\end{equation}
as $\vb*{\nabla}\delta\times\vb*{\nabla}\delta=0$. Similarly, the term $\partial_i\delta\partial_j\delta\partial_k\delta$ is symmetric under exchange of the indices, but $\epsilon_{ijk}$ is antisymmetric. Therefore, the product vanishes upon contraction. Non-standard operators, such as dot products with an angular momentum (a pseudo-vector), can be parity-violating and do not vanish under symmetrisation. Introducing such operators would, however, require strong physical motivation.

\subsection{Primordial parity violation}\label{sec: Primordial parity violation}

One of the most physically compelling routes to PV in the 4PCF is a primordial parity-odd trispectrum seeded during inflation. Under this scenario, PV would already be present in the initial conditions of the universe, implying that some mechanism broke parity symmetry in the high-energy regime. Such a feature cannot arise in standard single-field slow-rolling inflation with GR, which is a parity-conserving theory of gravity (unless one considers parity-breaking effects arising from non-trivial topologies, see \cite{COMPACT:2024cud}).  Several well-motivated examples introduce a parity-violating term into the Lagrangian such as Chern-Simons gravity \cite{Lue:1998mq,Jackiw:2003pm} (see \cite{Salvarese} for the first attempt) more general scalar-tensor chiral gravity models \cite{Moretti:2024fzb}, or axion inflation with a gauge-field coupling \cite{Maresuke2016PhRvD..94h3503S,Liu:2019fag,Niu:2022fki,Fujita2024JCAP...05..127F,Reinhard:2024evr} (see also, e.g., \cite{Coulton:2023oug} and \cite{Cabass:2022rhr}, for the discussion of other possibilities). The results of this paper would contribute to support the interpretation of the detections in \cite{Hou:2022wfj, Philcox:2022hkh, Slepian:2025kbb} as a primordial signal: we have demonstrated that RSD, including leading-order GR corrections, cannot contribute a parity-odd mode to the 4PCF once the estimator is averaged over all LOS. However, those parity-odd contributions can still be introduced at a subdominant level if the necessary conditions mentioned in this work are not met. For example, the non-zero odd multipole part of the survey window function can leak those effects into the estimator (see Section~\ref{sec: sky coverage}).

\section{Conclusions}
In this paper, we demonstrate theoretically that the parity-odd contribution to the observed 4PCF from the RSD effect, including the leading-order general relativistic correction at order $\mathcal{H}/k$, is suppressed upon averaging over the LOS. Consequently, the detections reported in \cite{Hou:2022wfj, Philcox:2022hkh, Slepian:2025kbb} cannot be attributed to this effect and must be explained by other phenomena.

We argue this parity conservation by writing the observed number density contrast $\Delta_g$ in integral form~\eqref{eqn: Fourier observed galaxy number density constrast} with polynomial structure kernels~\eqref{eqn: polynomial form of redshift kernel}. This is possible as all contributions considered in $\Delta_g^{(1)}$, $\Delta_g^{(2)}$, $\Delta_g^{(3)}$ are local terms. Therefore, the leading contribution to the theoretical 4PCF can be expressed in integral forms~\eqref{eqn: first observed 4PCF leading contribution} and \eqref{eqn: second observed 4PCF leading contribution}. The key insight driving our conclusion is that the isotropic 4PCF estimator used in \cite{Hou:2022wfj, Philcox:2022hkh, Slepian:2025kbb} measures galaxy statistics by averaging over the survey volume, including all LOS directions. We compute the kernel solutions of the observed galaxy number overdensity $\Delta_g$ by combining the perturbative bias expansion of the galaxy number overdensity up to third order in density field perturbation, including tidal effects \cite{McDonald2009JCAP...08..020M}; the dark matter density contrast and velocity divergence from EPT on an arbitrary cosmological background, accounting only for gravitational evolution on large scales \cite{Bernardeau2002PhR...367....1B, Takahashi2008PThPh.120..549T, Fasiello2016PhRvD..94f3516F}; and the RSD effect with the GR correction to linear order in $\mathcal{H}/k$ \cite{DiDio2020JCAP...09..058D, DiDio2019JCAP...04..050D}. Subdominant contributions, including lensing, the ISW effect, and the Shapiro time delay, are neglected. We find that the parity-odd contribution arises purely from the LOS projection factors $\mu_{\vb*{k_\sigma}} = \vb*{\hat{k}_\sigma}\cdot\vb*{\hat{n}}$ (see Section~\ref{sec: Parity Properties of 4PCF}) in the kernel~\eqref{eqn: polynomial form of redshift kernel}.

We then assess the parity properties of the isotropic 4PCF estimator $\hat{\zeta}$ computed from $\Delta_g$. Although RSD introduces dot products of Fourier mode vectors with the LOS to odd powers, these contributions vanish upon integration over all possible LOS directions on the sphere. We demonstrate this by expanding each contribution to the 4PCF in spherical harmonics of the elevation angle; angular momentum selection rules then enforce that only parity-even contributions survive the integration.

In the plane-parallel approximation, where all galaxies in the 4PCF share a common LOS, our analysis reproduces and generalises the result of \cite{Leonard:2024xvy}. In particular, the isotropic estimator reduces to a linear combination of isotropic basis functions $\mathcal{P}_{j_1,j_2,j_3}$ \eqref{eqn: isotropic basis} with coefficients $\mathcal{C}^{\alpha_2,\alpha_1,\alpha_3}_{j_1,j_2,j_3}$ \eqref{eqn: plane parallel coefficient of isotropic basis} that vanish for $j_1+j_2+j_3 = \text{odd}$ due to the first Wigner 3-j symbol. Therefore, all parity-odd contributions ($j_1+j_2+j_3 = \text{odd}$) are removed from the estimator. This result generalises the conclusion of the previous works \cite{Leonard:2024xvy, Paul:2024xff} to a case with the tidal operator in the bias expansion and on arbitrary cosmological background evolution rather than just an EdS background. We then extend this result to the wide-angle case by assigning each galaxy its own local LOS, defined by the arm vector $r_\sigma$ giving the position of each secondary galaxy relative to the reference. Our analysis confirms that the suppression of parity-odd modes persists in this more general setting. In particular, we project all cosine factors in the wide-angle 4PCF onto spherical harmonic functions. Then, we show that the $S$ parameter~\eqref{eqn: S parameter} controls the parity property of each contribution in the wide-angle case. Finally, the simultaneous constraints \eqref{eqn: s constraint} and \eqref{eqn: k constraints} force $S$ to always be even. Therefore, we show that the parity-odd contribution ($S=\text{odd}$) vanishes even upon introduction of the wide-angle effect.

A caveat of our results is that the parity-odd contribution may persist through averaging over LOS if the conditions mentioned in this paper are relaxed (see Section~\ref{sec: Discussion and Potential Sources of Violation}). In particular, the completeness of the sky coverage determines how much of the parity-odd contribution is suppressed. The derivation of the theoretical 4PCF assumes equal redshift for every galaxy in the tetrahedron (see the end of Section~\ref{sec: Probing Parity Violation With 4PCF}). This work also ignores the GR correction from the lensing effect, ISW effect, Shapiro delay effect, and other sub-leading corrections beyond $\mathcal{O}(\epsilon^{1/2}_\mathcal{H})$. Finally, the suppression of vorticity is assumed for the dark matter dynamics. This is crucial as vorticity is a pseudo-vector, which can be used to construct a parity-violating operator in the galaxy bias expansion. These are all conditions that deserve a careful investigation in the future. Apart from those, potential physical explanations should be sought elsewhere, for example, the primordial Parity Violation generated from non-standard inflation models. Examples of such models are inflation with Chern-Simons gravity, more general scalar-tensor chiral gravity models, or axion inflation with gauge-field couplings (see Section~\ref{sec: Primordial parity violation} for references).

That being said, these results together support the conclusion that RSD effects cannot substantially account for the parity-violating galaxy 4PCF signal reported in \cite{Hou:2022wfj, Philcox:2022hkh, Slepian:2025kbb}. This strengthens the case that a robust detection in future data would reflect a genuine physical signal, whether primordial or arising from some non-standard mechanism. Otherwise, it must be an artefact from other systematic effects than the RSD effect considered in this work. Future surveys such as Euclid, with near full-sky coverage, will provide a cleaner test of Parity Violation in the large-scale structure.

\acknowledgments

N.B. acknowledges support from the MUR PRIN2022 Project “BROWSEPOL: Beyond standaRd mOdel With coSmic microwavE background POLarization”-2022EJNZ53 financed by the European Union -- Next Generation EU. N.B., D.B., and S.M. acknowledge support from the INFN InDark Initiative. N.B., D.B., and S.M. acknowledge support from the COSMOS network (www.cosmosnet.it) through the ASI (Italian Space Agency) Grants 2016-24-H.0, 2016-24-H.1-2018, and 2020-9-HH.0. The authors used ChatGPT (OpenAI) and Claude (Anthropic) for language editing and manuscript formatting. All AI-assisted outputs were reviewed and verified by the authors, who take full responsibility for the final manuscript.

\appendix
\section{Perturbative solution of the large-scale structure}\label{sec: Perturbative Solution of the Large-scale Structure}
We review how the galaxy distribution can be described in terms of the initial matter distribution of the universe. In particular, galaxies are understood to be biased tracers of the underlying matter distribution. This idea dates back to the analysis of the Abell cluster catalogue by N. Kaiser in 1984 \cite{Kaiser:1984sw_Abell}, who found that the matter and galaxy two-point correlation functions share the same scale dependence but differ in amplitude by a factor known as the \emph{bias parameter}. It has since been shown that the tracer population is also sensitive to other physical quantities, such as the tidal field (see the review \cite{Desjacques:2016bnm} and references therein). In the $\Lambda$CDM model, the matter density budget is dominated by Cold Dark Matter (CDM), which evolves nonlinearly from the initial conditions set at the beginning of the universe. In this appendix, we describe the evolution of the dark matter density using perturbation theory, following the work of \cite{Bernardeau2002PhR...367....1B}. In Appendix~\ref{sec: The Bias Expansion}, we describe the tracer population using the perturbative bias expansion, following \cite{McDonald2009JCAP...08..020M}.

The following discussion of the LSS rests on an important assumption: we work in the large-scale limit, where gravity dominates the dynamics, and approximate it as the sole source of motion. One subtlety is that we treat gravitational dynamics in the Newtonian limit. This may appear inconsistent with the framework of Section~\ref{sec: The Relativistic Galaxy Number Density: Redshift Space Distortion}, where we include GR corrections to the observational effects. However, the GR correction to the perturbative solutions of the matter density contrast enters at order $\epsilon_\mathcal{H}\equiv(\mathcal{H}/k)^2$, which is neglected in this work. Therefore, solving the dynamics in the Newtonian limit is accurate within this approximation.

\subsection{Dynamics of cold dark matter}\label{sec: Dynamics of Cold Dark Matter}

Let $\delta$ and $\vb*{v}$ denote the matter density contrast and the velocity of the dark matter fluid, respectively. To describe the kinematics of dark matter, we introduce additional assumptions. We consider a collisionless, self-gravitating dark matter fluid. These assumptions hold well on large scales as discussed above. We also neglect the vorticity component of the fluid velocity, and instead describe the velocity field by its divergence, $\theta=\vb*{\nabla} \cdot \vb*{v}$. This is justified if the primordial vorticity is negligible and the fluid is perfect and pressureless. Under these assumptions, dark matter obeys the following fluid equations \cite{Peebles1980lssu.book.....P, Bernardeau1994ApJ...433....1B, Bernardeau2002PhR...367....1B},
\begin{equation}\label{eqn: dark matter fluid equation}
    \begin{split}
        \frac{D}{D\tau}\delta &= \theta(1+\delta) \\
    \frac{D \theta}{D \tau} + \mathcal{H}\theta &= -\partial_iv^j\partial_jv^i + \nabla^2 \Psi, \\
    \end{split}
\end{equation}
which are the continuity and the Euler equations, where $\partial_i\equiv\partial/\partial x^i$ is the spatial derivative and $\frac{D}{D\tau}=\frac{\partial}{\partial \tau}+v^i\partial_i$ is the convective time derivative. The gravitational potential $\Psi$ drives the dynamics, where its constraint equation is the Poisson equation,
\begin{equation}\label{eqn: Poisson equation}
    \nabla^2 \Psi = -\frac{3}{2}\Omega_m\mathcal{H}^2\delta,
\end{equation}
where $\Omega_m$ is the matter density parameter.

\subsection{Standard Eulerian perturbation theory}\label{sec: (Standard) Eulerian Perturbation Theory}

The evolution of dark matter on large scales in the Newtonian limit can be solved perturbatively from the fluid equations \eqref{eqn: dark matter fluid equation} and the Poisson equation~\eqref{eqn: Poisson equation}. We adopt the Eulerian description of the fluid, in which the coordinate system is fixed and time-independent.

In Fourier space, we can write the density and velocity divergence as
\begin{equation}\label{eqn: Inverse Fourier transformation of density, velocity and velocity divergence}
    \begin{split}
        \delta(\vb*{r},\tau) &= \int \frac{d^3k}{(2\pi)^3} \delta_{\vb*{k}}(\tau) e^{-i\vb*{k}\cdot\vb*{r}}, \\
        \theta(\vb*{r},\tau) &= \int \frac{d^3k}{(2\pi)^3} \theta_{\vb*{k}}(\tau) e^{-i\vb*{k}\cdot\vb*{r}}, \\
        \theta(\vb*{r},\tau) &= \vb*{\nabla}\cdot\vb*{v}(\vb*{r},\tau) \implies \vb*{v}_{\vb*{k}}(\tau) = i\frac{\vb*{k}}{k^2}\theta_{\vb*{k}}(\tau). \\
    \end{split}
\end{equation}
Under this transformation, equations \eqref{eqn: dark matter fluid equation} and \eqref{eqn: Poisson equation} become
\begin{equation}
    \begin{split}\label{eqn: Fourier space CDM fluid equation}
        \frac{\partial \delta_{\vb*{k}}(\tau)}{\partial \tau} &+ \theta_{\vb*{k}}(\tau) = \\
        &-\int_{\vb*{k_1}} \int_{\vb*{k_2}} \delta^{(3)}_D(\vb*{k}-\vb*{k_1}-\vb*{k_2})\alpha(\vb*{k_1},\vb*{k_2})\theta_{\vb*{k_1}}(\tau)\delta_{\vb*{k_2}}(\tau), \\
        \frac{\partial \theta_{\vb*{k}}(\tau)}{\partial \tau} &+ \mathcal{H}(\tau)\theta_{\vb*{k}}(\tau) + \frac{3}{2}\Omega_m(\tau)\mathcal{H}^2(\tau)\delta_{\vb*{k}}(\tau) = \\
        &-\int_{\vb*{k_1}} \int_{\vb*{k_2}} \delta^{(3)}_D(\vb*{k}-\vb*{k_1}-\vb*{k_2})\beta(\vb*{k_1},\vb*{k_2})\theta_{\vb*{k_1}}(\tau)\theta_{\vb*{k_2}}(\tau),
    \end{split}
\end{equation}
The mode-coupling kernels are defined as
\begin{equation}\label{eqn: EPT source kernels}
    \alpha(\vb*{k_1},\vb*{k_2}) = \frac{\vb*{k_1}\cdot (\vb*{k_1} + \vb*{k_2})}{k_1^2} \quad \text{and} \quad \beta(\vb*{k_1},\vb*{k_2}) = \frac{\norm{\vb*{k_1} + \vb*{k_2}}^2(\vb*{k_1}\cdot \vb*{k_2})}{2k_1^2k_2^2},
\end{equation}
where we implicitly apply $\vb*{\nabla}$ to the second equation of \eqref{eqn: dark matter fluid equation} and substitute $\nabla^2
\Psi$ with the Poisson equation~\eqref{eqn: Poisson equation}. 

In the framework of perturbation theory, the general solutions are written as
\begin{equation}\label{eqn: perturbative expansion of CDM density and velocity divergence}
    \begin{split}
        \delta_{\vb*{k}}(\tau) &= \sum^\infty_{n=1} \delta^{(n)}_{\vb*{k}}(\tau), \\
        \theta_{\vb*{k}}(\tau) &= \sum^\infty_{n=1} \theta^{(n)}_{\vb*{k}}(\tau),
    \end{split}
\end{equation}
where $n$ is the perturbation order. These expansions are analogous to \eqref{eqn: perturbative expansion of galaxy number density, velocity, and potential}, but formulated in Fourier space. Crucially, the $n$th-order solution in Fourier space corresponds to the Fourier transform of the same-order solution in real space. The system \eqref{eqn: Fourier space CDM fluid equation} can then be solved perturbatively: lower-order solutions are substituted into the right-hand side of \eqref{eqn: Fourier space CDM fluid equation}, which hosts the nonlinear contributions, to obtain the next-order solution. We begin with the linear equations, in which the right-hand side is neglected
\begin{equation}\label{eqn: Fourier space CDM fluid equations}
    \begin{split}
        &\frac{\partial \delta^{(n)}_{\vb*{k}}(a)}{\partial \ln{a}} + \frac{\theta^{(n)}_{\vb*{k}}(a)}{\mathcal{H}} = \\
        &-\sum^{n-1}_{m=1}\int_{\vb*{k_1}} \int_{\vb*{k_2}} \delta^{(3)}_D(\vb*{k}-\vb*{k_1}-\vb*{k_2})\alpha(\vb*{k_1},\vb*{k_2})\frac{\theta^{(m)}_{\vb*{k_1}}(a)}{\mathcal{H}}\delta^{(n-m)}_{\vb*{k_2}}(a), \\
        &\frac{\partial}{\partial \ln{a}} \left( \frac{\theta^{(n)}_{\vb*{k}}(a)}{\mathcal{H}} \right) + \left(\frac{\mathcal{H}'}{\mathcal{H}^2}+1\right)\left( \frac{\theta^{(n)}_{\vb*{k}}(a)}{\mathcal{H}} \right) + \frac{3}{2}\Omega_m\delta^{(n)}_{\vb*{k}}(a) = \\
        &-\sum^{n-1}_{m=1}\int_{\vb*{k_1}} \int_{\vb*{k_2}} \delta^{(3)}_D(\vb*{k}-\vb*{k_1}-\vb*{k_2})\beta(\vb*{k_1},\vb*{k_2})\frac{\theta^{(m)}_{\vb*{k_1}}(a)}{\mathcal{H}}\frac{\theta^{(n-m)}_{\vb*{k_2}}(a)}{\mathcal{H}},
    \end{split}
\end{equation}
where we manipulate the equations such that $\theta^{(n)}_{\vb*{k}}(a)$ is normalised by $\mathcal{H}$ using the chain rule, $\frac{1}{\mathcal{H}}\frac{\partial}{\partial\tau}=\frac{\partial}{\partial \ln{a}}$. We are interested in the solutions of this set of equations on an arbitrary cosmological background. An approximate solution was presented in \cite{Bernardeau2002PhR...367....1B}. Then, the solution up to the third order was computed by \cite{Takahashi2008PThPh.120..549T}. Later, \cite{Fasiello2016PhRvD..94f3516F} introduced a recursive formula for computing the time-dependent integral kernel of the solution at arbitrary order.

\subsubsection{Linear solution}\label{sec: Linear Solution}

The integral terms are nonlinear, making these equations nontrivial to solve. However, at the linear level, those terms are neglected,
\begin{equation}\label{eqn: First order Fourier space CDM fluid equations}
    \begin{split}
        \frac{\partial \delta^{(1)}_{\vb*{k}}(a)}{\partial \ln{a}} &+ \frac{\theta^{(1)}_{\vb*{k}}(a)}{\mathcal{H}} = 0, \\
        \frac{\partial}{\partial \ln{a}} \left( \frac{\theta^{(1)}_{\vb*{k}}(a)}{\mathcal{H}} \right) &+ \left(\frac{\mathcal{H}'}{\mathcal{H}^2}+1\right)\left( \frac{\theta^{(1)}_{\vb*{k}}(a)}{\mathcal{H}} \right) + \frac{3}{2}\Omega_m\delta^{(1)}_{\vb*{k}}(a) = 0.
    \end{split}
\end{equation}
This system of equations can be combined into
\begin{equation}\label{eqn: First order Fourier space CDM EOM}
    \frac{\partial^2 \delta^{(1)}_{\vb*{k}}(a)}{\partial \ln{a}^2} + \left(\frac{\mathcal{H}'}{\mathcal{H}^2}+1\right)\frac{\partial \delta^{(1)}_{\vb*{k}}(a)}{\partial \ln{a}} - \frac{3}{2}\Omega_m\delta^{(1)}_{\vb*{k}}(a) = 0.
\end{equation}
This Partial Differential Equation (PDE) is \emph{homogeneous} in $a$, and thus in time $\tau$. Therefore, it can be solved by the separation of variables method. Namely, we can write the ansatz of the linear solution of density as
\begin{equation}\label{eqn: first order density}
    \delta^{(1)}_{\vb*{k}}(a) = D_1(a)\delta^{(1)}_{\vb*{k}},
\end{equation}\label{eqn: linear ansatz of dm}
where we introduce the linear growth factor,
\begin{equation}
    D_1(a) \propto a^f \quad \text{where} \quad f(a) \equiv \frac{d\ln{D_1}}{d\ln{a}} = \frac{1}{\mathcal{H}}\frac{d\ln{D_1}}{d\tau} \quad \text{is the growth rate \cite{Kaiser:1987qv_RSD,Linder:2005in,linder2008APh....29..336L}.}
\end{equation}
This function describes the time evolution of the linear solution. The choice of cosmological background determines the time evolution of these functions. For example, in an Einstein-de Sitter (matter-dominated, $\Omega_m=1$) universe, $D_1(a)\propto a$ and $f=1$. In $\Lambda$CDM, the functions are more complicated and depend on the dark energy equation of state and the matter density parameter. Since we do not assume any particular cosmological background, we leave $D_1$ and $f$ as arbitrary functions of time. The first-order velocity divergence can be solved by substituting the ansatz \eqref{eqn: first order density} into the continuity equation,
\begin{equation}\label{eqn: first order velocity divergence}
    \theta^{(1)}_{\vb*{k}}(a) = -\mathcal{H}(a)f(a)D_1(a)\theta^{(1)}_{\vb*{k}}=E_1(a)\delta^{(1)}_{\vb*{k}},
\end{equation}
where $\delta^{(1)}_{\vb*{k}}=\theta^{(1)}_{\vb*{k}}$. The time-dependent part is factored out from the mode-dependent part for the linear solution. Then, the second equation gives the time evolution equation for the growth rate,
\begin{equation}
    \frac{\mathcal{H}'}{\mathcal{H}^2}+1 = \frac{3}{2}\frac{\Omega_m}{f}-f-\frac{\partial \ln{f}}{\partial \ln{a}}.    
\end{equation}
In this work, we are not interested in the formal expression of the growth factor. Instead, we remind the reader that the first-order solution can be factored into time- and mode-dependent parts.
\subsubsection{Second-, third-, and higher-order solutions}\label{sec: Second-, Third-, and Higher-order Solutions}
We now turn to the second-order equations,
\begin{equation}\label{eqn: Second order Fourier space CDM fluid equations}
    \begin{split}
        \frac{\partial \delta^{(2)}_{\vb*{k}}(a)}{\partial \ln{a}} &+ \frac{\theta^{(2)}_{\vb*{k}}(a)}{\mathcal{H}} = \\
        -f(a)D_1^2(a)\int_{\vb*{k_1}} \int_{\vb*{k_2}} &\delta^{(3)}_D(\vb*{k}-\vb*{k_1}-\vb*{k_2})\alpha(\vb*{k_1},\vb*{k_2})\theta^{(1)}_{\vb*{k_1}}\delta^{(1)}_{\vb*{k_2}}, \\
        \frac{\partial}{\partial \ln{a}} \left( \frac{\theta^{(2)}_{\vb*{k}}(a)}{\mathcal{H}} \right) &+ \left(\frac{\mathcal{H}'}{\mathcal{H}^2}+1\right)\left( \frac{\theta^{(2)}_{\vb*{k}}(a)}{\mathcal{H}} \right) + \frac{3}{2}\Omega_m\delta^{(2)}_{\vb*{k}}(a) = \\
        -f^2(a)D_1^2(a)\int_{\vb*{k_1}} \int_{\vb*{k_2}} &\delta^{(3)}_D(\vb*{k}-\vb*{k_1}-\vb*{k_2})\beta(\vb*{k_1},\vb*{k_2})\theta^{(1)}_{\vb*{k_1}}\theta^{(1)}_{\vb*{k_2}}.
    \end{split}
\end{equation}
At this order, the equations become inhomogeneous and nonlinear, so the second-order solutions cannot be factored as simply as the first-order solution. However, they can be written as a linear combination of two factorisable solutions, since the source terms can be separated into time- and mode-dependent parts. For example, in the first equation, $f(a)D^2_1(a)$ is separated from the integral. Therefore, the separation of variables can still be applied, with the ansatz being a linear combination of two solutions,
\begin{equation}\label{eqn: Second-order density}
    \delta^{(2)}_{\vb*{k}} (a)=D_{A}(a)\delta^{A}_{\vb*{k}} +D_{B}(a)\delta^{B}_{\vb*{k}}.
\end{equation}
Substituting the ansatz into the equations \eqref{eqn: Second order Fourier space CDM fluid equations} gives us the solutions,
\begin{equation}\label{eqn: second-order mode-dependent bases}
    \begin{split}
        \delta^{A}_{\vb*{k}}&=\int_{\vb*{k_1}} \int_{\vb*{k_2}}\delta^{(3)}_D(\vb*{k}-\vb*{k_1}-\vb*{k_2})\alpha(\vb*{k_1},\vb*{k_2})\delta^{(1)}_{\vb*{k_1}}\delta^{(1)}_{\vb*{k_2}}, \\
        \delta^{B}_{\vb*{k}}&=\int_{\vb*{k_1}} \int_{\vb*{k_2}}\delta^{(3)}_D(\vb*{k}-\vb*{k_1}-\vb*{k_2})\beta(\vb*{k_1},\vb*{k_2})\delta^{(1)}_{\vb*{k_1}}\delta^{(1)}_{\vb*{k_2}}, \\
    \end{split}
\end{equation}
and the time-dependent components are solved from
\begin{equation}\label{eqn: second order time evolution equation}
    \begin{split}
        \left[\frac{\partial^2}{\partial \ln{a}^2}+\left( \frac{\mathcal{H}'}{\mathcal{H}^2}+1\right)\frac{\partial}{\partial \ln{a}}-\frac{3}{2}\Omega_m \right]D_{A}(a)&=-\left[\frac{\partial}{\partial \ln{a}}+\left( \frac{\mathcal{H}'}{\mathcal{H}^2}+1\right) \right]\left[f(a)D^2_1(a)\right], \\
        \left[\frac{\partial^2}{\partial \ln{a}^2}+\left( \frac{\mathcal{H}'}{\mathcal{H}^2}+1\right)\frac{\partial}{\partial \ln{a}}-\frac{3}{2}\Omega_m \right]D_{B}(a)&=-f^2(a)D^2_1(a).
    \end{split}
\end{equation}
The solution for the velocity divergence can be obtained by substituting the density solution into the first equation of \eqref{eqn: Second order Fourier space CDM fluid equations},
\begin{equation}\label{eqn: second order velocity divergence}
    \begin{split}
        \theta^{(2)}_{\vb*{k}}(a) &\equiv E_{A}(a)\theta^{A}_{\vb*{k}} +E_{B}(a)\theta^{B}_{\vb*{k}} \\
        \frac{\theta^{(2)}_{\vb*{k}}(a)}{\mathcal{H}(a)f(a)}&=\left( D_1^2(a)-\frac{\partial D_{A}}{\partial\ln{D_1}} \right)\delta^{A}_{\vb*{k}} - \frac{\partial D_{B}}{\partial\ln{D_1}} \delta^{B}_{\vb*{k}}.
    \end{split}
\end{equation}
The mode-dependent parts are identical to those of the density solution. Since all derivatives in the PDEs are with respect to the time variable, the mode-dependent parts are non-dynamical, and their expressions follow directly from the source terms in equation~\eqref{eqn: Second order Fourier space CDM fluid equations}. This property holds for higher-order solutions as well, as it is an intrinsic feature of our PDEs.

Now that we have the second-order solutions, let us proceed to the third-order equations,
\begin{equation}\label{eqn: Third order Fourier space CDM fluid equations}
    \begin{split}
        \frac{\partial \delta^{(3)}_{\vb*{k}}(a)}{\partial \ln{a}} &+ \frac{\theta^{(3)}_{\vb*{k}}(a)}{\mathcal{H}} = \\
        -\int_{\vb*{k_1}} \int_{\vb*{k_2}} &\delta^{(3)}_D(\vb*{k}-\vb*{k_1}-\vb*{k_2})\alpha(\vb*{k_1},\vb*{k_2})
        \left[\theta^{(2)}_{\vb*{k_1}}(a)\delta^{(1)}_{\vb*{k_2}}(a)+\theta^{(1)}_{\vb*{k_1}}(a)\delta^{(2)}_{\vb*{k_2}}(a)\right], \\
        \frac{\partial}{\partial \ln{a}} \left( \frac{\theta^{(3)}_{\vb*{k}}(a)}{\mathcal{H}} \right) &+ \left(\frac{\mathcal{H}'}{\mathcal{H}^2}+1\right)\left( \frac{\theta^{(3)}_{\vb*{k}}(a)}{\mathcal{H}} \right) + \frac{3}{2}\Omega_m\delta^{(3)}_{\vb*{k}}(a) = \\
        -2 \int_{\vb*{k_2}} &\delta^{(3)}_D(\vb*{k}-\vb*{k_1}-\vb*{k_2})\beta(\vb*{k_1},\vb*{k_2})\theta^{(1)}_{\vb*{k_1}}(a)\theta^{(2)}_{\vb*{k_2}}(a).
    \end{split}
\end{equation}
The product $\theta^{(1)}\theta^{(2)}$ is symmetrised by the kernel $\beta$. Since the formal expression of the mode-dependent basis is determined by the mode mixing in the source term, we can guess the basis of the solution from the source terms. Accordingly, the third-order solutions are
\begin{equation}
    \begin{split}
        \delta^{(3)}_{\vb*{k}}(a)=&D_{AA,1}(a)\delta^{AA,1}_{\vb*{k}}+D_{AA,2}(a)\delta^{AA,2}_{\vb*{k}}+D_{AB,1}(a)\delta^{AB,1}_{\vb*{k}}+D_{AB,2}(a)\delta^{AB,2}_{\vb*{k}} \\
        &+D_{BA}(a)\delta^{BA}_{\vb*{k}}+D_{BB}(a)\delta^{BB}_{\vb*{k}}, \\
        \theta^{(3)}_{\vb*{k}}(a)=&E_{AA,1}(a)\delta^{AA,1}_{\vb*{k}}+E_{AA,2}(a)\delta^{AA,2}_{\vb*{k}}+E_{AB,1}(a)\delta^{AB,1}_{\vb*{k}}+E_{AB,2}(a)\delta^{AB,2}_{\vb*{k}} \\
        &+E_{BA}(a)\delta^{BA}_{\vb*{k}}+E_{BB}(a)\delta^{BB}_{\vb*{k}}, \\
    \end{split}
\end{equation}
where
\begin{equation}\label{eqn: third-order mode-dependent bases}
    \begin{split}
        \delta^{AA}_{\vb*{k}}&=\int_{\vb*{k_1}} \int_{\vb*{k_2}}\delta^{(3)}_D(\vb*{k}-\vb*{k_1}-\vb*{k_2})\alpha(\vb*{k_1},\vb*{k_2})\delta^{A}_{\vb*{k_1}}\delta^{(1)}_{\vb*{k_2}}, \\
        \delta^{A'A}_{\vb*{k}}&=\int_{\vb*{k_1}} \int_{\vb*{k_2}}\delta^{(3)}_D(\vb*{k}-\vb*{k_1}-\vb*{k_2})\alpha(\vb*{k_1},\vb*{k_2})\delta^{(1)}_{\vb*{k_1}}\delta^{A}_{\vb*{k_2}}, \\
        \delta^{AB}_{\vb*{k}}&=\int_{\vb*{k_1}} \int_{\vb*{k_2}}\delta^{(3)}_D(\vb*{k}-\vb*{k_1}-\vb*{k_2})\alpha(\vb*{k_1},\vb*{k_2})\delta^{B}_{\vb*{k_1}}\delta^{(1)}_{\vb*{k_2}}, \\
        \delta^{A'B}_{\vb*{k}}&=\int_{\vb*{k_1}} \int_{\vb*{k_2}}\delta^{(3)}_D(\vb*{k}-\vb*{k_1}-\vb*{k_2})\alpha(\vb*{k_1},\vb*{k_2})\delta^{(1)}_{\vb*{k_1}}\delta^{B}_{\vb*{k_2}}, \\
        \delta^{BA}_{\vb*{k}}&=\int_{\vb*{k_1}} \int_{\vb*{k_2}}\delta^{(3)}_D(\vb*{k}-\vb*{k_1}-\vb*{k_2})\beta(\vb*{k_1},\vb*{k_2})\delta^{A}_{\vb*{k_1}}\delta^{(1)}_{\vb*{k_2}}, \\
        \delta^{BB}_{\vb*{k}}&=\int_{\vb*{k_1}} \int_{\vb*{k_2}}\delta^{(3)}_D(\vb*{k}-\vb*{k_1}-\vb*{k_2})\beta(\vb*{k_1},\vb*{k_2})\delta^{B}_{\vb*{k_1}}\delta^{(1)}_{\vb*{k_2}}. \\
    \end{split}
\end{equation}
We assign the indices A and B to represent the solution with integral kernels, $\alpha$ and $\beta$. Since the kernel $\alpha(\vb*{k_1},\vb*{k_2})$ depends on the order of its arguments, the ordering of $\delta$'s is important. $A'$ indicates the case where $\vb*{k_2}$ belongs to the higher-order $\delta$. As before, the time-dependent components can be solved from the PDEs obtained by substituting the ansatzes into equations \eqref{eqn: Third order Fourier space CDM fluid equations}. The second- and third-order solutions can be found in \cite{Takahashi2008PThPh.120..549T}.

We are interested in the mode-dependent basis since the parity transformation directly changes $\vb*{k}\longrightarrow-\vb*{k}$. Up to the third order, any basis can be written in the integral form,
\begin{equation}\label{eqn: n-order mode-dependent basis}
    \delta^{X_n}_{\vb*{k}} = \int_{\vb*{k_1}} ... \int_{\vb*{k_n}}\delta^{(3)}_D\left(\vb*{k}-\sum^n_{\sigma=1}\vb*{k_\sigma}\right)\Delta^S_{X_n}(\vb*{k_1},\dots ,\vb*{k_n})\prod^n_{\sigma=1}\delta^{(1)}_{\vb*{k_\sigma}},
\end{equation}
where $n$ is the order of the solution that the basis belongs to, and $X_n$ is a string of $n-1$ letters chosen from A, A', and B, corresponding to each basis index. $\Delta^S_{X_n}$ is the kernel for the basis $X_n$. The superscript "S" indicates that the kernel has been symmetrised in $\vb*{k_\sigma}$'s by the product $\prod^n_{\sigma=1}\delta^{(1)}_{\vb*{k_\sigma}}$. The first-order kernel is $\Delta_1=1$ since the basis for the first-order solution is simply $\delta^{(1)}_{\vb*{k}}$. The second-order kernels are $\Delta_A(\vb*{k_1},\vb*{k_2})=\alpha(\vb*{k_1},\vb*{k_2})$ and $\Delta_B(\vb*{k_1},\vb*{k_2})=\beta(\vb*{k_1},\vb*{k_2})$, matching from the equations \eqref{eqn: second-order mode-dependent bases}. The formal expression of the third-order kernels can be found by unpacking the second-order bases, $\delta^A$ and $\delta^B$, in the equations \eqref{eqn: third-order mode-dependent bases}. For example, the second basis can be written as,
\begin{equation}
    \begin{split}
        \delta^{A'A}_{\vb*{k}} &=\int_{\vb*{k_1}} \int_{\vb*{k'}}\delta^{(3)}_D(\vb*{k}-\vb*{k_1}-\vb*{k'})\alpha(\vb*{k_1},\vb*{k'})\delta^{(1)}_{\vb*{k_1}}\delta^{A}_{\vb*{k'}} \\
        &= \int_{\vb*{k_1}} \int_{\vb*{k_2}} \int_{\vb*{k_3}}\delta^{(3)}_D(\vb*{k}-\vb*{k_1}-\vb*{k_2}-\vb*{k_3})\underbrace{\alpha(\vb*{k_1},\vb*{k_{23}})\alpha(\vb*{k_2},\vb*{k_3})}_{\equiv \Delta_{A'A}(\vb*{k_1},\vb*{k_2},\vb*{k_3})}\delta^{(1)}_{\vb*{k_1}}\delta^{(1)}_{\vb*{k_2}}\delta^{(1)}_{\vb*{k_3}}.
    \end{split}
\end{equation}
Hence, the kernel is $\Delta_{A'A}(\vb*{k_1},\vb*{k_2},\vb*{k_3})=\alpha(\vb*{k_1},\vb*{k_{23}})\alpha(\vb*{k_2},\vb*{k_3})$. Similarly, the rest are
\begin{equation}\label{eqn: third-order dark matter mode kernels}
    \begin{split}
        \Delta_{AA}(\vb*{k_1},\vb*{k_2},\vb*{k_3})&=\alpha(\vb*{k_{12}},\vb*{k_{3}})\alpha(\vb*{k_1},\vb*{k_2}) , 
        \quad \Delta_{A'A}(\vb*{k_1},\vb*{k_2},\vb*{k_3})=\alpha(\vb*{k_1},\vb*{k_{23}})\alpha(\vb*{k_2},\vb*{k_3}), \\
        \Delta_{AB}(\vb*{k_1},\vb*{k_2},\vb*{k_3})&=\alpha(\vb*{k_{12}},\vb*{k_{3}})\beta(\vb*{k_1},\vb*{k_2}), 
        \quad \Delta_{A'B}(\vb*{k_1},\vb*{k_2},\vb*{k_3})=\alpha(\vb*{k_1},\vb*{k_{23}})\beta(\vb*{k_2},\vb*{k_3}), \\
        \Delta_{BA}(\vb*{k_1},\vb*{k_2},\vb*{k_3})&=\beta(\vb*{k_1},\vb*{k_{23}})\alpha(\vb*{k_2},\vb*{k_3}), 
        \quad \Delta_{BB}(\vb*{k_1},\vb*{k_2},\vb*{k_3})=\beta(\vb*{k_1},\vb*{k_{23}})\beta(\vb*{k_2},\vb*{k_3}).
    \end{split}
\end{equation}
Lastly, the kernel must be symmetrised by the exchange of $\vb*{k_\sigma}$'s in the product of $\delta^{(1)}_{\vb*{k_\sigma}}$'s in the integral \eqref{eqn: n-order mode-dependent basis}. In particular, the commutative property of the product enforces the symmetry in the kernel. The symmetrised kernel is
\begin{equation}
    \Delta^S_{X_n}(\vb*{k_1},\dots ,\vb*{k_n}) = \frac{1}{n!}\sum_{\{\vb*{k}_n\}}\Delta_{X_n}(\vb*{k_1},\dots ,\vb*{k_n}),
\end{equation}
where the set $\{\vb*{k}_n\}$ contains all possible orderings of $\vb*{k}_1,...,\vb*{k}_n$ giving $n!$ combinations in total. We have now obtained all mode-dependent parts of the solutions up to third order. In fact, expression \eqref{eqn: n-order mode-dependent basis} holds to arbitrary order: all higher-order kernels can be obtained by iterating the procedure described here, i.e., solving the equations \eqref{eqn: Fourier space CDM fluid equations} order by order. The solutions always take the form,
\begin{equation}\label{eqn: n-order solutions}
    \begin{split}
        \delta^{(n)}_{\vb*{k}}(\tau)&=\sum_{\{X_n\}}D_{X_n}(\tau)\delta^{X_n}_{\vb*{k}}, \\
        \theta^{(n)}_{\vb*{k}}(\tau)&=\sum_{\{X_n\}}E_{X_n}(\tau)\delta^{X_n}_{\vb*{k}}, \\
    \end{split}
\end{equation}
where $D_{X_n}$ and $E_{X_n}$ are the growth factors of each basis. For example, the second-order growth factors are solved from \eqref{eqn: second order time evolution equation} mentioned earlier. The set $\{X_n\}$ contains all possible strings $X_n$. Since the string has $n-1$ length constructed from the choice of A, A', and B at each position except the last one (only A or B), the number of members in the set $\{X_n\}$ is $2\times3^{n-2}$ for $n\geq 2$. For example, the third-order solutions have 6 kernels \eqref{eqn: third-order dark matter mode kernels}.

\section{The bias expansion}\label{sec: The Bias Expansion}
On a large scale, we write the \emph{bias expansion} of the tracer as
\begin{equation}\label{eqn: general bias expansion}
     \delta_g(\vb*{r},\tau) = \sum_{\mathcal{O}}(b_{\mathcal{O}}(\tau)+\varepsilon_{\mathcal{O}}(\vb*{r},\tau))\mathcal{O}(\vb*{r},\tau)+\varepsilon(\vb*{r},\tau),
\end{equation}
where $\mathcal{O}$ is a physical operator representing a quantity on which the tracer number density depends. $b_{\mathcal{O}}$ is the bias parameter associated with the operator $\mathcal{O}$. As discussed at the beginning of Appendix~\ref{sec: Perturbative Solution of the Large-scale Structure}, gravity is assumed to be the only source of dynamics on large scales, so that variations in the tracer density across space are predominantly gravitational in origin. All small-scale physical effects are captured by the stochastic terms $\varepsilon_{\mathcal{O}}$ and $\varepsilon$. Although we write them down for completeness, they are uncorrelated with the deterministic large-scale part, e.g., $\expval{\varepsilon \mathcal{O}}=\expval{\varepsilon_\mathcal{O} \mathcal{O}}=0$ \cite{Dekel1999ApJ...520...24D, Taruya1999ApJ...522...46T,Desjacques:2016bnm}. Therefore, we will ignore them in the following discussion.

There are generally two approaches to writing the perturbative bias expansion, each corresponding to a different picture in fluid dynamics. The first is the Lagrangian bias expansion \cite{Matsubara2008PhRvD..78h3519M}, defined on the initial Lagrangian coordinates and evolved using Lagrangian perturbation theory \cite{Tatekawa2004astro.ph.12025T}. The second is the Eulerian bias expansion \cite{Kaiser:1984sw_Abell, Fry1993ApJ...413..447F}, defined on Eulerian coordinates. We adopt the latter, as it is more naturally related to the observational frame. By contrast, the Lagrangian approach requires tracking galaxies along their trajectories to arrive at the observed coordinates, which is more involved. For further details, we refer the reader to the review \cite{Desjacques:2016bnm}.

This bias modelling framework is systematic, as one can enumerate all possible operators to include in the expansion. For instance, the matter density contrast $\delta$ was identified as one such operator in Kaiser's study \cite{Kaiser:1984sw_Abell}, leading to linear bias theory, $\delta_g=b_1\delta$. The local bias expansion, $\delta_g=\sum_{n=0}^\infty (b_n/n!)\delta^n$, was subsequently introduced by \cite{Fry1993ApJ...413..447F}, extending the list of operators to arbitrary powers of $\delta$. As an improvement to the local model, \cite{McDonald2009JCAP...08..020M} incorporated the tidal field and other operators into the Eulerian bias expansion up to third order. The expansion was constructed from the Eulerian perturbation theory on the EdS universe, serving as a good approximation to the real background expansion. \cite{Fujita2020JCAP...10..059F} devised an algorithm to build the bias expansion in a generalised $\Lambda$CDM universe. Up to the third order, which is what we are interested in, the results of \cite{Fujita2020JCAP...10..059F} under the EdS condition are consistent not only with \cite{McDonald2009JCAP...08..020M}, but also with other models like \cite{Mirbabayi:2014zca} and the Effective Field Theory approach \cite{Senatore:2014eva, Angulo:2015eqa, Fujita2020JCAP...01..009F}.

With a caveat, we adopt the bias operators from \cite{McDonald2009JCAP...08..020M},
\begin{equation}
    \begin{split}
        \delta_g(\vb*{r},\tau)&=b_\delta \delta \\
        &+b_{\delta^2} \delta^2+b_{s^2} s^{ij}s_{ij} \\
        &+b_{\delta^3} \delta^3+b_{\delta s^ 2} \delta s^{ij}s_{ij}+b_{s t} s t+b_{s^3} s^i_j s^j_k s^k_i +b_\psi \psi,
    \end{split}
\end{equation}
where
\begin{equation}
    \begin{split}
         &s_{i j} \equiv\left(\frac{\partial_i \partial_j}{\partial^2}-\frac{1}{3} \delta_{i j}\right) \delta, \quad \eta \equiv -\frac{\theta}{\mathcal{H}f}-\delta \\
         &t_{i j} \equiv\left(\frac{\partial_i \partial_j}{\partial^2}-\frac{1}{3} \delta_{i j}\right) \eta, \quad \psi \equiv \eta-\eta^{(2)},
    \end{split}
\end{equation}
where the time and position arguments are omitted for tidiness. $i$, $j$, and $k$ are the spatial indices. $s_{ij}$ and $t_{ij}$ are the traceless tidal field and velocity gradient, respectively.\footnote{$s_{ij}$ and $t_{ij}$ are defined to be traceless since their trace would degenerate with $\delta$ and $\eta$, respectively.} $\eta$ is zero at the first order of perturbation according to the first order solutions \eqref{eqn: first order density} and \eqref{eqn: first order velocity divergence}.

There is a subtlety in the definition of $\psi$. In \cite{McDonald2009JCAP...08..020M}, this operator is constructed starting from third order, because the second-order velocity gradient degenerates with the density and tidal field: $\eta^{(2)}_{\text{EdS}} = \frac{2}{7} s^{(1)\,ij}s^{(1)}_{ij}-\frac{4}{21} {(\delta^{(1)})}^2$ in the EdS universe. Therefore, $\psi$ is constructed in place by subtracting out this redundant second-order contribution. This is no longer the case for an arbitrary cosmological background, where $\eta^{(2)}$ cannot be expressed in terms of the density and tidal field. Hence, the operator $\eta$ must be introduced into the expansion in place of $\psi$. Since the density part of $\eta$ is already redundant with the existing operator $\delta$, we use the velocity divergence $\theta$ directly. The bias expansion at each order then takes the form,
\begin{equation}
    \begin{split}
        \delta_g^{(1)}&=b_\delta \delta^{(1)}, \\
        \delta^{(2)}_g&=b_\delta \delta^{(2)}+b_\theta \theta^{(2)}+b_{\delta^2} (\delta^{(1)})^2+b_{s^2} s^{(1)\,ij}s^{(1)}_{ij}, \\
        \delta^{(3)}_g&=b_\delta \delta^{(3)}+b_\theta \theta^{(3)} +b_{\delta^2} \left[\delta^2\right]^{(3)}+b_{s^2} \left[s^{ij}s_{ij}\right]^{(3)}, \\
        &+b_{\delta^3} (\delta^{(1)})^3+b_{\delta s^ 2} \delta^{(1)} s^{(1)\,ij}s^{(1)}_{ij}+b_{s t} s^{(1)\,ij} t^{(2)}_{ij}+b_{s^3} (s^{(1)})^i_j (s^{(1)})^j_k (s^{(1)})^k_i,
    \end{split}
\end{equation}
where the square brackets represent all possible combinations that have the indicated order of perturbation, as introduced in equation~\eqref{eqn: third-order observed galaxy number density contrast}. This choice of bias expansion does not restrict the generality of our analysis. As noted in \cite{Fujita2020JCAP...10..059F}, the difference between various expansion schemes lies only in the definition of the bias parameters, while the mode-dependent factors that enter the operators are the same. Finally, by the equivalence principle on large scales, the galaxy velocity is assumed to equal the dark matter velocity
\begin{equation}\label{eqn: no velocity bias}
    \theta_g \equiv \vb*{\nabla}\cdot \vb*{v}=\theta.
\end{equation}
Namely, the galaxy is comoving with the dark matter fluid under the dominant influence of gravity on a large scale.\footnote{Beyond this approximation is the velocity bias, which is associated with the higher derivative operator in the bias expansion \cite{Mirbabayi:2014zca}. However, the effect is suppressed on a large scale. \cite{Zheng2015PhRvD..91l3512Z} shows that the comoving condition deviates by less than 2\% on the scale $k\leq0.1h\,\text{Mpc}^{-1}$, which is smaller than the scale of interest ($k\leq0.05h\,\text{Mpc}^{-1}$).} In Fourier space, the tidal field and velocity gradient take the forms,
\begin{equation}
    s_{\vb*{k},ij}(\tau)=\mathcal{K}_{\vb*{k},ij}\delta_{\vb*{k}}(\tau) \; , \quad t_{\vb*{k},ij}(\tau)=\mathcal{K}_{\vb*{k},ij}\left(-\frac{\theta_{\vb*{k}}(\tau)}{\mathcal{H}(\tau)f(\tau)}-\delta_{\vb*{k}}(\tau)\right)\; ,
\end{equation}
where
\begin{equation}
    \mathcal{K}_{\vb*{k},ij}= \frac{k_i k_j}{k^2}-\frac{1}{3}\delta_{ij}.
\end{equation}
Hence, the Fourier transform of the bias expansion at each order can be written as
\begin{equation}
    \begin{split}
        \delta_{g,\vb*{k}}^{(1)}(\tau)&=b_\delta \delta^{(1)}_{\vb*{k}}(\tau), \\
        \delta^{(2)}_{g,\vb*{k}}(\tau)&=b_\delta \delta^{(2)}_{\vb*{k}}(\tau)+b_\theta \theta^{(2)}_{\vb*{k}}(\tau) \\
        &+\int_{\vb*{k_1}} \int_{\vb*{k_2}}\delta^{(3)}_D\left( \vb*{k}-\vb*{k_1}-\vb*{k_2}\right) \left[ b_{\delta^2}+b_{s^2} S(\vb*{k_1},\vb*{k_2})\right]\delta^{(1)}_{\vb*{k_1}}(\tau)\delta^{(1)}_{\vb*{k_2}}(\tau), \\
        \delta^{(3)}_{g,\vb*{k}}&=b_\delta \delta^{(3)}_{\vb*{k}}(\tau)+b_\theta \theta^{(3)}_{\vb*{k}}(\tau) \\
        &+\int_{\vb*{k_1}} \int_{\vb*{k_2}}\delta^{(3)}_D\left( \vb*{k}-\vb*{k_1}-\vb*{k_2}\right) \left[ b_{\delta^2}+(b_{s^2}-b_{st}) S(\vb*{k_1},\vb*{k_2})\right]\delta^{(1)}_{\vb*{k_1}}(\tau)\delta^{(2)}_{\vb*{k_2}}(\tau) \\
        &-\int_{\vb*{k_1}} \int_{\vb*{k_2}}\delta^{(3)}_D\left( \vb*{k}-\vb*{k_1}-\vb*{k_2}\right) \frac{b_{st}}{\mathcal{H}(\tau)f(\tau)} S(\vb*{k_1},\vb*{k_2})\delta^{(1)}_{\vb*{k_1}}(\tau)\theta^{(2)}_{\vb*{k_2}}(\tau) \\
        &+\int_{\vb*{k_1}} \int_{\vb*{k_2}} \int_{\vb*{k_3}} \delta^{(3)}_D\left( \vb*{k}-\vb*{k_1}-\vb*{k_2}-\vb*{k_3}\right)\Bigg\{b_{\delta^3}+\frac{b_{\delta s^ 2}}{3}\left[S(\vb*{k_1},\vb*{k_2})+S(\vb*{k_2},\vb*{k_3})+S(\vb*{k_1},\vb*{k_3})\right]\\
        &+F(\vb*{k_1},\vb*{k_2},\vb*{k_3})\Bigg\}\delta^{(1)}_{\vb*{k_1}}(\tau)\delta^{(1)}_{\vb*{k_2}}(\tau)\delta^{(1)}_{\vb*{k_3}}(\tau), \\
    \end{split}
\end{equation}
where
\begin{equation}\label{eqn: tidal combinations in Fourier space}
    \begin{split}
        &S(\vb*{k_a},\vb*{k_b})=\mathcal{K}_{\vb*{k_a},ij}\mathcal{K}_{\vb*{k_b}}^{ij}=\frac{(\vb*{k_a}\cdot \vb*{k_b})^2}{k_a^2k_b^2}+\frac{1}{3}, \\
        &F(\vb*{k_a},\vb*{k_b},\vb*{k_c})=(\mathcal{K}_{\vb*{k_a}})^i_j(\mathcal{K}_{\vb*{k_b}})^j_k(\mathcal{K}_{\vb*{k_c}})^k_i \\
        &=\frac{(\vb*{k_a}\cdot \vb*{k_b})(\vb*{k_b}\cdot \vb*{k_c})(\vb*{k_a}\cdot \vb*{k_c})}{k_a^2k_b^2k_c^2}-\frac{1}{3}\left[\frac{(\vb*{k_a}\cdot \vb*{k_b})^2}{k_a^2k_b^2}+\frac{(\vb*{k_b}\cdot \vb*{k_c})^2}{k_b^2k_c^2}+\frac{(\vb*{k_a}\cdot \vb*{k_c})^2}{k_a^2k_c^2}\right]+\frac{2}{9}.
    \end{split}
\end{equation}

The perturbative solutions of dark matter density and velocity divergence in this expansion have already been derived in Appendices~\ref{sec: Linear Solution} and \ref{sec: Second-, Third-, and Higher-order Solutions}. Hence, we may substitute the general form of $n$th-order solution \eqref{eqn: n-order solutions} into the bias expansion at each order and manipulate them into the integral form,
\begin{equation}\label{eqn: explicit integral bias expansion}
    \begin{split}
        \delta_{g,\vb*{k}}^{(1)}(\tau)&=b_\delta D_1(\tau)\delta^{(1)}_{\vb*{k}}, \\
        \delta_{g,\vb*{k}}^{(2)}(\tau)&=\int_{\vb*{k_1}} \int_{\vb*{k_2}}\delta^{(3)}_D\left( \vb*{k}-\vb*{k_1}-\vb*{k_2}\right)\Bigg\{ \sum_{\{X_2\}}\Big[b_\delta D_{X_2}(\tau)+b_\theta E_{X_2}(\tau)\Big]\Delta^S_{X_2}(\vb*{k_1},\vb*{k_2}) \\
        &+D_1^2(\tau)\Big[ b_{\delta^2}+b_{s^2} S(\vb*{k_1},\vb*{k_2})\Big]\Bigg\}\delta^{(1)}_{\vb*{k_1}}\delta^{(1)}_{\vb*{k_2}}, \\
        \delta_{g,\vb*{k}}^{(3)}(\tau)&=\int_{\vb*{k_1}} \int_{\vb*{k_2}} \int_{\vb*{k_3}} \delta^{(3)}_D\left( \vb*{k}-\vb*{k_1}-\vb*{k_2}-\vb*{k_3}\right)\Bigg\{  \sum_{\{X_3\}}\Big[ b_\delta D_{X_3}(\tau)+b_\theta E_{X_3}(\tau) \Big]\Delta^S_{X_3}(\vb*{k_1},\vb*{k_2},\vb*{k_3}) \\
        &+\sum_{\{X_2\}}\frac{D_1(\tau)}{3}\Big\{ D_{X_2}(\tau)\Big[ b_{\delta^2}+\big(b_{s^2}-b_{st}\big)S(\vb*{k_1},\vb*{k_{23}}) \Big] \\
        &-E_{X_2}(\tau)\frac{b_{st}}{\mathcal{H}(\tau)f(\tau)}S(\vb*{k_1},\vb*{k_{23}})\Big\}\Delta^S_{X_2}(\vb*{k_2},\vb*{k_3})+2\,\text{perm.} \\
        &+D_1^3(\tau)\Big\{ b_{\delta^3}+\frac{b_{\delta s^2}}{3}\big[ S(\vb*{k_1},\vb*{k_2})+2\,\text{perm.} \big]+F(\vb*{k_1},\vb*{k_2},\vb*{k_3}) \Big\}\Bigg\}\delta^{(1)}_{\vb*{k_1}}\delta^{(1)}_{\vb*{k_2}}\delta^{(1)}_{\vb*{k_3}},
    \end{split}
\end{equation}
where $\vb*{k_{ab}}=\vb*{k_a}+\vb*{k_b}$. "2~perm." are the $\sum_{\{X_2\}}$ expression repeated with the argument switching between 3 modes, $\vb*{k_1}$, $\vb*{k_2}$, and $\vb*{k_3}$. Notice that the integrands are symmetric under the permutation of $\vb*{k_\sigma}$'s.

All of these solutions are structured as integrals with time-dependent kernels and can be written in general as
\begin{equation}\label{eqn: Perturbative solution of the theoretical galaxy number density and velocity divergence}
    \begin{split}
        \delta_{g,\vb*{k}}^{(n)}(\tau)&=\int_{\vb*{k_1}} \dots \int_{\vb*{k_n}}\delta^{(3)}_D\left(\vb*{k}-\sum^n_{\sigma=1}\vb*{k_\sigma}\right)\Delta^S_{g,n}(\vb*{k_1},\dots ,\vb*{k_n};\tau)\prod^n_{\sigma=1}\delta^{(1)}_{\vb*{k_\sigma}}, \\
        \theta_{g,\vb*{k}}^{(n)}(\tau)&=\int_{\vb*{k_1}} \dots \int_{\vb*{k_n}}\delta^{(3)}_D\left(\vb*{k}-\sum^n_{\sigma=1}\vb*{k_\sigma}\right)\Theta^S_{g,n}(\vb*{k_1},\dots ,\vb*{k_n};\tau)\prod^n_{\sigma=1}\delta^{(1)}_{\vb*{k_\sigma}},
    \end{split}
\end{equation}
where the time-dependent kernels can always be separated,
\begin{equation}\label{eqn: kernel of the theoretical galaxy number density and velocity divergence}
    \begin{split}
        \Delta^S_{g,n}(\vb*{k_1},\dots ,\vb*{k_n};\tau)&=\sum^{N_n}_{\lambda=1}D_{g,n,\lambda}(\tau)\Delta^S_{g,n,\lambda}(\vb*{k_1},\dots ,\vb*{k_n}), \\
        \Theta^S_{g,n}(\vb*{k_1},\dots ,\vb*{k_n};\tau)&=\sum_{\{X_n\}}E_{X_n}(\tau)\Delta^S_{X_n}(\vb*{k_1},\dots ,\vb*{k_n}).
    \end{split}
\end{equation}
$N_n$ is the number of terms in the $n$th order integral form \eqref{eqn: explicit integral bias expansion}. The function $D_{g,n,\lambda}$ encapsulates the time dependence introduced by the growth factors from the EPT solutions, and the function $\Delta^S_{g,n,\lambda}$ holds the mode dependence from the EPT solutions and the bias operators. For example, the first-order kernel is constructed from $D_{g,1,1}(\tau)=b_\delta D_1(\tau)$ and $\Delta^S_{g,1,1}(\vb*{k_1})=1$. Since we assume the equivalence principle, the kernels for $\theta^{(n)}_{g,k}$ are the same as those for the dark matter $\theta^{(n)}_{k}$, which are $\Delta^S_{X_n}$.
\section{Leading contribution of the connected 4PCF}\label{apx: Leading Contribution of the Connected 4PCF}
The leading contribution of the connected 4PCF is the following,
\begin{equation}
    \begin{split}
        &\zeta(\vb*{s};\vb*{r_1},\vb*{r_2},\vb*{r_3}) = \expval{\Delta_g(\vb*{s})\Delta_g(\vb*{R_1})\Delta_g(\vb*{R_2})\Delta_g(\vb*{R_3})} \\
        & =\Big\{\expval{\Delta_g^{(1)}(\vb*{s})\Delta_g^{(1)}(\vb*{R_1})\Delta_g^{(1)}(\vb*{R_2})\Delta_g^{(3)}(\vb*{R_3})}+\expval{\Delta_g^{(1)}(\vb*{s})\Delta_g^{(1)}(\vb*{R_1})\Delta_g^{(3)}(\vb*{R_2})\Delta_g^{(1)}(\vb*{R_3})} \\
        &+\expval{\Delta_g^{(1)}(\vb*{s})\Delta_g^{(3)}(\vb*{R_1})\Delta_g^{(1)}(\vb*{R_2})\Delta_g^{(1)}(\vb*{R_3})}+\expval{\Delta_g^{(3)}(\vb*{s})\Delta_g^{(1)}(\vb*{R_1})\Delta_g^{(1)}(\vb*{R_2})\Delta_g^{(1)}(\vb*{R_3})}\Big\} \\
        &+\Big\{\expval{\Delta_g^{(1)}(\vb*{s})\Delta_g^{(1)}(\vb*{R_1})\Delta_g^{(2)}(\vb*{R_2})\Delta_g^{(2)}(\vb*{R_3})}+\expval{\Delta_g^{(2)}(\vb*{s})\Delta_g^{(2)}(\vb*{R_1})\Delta_g^{(1)}(\vb*{R_2})\Delta_g^{(1)}(\vb*{R_3})} \\
        &+\expval{\Delta_g^{(2)}(\vb*{s})\Delta_g^{(1)}(\vb*{R_1})\Delta_g^{(2)}(\vb*{R_2})\Delta_g^{(1)}(\vb*{R_3})}+\expval{\Delta_g^{(1)}(\vb*{s})\Delta_g^{(2)}(\vb*{R_1})\Delta_g^{(1)}(\vb*{R_2})\Delta_g^{(2)}(\vb*{R_3})} \\
        &+\expval{\Delta_g^{(2)}(\vb*{s})\Delta_g^{(1)}(\vb*{R_1})\Delta_g^{(1)}(\vb*{R_2})\Delta_g^{(2)}(\vb*{R_3})}+\expval{\Delta_g^{(1)}(\vb*{s})\Delta_g^{(2)}(\vb*{R_1})\Delta_g^{(2)}(\vb*{R_2})\Delta_g^{(1)}(\vb*{R_3})}\Big\} \\
        &+\mathcal{O}((\delta^{(1)})^8),
    \end{split}
\end{equation}
where we use a shorthand notation $\vb*{R_\sigma}=\vb*{s}+\vb*{r_\sigma}$ for clarity. The first and second brackets contain the $\expval{\Delta_g^{(1)}\Delta_g^{(1)}\Delta_g^{(1)}\Delta_g^{(3)}}$'s and $\expval{\Delta_g^{(1)}\Delta_g^{(1)}\Delta_g^{(2)}\Delta_g^{(2)}}$'s contribution, respectively.
\section{Alternative LOS definition for wide angle effect: centroid LOS}\label{apx: Alternative LOS definition for Wide Angle Effect}
Throughout this paper, we adopt the end-point LOS definition, in which the LOS towards one galaxy is chosen as the reference, and the other three LOS's are defined relative to it. This definition has been found to introduce a spurious asymmetry into the point correlation function, which contaminates the intrinsic dipole and octupole \cite{Beutler2019JCAP...03..040B}. A better definition that does not introduce such an artificial signal is the centroid definition, where the main LOS is the centroid position of the point correlation function in the configuration space,
\begin{equation}\label{eqn:centroid_LOS}
    \vb*{\hat{n}_C}=\frac{\vb*{l_C}}{\norm{\vb*{l_C}}}\quad;\quad \vb*{l_C}=\frac{1}{4}\left(\vb*{s}+\vb*{R_1}+\vb*{R_2}+\vb*{R_3}\right)=\vb*{s}+\frac{1}{4}\left(\vb*{r_1}+\vb*{r_2}+\vb*{r_3}\right).
\end{equation}
$\vb*{l_C}$ and $\vb*{R_\sigma}=\vb*{s}+\vb*{r_\sigma}$ are the position of the 4PCF centroid and the positions of the secondary galaxies relative to the observer, respectively. In other words, we integrate the 4PCF over the direction $\vb*{\hat{n}_C}$ instead of the LOS of the reference galaxy $\vb*{\hat{s}}$,
\begin{equation}
    \hat{\zeta}(\vb*{r_1}, \vb*{r_2}, \vb*{r_3})=\int d\Omega_{n_C}\,\zeta(\vb*{n_C};\vb*{r_1}, \vb*{r_2}, \vb*{r_3}).
\end{equation}
In this case, we can write the position of each point in 4PCF relative to the observer as
\begin{equation}
    \begin{split}
        \vb*{s}&=\vb*{l_C}-\frac{1}{4}\sum^3_{\sigma=1}\vb*{r_\sigma}, \\
        \vb*{R_\sigma}&=\vb*{l_C}-\frac{1}{4}\left(\sum^2_{\rho\neq\sigma}\vb*{r_\rho}-3\vb*{r_\sigma}\right);\;\sigma=\{1,2,3\}.
    \end{split}
\end{equation}
Then, the LOS can be written as $\vb*{\hat{n}_1}=\vb*{\hat{s}}=\frac{\vb*{s}}{\norm{\vb*{s}}}$, and $\vb*{\hat{n}_\sigma}=\frac{\vb*{R_\sigma}}{\norm{\vb*{R_\sigma}}}$. As in Section~\ref{sec: Wide-angle Effect: Theoretical Comeback}, we can express these local LOS as an infinite sum of Legendre polynomials using the generating function. For instance,
\begin{equation}
    \frac{1}{\norm{\vb*{l_C}-\frac{1}{4}\sum^4_{\sigma=1}\vb*{r_\sigma}}}=\frac{1}{l_C}\sum^\infty_{n=1}\left(\frac{r_3}{4\norm{\vb*{l_C}-\frac{1}{4}\sum^2_{\sigma=1}\vb*{r_\sigma}}}\right)^n\mathcal{P}_n\left(\vb*{\hat{n}_C}\cdot\vb*{\hat{r}_3}\right),
\end{equation}
where $l_C<r_\sigma$. So,
\begin{equation}\label{eqn:endpoint_LOS_in_centroid_LOS}
    \vb*{\hat{n}_1}=\vb*{\hat{s}}=\left(\vb*{\hat{n}_C}-\frac{1}{4}\sum^3_{\sigma=1}\left(\frac{r_\sigma}{l_C}\right)\vb*{\hat{r}_\sigma}\right)\sum^\infty_{n=1}\left(\frac{r_3}{4\norm{\vb*{l_C}-\frac{1}{4}\sum^2_{\sigma=1}\vb*{r_\sigma}}}\right)^n\mathcal{P}_n\left(\vb*{\hat{n}_C}\cdot\vb*{\hat{r}_3}\right)
    .
\end{equation}
Iteratively, we can apply the generating function to the inverse norm until we end up with an expression of $\vb*{\hat{n}_1}$ that is an infinite polynomial of
\begin{equation}
    l_C,r_1,r_2,r_3,\text{ and dot products between }\vb*{\hat{n}_C},\vb*{\hat{r}_1},\vb*{\hat{r}_2},\vb*{\hat{r}_3}.
\end{equation}
The same procedure can be applied to $\vb*{\hat{n}_\sigma}$, yielding the same infinite polynomials of terms above with different coefficients. So, the cosine factor between the local LOS and mode vector,
\begin{equation}
    \mu_{\rho\sigma,C}=\vb*{\hat{k}_\rho}\cdot\vb*{\hat{n}_\sigma}\; ; \; \rho,\sigma\in\{1,2,3,4\},
\end{equation}
can be expressed as an infinite polynomial of the following projection factors,
\begin{equation}
    \begin{split}
        \mu_{\vb*{k_\rho}C}=\vb*{\hat{k}_\rho}\cdot\vb*{\hat{n}_C}&\;,\; \nu_{\rho\sigma,C}=\vb*{\hat{k}_\rho}\cdot\vb*{\hat{r}_\sigma}\;,\;\rho_{\sigma,C}=\vb*{\hat{n}_C}\cdot\vb*{\hat{r}_\sigma}\; ;\rho,\sigma\in\{1,2,3,4\} , \\
        &\text{ and dot products between }\vb*{\hat{r}_1},\vb*{\hat{r}_2},\vb*{\hat{r}_3},
    \end{split}
\end{equation}
ignoring the norms, $l_C,r_1,r_2,r_3$. This set of projections is analogous to what we have obtained with the end-point LOS definition (see expression~\eqref{eqn: wide-angle projection factors}), substituting the reference galaxy LOS, $\vb*{\hat{s}}$, with the centroid LOS, $\vb*{\hat{n}_C}$. Once we substitute this expression of LOS into $\zeta$, and integrate over $d\Omega_{n_C}$, we can factor the dot products between $\vb*{\hat{r}_1}$, $\vb*{\hat{r}_2}$, and $\vb*{\hat{r}_3}$ out of the integral. Ultimately, we are left with the same integral of the analogous geometrical factors, $\mu_{\vb*{k_\rho}C}$, $\nu_{\rho\sigma,C}$, and $\rho_{\sigma,C}$, as in Section~\ref{sec: Wide-angle Effect: Theoretical Comeback}. Therefore, the same calculation can be used to assess the PV of $\hat{\zeta}$, yielding the same angular momentum selection rule~\eqref{eqn: s constraint}.

Intuitively, the reason why the conclusion remains the same is that the difference between the end-point and centroid LOS is in the form of dot products between the LOS $\vb*{\hat{n}_C}$ and $\vb*{\hat{r}_\sigma}$'s. This can be seen in equation~\eqref{eqn:endpoint_LOS_in_centroid_LOS}, which is the end-point LOS, $\vb*{\hat{s}}$, expanded from the centroid LOS, $\vb*{\hat{n}_C}$. These dot products induce parity-odd modes of $\zeta$ that will be washed out by integrating over all $\vb*{\hat{n}_C}$, in a similar way as we demonstrated in Section~\ref{sec: Wide-angle Effect: Theoretical Comeback} with $\vb*{\hat{s}}$. Therefore, the correction terms arising from changing the LOS choice, $\vb*{\hat{n}_C}\leftrightarrow \vb*{\hat{s}}$, ultimately do not induce any spurious PV in $\hat{\zeta}$.
\section{Integral over N spherical harmonics}\label{apx: Integrals of Spherical Harmonics}

This appendix derives the selection rule governing the integral of a product of N spherical harmonics over the sphere,
\begin{equation}
    \int d\Omega_s \prod^N_{n=1}Y_{j_n,m_{j_n}}(\vb*{\hat{s}}).
\end{equation}
Writing the spherical harmonic explicitly as
\begin{equation}
    Y_{j,m_j}(\theta,\varphi) = \sqrt{\frac{2j+1}{4\pi}\frac{(j-m_j)!}{(j+m_j)!}} \mathcal{P}^{m_j}_j(\cos{\theta})e^{im_j\varphi},
\end{equation}
where $\mathcal{P}^{m_j}_j$ is the \emph{associated Legendre polynomial}. Then, the integral is proportional to
\begin{equation}
    \int d\Omega_s \prod^N_{n=1}Y_{j_n,m_{j_n}}(\vb*{\Hat{s}}) \propto \int_0^{2\pi} d\varphi \mathbf{exp}\left[i\sum_{n=1}^N m_{j_n} \varphi \right] \int_{-1}^1 d\cos{\theta} \prod^N_{n=1} \mathcal{P}^{m_{j_n}}_{j_n}(\cos{\theta}).
\end{equation}
The conditions for non-vanishing follow from each factor in turn. The azimuthal integral over $\varphi$ requires
\begin{equation}\label{eqn: first constraint}
    \sum_{n=1}^N m_{j_n} = 0.
\end{equation}
The polar integral over $\cos{\theta}$ is non-vanishing only if the integrand is an even function of $\cos{\theta}$, since the integration domain $[-1,1]$ is symmetric. Using the parity property of the associated Legendre polynomials,
\begin{equation}\label{eqn: lengendre polynomials parity}
    \mathcal{P}^{m_j}_j(-x) = (-1)^{j-m_j}\mathcal{P}^{m_j}_j(x),
\end{equation}
the product $\prod^N_{n=1} \mathcal{P}^{m_{j_n}}_{j_n}(\cos{\theta})$ is even if and only if
\begin{equation}\label{eqn: cosine integral constraint}
    \sum^N_{n=1} \left( j_n-m_{j_n} \right) = \text{even}.
\end{equation}
Combining this with constraint \eqref{eqn: first constraint}, the two conditions together imply
\begin{equation}\label{eqn: second constraint}
    \sum^N_{n=1}j_n = \text{even}.
\end{equation}
The integral over $N$ spherical harmonics is therefore non-vanishing only when conditions \eqref{eqn: first constraint} and \eqref{eqn: second constraint} are satisfied simultaneously.

Two remarks are in order. First, the constraint \eqref{eqn: second constraint} is unchanged when some of the spherical harmonics are complex-conjugated. The conjugate flips the sign of the projection index $m_{j_n}$ and introduces just an extra constant factor,
\begin{equation}
    Y^*_{j_n,m_{j_n}} =(-1)^{m_{j_n}}Y_{j_n,-m_{j_n}}.
\end{equation}
$(-1)^{m_{j_n}}$ is a constant and not a parity factor under the reflection of $\cos{\theta}$. Hence, it only changes the signs of $m_{j_n}$ in condition \eqref{eqn: first constraint} and property~\eqref{eqn: lengendre polynomials parity} simultaneously. Therefore, the retrieval of constraint~\eqref{eqn: second constraint} is still protected. Second, condition  \eqref{eqn: cosine integral constraint}, and therefore \eqref{eqn: second constraint}, holds for any symmetric integration domain $[-a,a]$ with $0<a \leq 1$, not only for the full domain $[-1,1]$ of the cosine of elevation angle $\cos{\theta}$. Although the full revolution of the azimuthal angle is still required, i.e., the integration measure of $\varphi$ must be $[0,2\pi]$, this insight relaxes the full-sky condition considerably; we just need the survey region to be symmetric in elevation angle to remove PV in 4PCF from the RSD effect.

\bibliographystyle{JHEP}
\bibliography{references}

\end{document}